\documentstyle[epsfig,bezier]{ioplppt}
\jl{6}
\eqnobysec

\def\e{{\bf e}}
\def\g{\gamma}

\def\ab{\overline{\cal A}}

\def\agb{{\overline {{\cal A}/{\cal G}}}}

\def\G{{\cal G}}
\def\Gb{{\overline \G}}
\def\H{{\cal H}}

\def\Ho{{\H}^o}

\def\f{\begin{equation}}

\def\uno{{\scriptscriptstyle (\!1\!)}}
\def\due{{\scriptscriptstyle (\!2\!)}}
\def\unodue{{\scriptscriptstyle (\!i\!)}}

\newcommand{\ket}[1]{\mid #1 \rangle}

\begin{document}

\begin{flushright}\mbox{
\vbox{{\small \hbox{CGPG-97/4-3}
              \hbox{U.Parma UPRF 97-04}
              \hbox{gr-qc/9703090}
}}}
\end{flushright}
\vspace{-1cm}


\title{Matrix Elements of Thiemann's Hamiltonian 
       Constraint in Loop Quantum Gravity
}[Matrix elements of Thiemann's Hamiltonian]
\author{
Roumen Borissov \dag\ftnote{4}{E-mail address: borissov@phys.psu.edu}, 
Roberto De Pietri \ddag\ftnote{5}{E-mail address: depietri@pr.infn.it}   
and Carlo Rovelli \S\dag\ftnote{8}{E-mail address: rovelli@pitt.edu}}
\address{
\dag\ Center for Gravitational Physics and Geometry,  
      The Pennsylvania State University, University Park, 
      PA 16802, USA.}
\address{
\ddag\  Dipartimento Di Fisica, Universit\`a di Parma and 
        I.N.F.N. Gruppo Collegato di Parma,
        I-43100 Parma, Europe.}
\address{
\S\  Department of Physics and Astronomy,  
     University of Pittsburgh, Pittsburgh, PA 15260, USA;\\
     Erwin Schr\"odinger Institute, A-1090 Vienna, Europe.}


\begin{abstract} 

We present an explicit computation of matrix elements of the
hamiltonian constraint operator in non-perturbative quantum
gravity.  In particular, we consider the euclidean term of
Thiemann's version of the constraint and compute its action
on trivalent states, for all its natural orderings.  The
calculation is performed using graphical techniques from the
recoupling theory of colored knots and links.  We exhibit
the matrix elements of the hamiltonian constraint operator
in the spin network basis in compact algebraic form.

\end{abstract}

\maketitle




\section{Introduction}

In general relativity, the local dynamics of the
gravitational field is governed by a constraint rather than
by a hamiltonian \cite{PAMD_ADM}.  Accordingly, in
non-perturbative quantum gravity the quantum dynamics is
governed by a constraint operator \cite{WheelerDeWitt},
usually denoted the ``hamiltonian constraint operator''
(HCO) or the Wheeler-DeWitt operator.  The construction and
the analysis of this operator is a central problem in
quantum gravity.  For a long time, only formal or seriously
ill-defined versions of the HCO were known.  Concrete
computations were thus impossible beyond semiclassical or
minisuperspace approximations, and a predictive theory of
the non-perturbative quantum dynamics of gravity was
consequently lacking.

The situation is perhaps improving with the development of
the loop version of non-perturbative quantum gravity
\cite{loops}, derived from the introduction of the
Sen-Ashtekar variables \cite{Sen,AbhayVAR} (now evolved into
the real Barbero variables \cite{Barbero,Real}).  This
approach, denoted ``loop quantum gravity'', has provided a
rigorous framework for describing quantum geometry (for
recent reviews, and complete references see
\cite{Ascona,DePiRo96,Reviews}).  As far as kinematics is
concerned, the approach has yielded striking predictions: in
particular, geometric operators corresponding to volumes and
areas have been shown to have computable discrete
spectra. This result was first obtained by Rovelli and
Smolin in \cite{RovelliSmolin95} and later re-derived and
sharpened with various alternative techniques
(\cite{Loll,VolumeAL,LeFrRo}; see \cite{Lewandowski97} and
references therein); its physical significance was first
asserted in \cite{Rovelli93}.

As far as dynamics is concerned, an extensive array of
investigations of the HCO, starting from the pioneering work
of Jacobson and Smolin \cite{JacSmo} and continuing in
\cite{loops,Hamilt,RovelliSmolin94,Bor96,Ezawa}, has
recently culminated with the construction, due to Thiemann,
of a mathematically well-defined operator \cite{Thiemann}.
This construction is already being used in various
directions of research, for instance in the search for exact
solutions to the constraints \cite{Pullin}, or as a key
ingredient of a spacetime formulation of the theory
\cite{RovReis,spacetime}.  In this last context, the matrix
elements of the HCO play a role analogous to the
coefficients of the vertices in the Feymann
diagrams. Modifications of the operator have been considered
in \cite{Smo96}.

Thiemann's operator, however, is defined in a rather
involved manner: its matrix elements and explicit action are
far from easy to evaluate.  This difficulty obstructs the
research directions mentioned and renders the analysis of
the operator quite difficult.  In this work, we address this
difficulty directly by studying Thiemann's operator in
detail, and showing how its matrix elements can be computed
explicitly.

To perform the computation, we take advantage of two key
techniques.  First, we work in the spin network basis,
introduced in quantum gravity in \cite{Rovelli95a,LewSpin},
and fully developed in \cite{Baez} (see \cite{Lee} for an
interesting historical reconstruction of these
developments).  Second, we use the technique developed in
\cite{DePiRo96}, based on the (Kauffman-Lins) tangle
theoretical version of recoupling theory \cite{Kauffman94}
(whose utility for quantum gravity was first suggested in
\cite{LeeRoumSeth}).  Since Thiemann's operator is gauge
invariant but is defined in terms of gauge non-invariant
operators, we extend here these techniques to gauge
non-invariant states \cite{Thiemann96,AshLe97a}.

Furthermore, we take full advantage of the complete
equivalence between the ``loop representation''
\cite{loops,DePiRo96} and the ``connection representation''
\cite{AshIsh,ALewKnots,Thiemann} of loop quantum gravity
(proven in detail in \cite{DePietri97}; see also
\cite{Lewandowski97,Thiemann96}).  It is this equivalence
that allows us, for example, to use graphical calculus in a
rigorous way in dealing with states that are functions of
generalized connections.

We study here only the euclidean term of the HCO, but we
expect that the techniques we use can be extended to the
lorentzian term as well.  Also, we restrict our attention to
trivalent vertices: extension to any valence is
straightforward.  Thiemann defines two versions of the HCO:
symmetric and non-symmetric.  Here, we consider both
operators, and show that they exhaust all ``natural''
orderings of the constraint.

We do not discuss the physical correctness of Thiemann's
operator. In particular, we do not address the objections
that have been raised against it, concerning for instance
the state dependence of the regularization, the strong
commutativity (which shows consistency, but seems to fail to
reproduce the classical constraint algebra \cite{Marolf}) or
the possible need for adding terms to the operator suggested
for instance in \cite{Smo96,RovReis}. Rather, we simply
study the operator as defined by Thiemann, and show how its
matrix elements can be computed, using recoupling theory.
This analysis is likely to be needed before attempting a
physical evaluation and, in any case, it is relevant even if
modifications to the Thiemann operator are to be made.

In \sref{sec:StateSpace} we define the mathematical
framework.  In particular, we discuss the generalization of
the spin networks to the gauge non-invariant case.  In
\sref{sec:LoopRepr} we show in detail why graphical methods
can be used in the present context, following
\cite{DePietri97}.  \Sref{sec:Thiemann} reviews Thiemann's
construction of the hamiltonian constraint. Section 5
presents our main calculations.  For completeness, we have
included three appendices: A discussion of the volume
operator (Appendix A); technical details on the application
of the graphical calculus in the context of the connection
representation (Appendix B); and the basic formulae and
concepts of recoupling theory (Appendix C).


\section{Loop quantum gravity.
\label{sec:StateSpace}}

We start with a 4-manifold ${\cal M}$ diffeomorphic to
$\Sigma \times R$, where $\Sigma$ is a compact 3-manifold.
The classical phase space of general relativity can be taken
as the space of the canonical pairs $(A_{a}^{i},
\widetilde{E}^{a}_{i})$ of smooth fields over $\Sigma$.
$A_{a}^{i}$ is a connection 1-form on $\Sigma$ taking values
in the Lie algebra of $SU(2)$.  Its conjugate momentum
$\widetilde{E}^{a}_{i}$ is a vector density of weight one on
$\Sigma$ which takes values in the dual of $su(2)$.  We use
the dimensional constants
\begin{eqnarray*}
  G &=& \frac{16 \pi G_{\rm Newton}}{c^3}, \\ l_0^2 &=& \hbar G =
  16 \pi l^2_{\rm Plank} =\frac{16 \pi \hbar G_{\rm
  Newton}}{c^3} ,
\end{eqnarray*}
where $G_{\rm Newton}$ is Newton's gravitational constant
and $l_{\rm Planck}$ is the Planck's length. The fundamental
Poisson bracket is:
\begin{equation}
 \{ A_{a}^{i}(x), \widetilde{E}^{b}_{j} (y) \} = G\ \delta^i_j
   \delta^b_a ~\delta^3\!(x,y).
\end{equation}
The canonical variables are subject to three sets of
constraints -- the $SU(2)$ Gauss constraint, the
diffeomorphism constraint and the hamiltonian constraint.
The space $\cal A$ of the $SU(2)$ connections on $\Sigma$ is
the classical configuration space of general relativity
\cite{Barbero,Real}.

The quantum theory is defined as a linear representation of
an algebra of classical observables on a Hilbert space
$\Ho$.  There exist now a number of constructions of the
Hilbert space $\Ho$ needed for defining this representation
(see \cite{Ascona}), which have all turned out to be unitary
equivalent.  We will start here from the Schr\"odinger-like
representation of $\Ho$ as an $L_{2}$ space of functionals
over the configuration space \cite{ALewKnots}.  More
precisely, as it is usually the case in quantum field theory
\cite{GlimmJaffe}, $\Ho$ is defined as a space of
functionals over a suitable completion $\ab$ of the
configuration space $\cal A$, formed by ``generalized
connections'', or ``distributional connections''.

\subsection{State space}

In order to define a generalized connection, we need some
definitions.  An {\it edge} $\e$ is an oriented,
1-dimensional sub-manifold of $\Sigma$ with two boundary
points, called {\it vertices}, which is analytic everywhere,
including the vertices.  A generalized connection $\bar{A}$
in $\ab$ is defined \cite{VolumeAL,Ba94} as a map that
assigns an element $\bar{A}(\e)$ of $SU(2)$ to each oriented
edge $\e$ in $\Sigma$, satisfying the following
requirements: i) $\bar{A}(\e^{-1}) = (\bar{A}(\e))^{-1}$;
and, ii) $\bar{A}(\e_2\circ \e_1) = \bar{A}(\e_2)\cdot
\bar{A}(\e_1)$.  Here $\e^{-1}$ is obtained from $\e$ by
reversing its orientation, $\e_2\circ \e_1$ denotes the
composition of the two edges (obtained by connecting the end
of $\e_1$ with the beginning of $\e_2$) and
$\bar{A}(\e_2)\cdot\bar{A}(\e_1)$ is the composition in
$SU(2)$.  Any smooth connection $A$ is also a generalized
connection: the group element that it assigns to an edge
$\e$ is the parallel transport along $\e$, that is
$\bar{A}(\e) = h_\e(A):= {\cal P}\exp(-\int_\e A)$.
Accordingly, we refer to $\bar{A}(\e)$ as ``the parallel
transport of the generalized connection $\bar A$'', in the
same spirit in which we write $\delta(f) = \int dx \delta(x)
f(x)$ for the Dirac $\delta$-``function''.  One can view the
space $\ab$ of the generalized connections as the closure of
the configuration space $\cal A$ in a suitable topology.

There exist a natural measure $\mu^o$ on $\ab$, induced by
the Haar measure on $SU(2)$, denoted the
Ashtekar-Lewandowski-Baez measure.  Its explicit form can be
given as follows.  A {\it graph} $\gamma$ in $\Sigma$ is a
collection of edges such that if two distinct edges meet,
they do so only at the vertices.  A vertex $v$ is called
$n$-valent or of valence $n$ if it has $n$ adjacent edges.
Fix a graph $\g$ with $N$ edges $\e_1, ..., \e_N$, and
consider the functions $\Psi_{\g,\psi}$ on $\ab$ of the form
$\Psi_{\g,\psi}(\bar{A}) = \psi (\bar{A}(\e_1),...,
\bar{A}(\e_N))$ for some smooth function $\psi$ on
$[SU(2)]^N$.  These are cylindrical functions, since they
depend only on a finite subset of the infinite
``coordinates'' of $\ab$.  The measure ${\mu}^o$ is defined
by (its cylindrical projection)
\begin{equation}\fl 
  \int_{\overline{\cal A}}d\mu^o(\bar{A})\ 
               \Psi_{\g,\psi} (\bar{A}) = 
  \int_{SU(2)^{n}}d^{n}\mu_{H}(g_1,\ldots,g_N) \ 
  \psi(g_1,\ldots ,g_n),
\end{equation}
where $\mu_{H}$ is the Haar measure on $SU(2)$.  We use the
Hilbert space $\Ho := L^2(\ab, d\mu^o)$ of square-integrable
functions on $\ab$ as the (kinematical) Hilbert space of the
theory.  (``Kinematical'' in the sense of Dirac -- the
physical Hilbert space is then obtained by imposing the
quantum constraints.)  The elements of $\Ho$ can be viewed
as wave functions of (generalized) connections, analogous to
the wave functions of distributional fields in the
Schr\"odinger representation of a free field theory. The
cylindrical functions are dense in $\Ho$; they represent
``typical'' normalizable states.

Let us now discuss the SU(2) gauge invariance.  A
generalized gauge transformation is a map $g$ which assigns
an $SU(2)$ element $g(x)$ to each point $x$ of $\Sigma$.  It
acts on $\bar{A}$ at the end points of edges:
$\bar{A}(\e)\rightarrow g(v_+)^{-1} \cdot\bar{A}(\e)\cdot
g(v_-)$, where $v_{-}$ and $v_+$ are the beginning and the
end point of $\e$.  The space of the equivalence classes of
generalized connections under generalized gauge
transformations is denoted $\agb$.  The measure $\mu^o$
induces a natural measure $\tilde{\mu}^o$ on $\agb$,
obtained by the push-forward of $\mu^o$ under the projection
map that sends $\ab$ to $\agb$.

The gauge invariant functions form the $SU(2)$ gauge
invariant Hilbert space $\tilde\Ho := L^2(\agb,
d\tilde{\mu}^o)$.  These are the solutions of the $SU(2)$
constraint.  Since the spaces under consideration are
compact and measure-normalized, we can regard $\tilde\Ho$ as
the gauge invariant sub-space of the Hilbert space $\Ho :=
L^2(\ab, d\mu^o)$ \cite{VolumeAL,Ba94,MaMo}.  Thus, the
$SU(2)$ gauge invariant quantum states can be expressed as
complex-valued, square-integrable functions on $\agb$, or,
equivalently, as $\Gb$-invariant square-integrable functions
on $\ab$.

\subsection{Spin network basis}

In order to work on a Hilbert space it is convenient to have
a basis. A very convenient basis in $\Ho$ is formed by the
spin network states.  These are obtained by using the fact
that functions on $SU(2)$ can be expanded in irreducible
representations.  Given $N$ irreducible representations
$j_1, ..., j_N$ of $SU(2)$, we define an invariant tensor
${c^{m_{k+1}...m_N}}_{m_1...m_k}$ as a multi-linear map from
$\bigotimes_{I=1}^{k} j_I$ to $\bigotimes_{I=k+1}^{N} j_I$
that transforms covariantly, namely such that:
\begin{eqnarray}\fl
{c^{n_{k+1}...n_N}}_{n_1...n_k} = \\
\lo{=}
j_{k+1}(g)_{m_{k+1}}^{n_{k+1}}\,... \,j_N (g)_{m_N}^{n_N} 
{c^{m_{k+1}...m_N}}_{m_1...m_k}\, j_1(g^{-1})^{m_1}_{n_1}\, ... 
\, j_N(g^{-1})_{m_k}^{n_k}
, \nonumber
\end{eqnarray}
for arbitrary $g \in SU(2)$, where $j_I(g)$ is the matrix
representing $g$ in the representation $j_I$.  Such an invariant
tensor is also called an {\it
intertwining tensor} from the representations $j_{1},...,j_k$
to $j_{k+1},...,j_N$.  All invariant tensors are given by the
standard Clebsch-Gordan theory.

An {\em extended spin network} $s$ is defined as a quintuple
$s=(\gamma,\vec{j},\vec{c}, \vec{\rho},\vec{M})$ consisting
of:
\begin{itemize}
\item A graph $\gamma$ in $\Sigma$;
\item A labeling $\vec{j}:=(j_1,..,j_N)$ of the edges
      $\e_1,...,\e_N$ of the graph $\gamma$ with non-trivial
      irreducible representations of $SU(2)$. We refer to
      $p_I$, equal to twice the spin of the representation
      $j_I$, as the {\em color\/} of the edge $\e_I$;
\item A labeling $\vec{\rho}:=(\rho_1,...,\rho_V)$ of the
    vertices $v_1,...,v_V$ of $\gamma$ with (possibly
    trivial) irreducible representations of $SU(2)$, with
    the constraint that for every vertex $v_{\alpha}$ the
    representation $\rho_{\alpha}$ sits in the tensor
    product of the representations assigned to the edges
    adjacent to $v_{\alpha}$;
\item A labeling $\vec{c}=(c_1,...,c_V)$ of the vertices
     $v_1,...,v_V$ of $\gamma$ with invariant tensors; more
     precisely, assigned to a vertex $v_{\alpha}$ is an
     intertwining tensor $c_{\alpha}$ from the
     representations assigned to incoming edges and
     $\rho_{\alpha}$ to the representations assigned to the
     outgoing edges;
\item A labeling $\vec{M} :=(M_1,...,M_V)$ of the vertices
     $v_1,...,v_V$ of $\gamma$ which assigns a vector
     $M_{\alpha}$ in the representation $\rho_{\alpha}$ to
     every vertex $v_{\alpha}$.
\end{itemize}

The (extended) spin network state $\Psi_{s}$ is defined as
follows.  Consider the ``parallel propagator'' $\bar
A(\e_I)$ of $\bar A$ along each edge $\e_I$ of $\gamma$, in
the representation $j_I$. Contract these parallel
propagators at each vertex $v_\alpha$, using the invariant
tensor $c_\alpha$.  Contract the leftover indices of
$c_\alpha$ at each vertex $v_\alpha$ with the vector
$M_\alpha$ assigned to the vertex.  Formally:
\begin{equation}
\label{2.1}
\Psi_{s}\,(\bar{A}):= \big[\bigotimes_{I=1}^N\,\,
j_I(\bar{A}(\e_I))\otimes\bigotimes_{\alpha=1}^V
M_{\alpha}\big]\,\cdot\, \big[\otimes_{\alpha=1}^V
c_{\alpha}\big].
\end{equation}
Here `$\cdot$' stands for contracting, at each vertex
$v_{\alpha}$ of $\gamma$, the upper-indices of the matrices
corresponding to all the incoming edges, the lower indices
of the matrices assigned to all the outgoing edges and the
upper index of the vector $M_{\alpha}$ with all the
corresponding indices of $c_{\alpha}$.  $\Psi_{s}$ is a
$C^\infty$ cylindrical function on $\ab$.

The introduction of the vectors $M$ has the sole purpose of
avoiding the choice of a preferred basis in the $SU(2)$
representations.  If we introduce such a a preferred basis,
we can replace the $M$'s with the specification of a basis
vector.  In other words, the extended spin network states
have a free index at each vertex $v_\alpha$, living in the
representation $\rho_{\alpha}$.  Below, we will find it more
convenient to use the physicist-style indices, rather than
the mathematician-style $M$ notation.

For any spin network $s$, $\Psi_{s}$ is a function on $\ab$
which is square-integrable with respect to the measure
$\mu^o$.  We can fix a basis in the (finite dimensional)
space of the invariant tensors at each vertex, and a basis
in each representation $\rho_{\alpha}$, and restrict
$\vec{c}$ and $\vec{M}$ to basis elements.  The resulting
set of spin network states form a basis in $\Ho$.

It is easy to see that a spin network state $\Psi_{s}$ is
gauge invariant if and only if all the representations in
$\vec\rho$ are trivial.  The gauge invariant spin network
states are therefore characterized solely by the graph
$\gamma$, the labelings $\vec{j}$ of the edges and the
intertwiners $\vec{c}$ at the vertices.  The inclusion of
the representations $\vec\rho$ and the vectors $\vec{M}$ at
the vertices yields the extension of the spin network
technology to the gauge non-invariant states, namely the
extension from $\tilde\Ho$ to $\Ho$.

\subsection{Loop states}

An {\it open loop\/} $\l=(\e_1,\ldots\e_{n},M_{-},M_{+})$ is
defined by an oriented open line in $\Sigma$ formed by a
sequence of edges $\e_1,\ldots\e_{n}$, and by two vectors
$(M_{-},M_{+})$ in the fundamental representation of
$SU(2)$.  A {\it loop\/} $\alpha=(\e_1,\ldots\e_{n})$ is
defined as a closed line formed by a sequence of edges.  An
{\it open loop state\/} is defined by $\Psi_l(\bar{A})=
M_{-}\cdot j(\bar{A}(\e_1))\cdot \ldots \cdot
j(\bar{A}(\e_n))\cdot M_{+}$.  A {\it loop state\/} is
defined as $\Psi_{\alpha}(\bar{A}) =
Tr[j(\bar{A}(\e_1))\cdot \ldots \cdot j(\bar{A}(\e_n))]$,
where $j$ is the fundamental (spin 1/2) representation of
$SU(2)$.  Sums of products of open loop states span $\Ho$,
and sums of products of loop states span $\tilde\Ho$.
Products of loop states are called {\it multiple loop
states}.  Thus the multiple loop states span $\tilde\Ho$;
however, they do not form a true basis, but only an
overcomplete basis.  This is the overcomplete basis that was
used in loop quantum gravity before the introduction of the
spin network true basis.

The relation between spin network states and loop states is
as follows.  Standard $SU(2)$ representation theory shows
that the spin network states can all be obtained as linear
combinations of products of open loop states. This is
because all representations of $SU(2)$ can be obtained by
tensoring the fundamental representation.  Explicitly, these
linear combinations are obtained by ``symmetrizing the
loops'' along each edge as illustrated in
\cite{Rovelli95a}. Conversely, every product of fundamental
representations can be decomposed in irreducible
representations, and therefore multiple loop states can be
expanded in the spin network basis.  Thus, the decomposition
of quantum states in the spin network basis is the
decomposition of functions on $SU(2)$ in irreducible
representations.  See \cite{DePiRo96,Rovelli95a} for
details.

\subsection{Operators}

We shall not give here a complete construction of the
operator algebra defined on $\Ho$, but only a few remarks on
the operators that enter the Thiemann's definition of the
HCO.  These are the parallel transport operator and the
volume operator.  For each edge $\e$, the parallel transport
operator $\hat{h}[\e]$ (with a free spin 1/2 index at each
end of $\e$) represents the classical quantity $h_\e(A):=
{\cal P}\exp(-\int_\e A)$, that is, the parallel propagator
of the connection along $\e$.  This operator is diagonal,
and is given by multiplication by $\bar A(\e)$:
\begin{equation}
  \left[\hat{h}[\e]\,\Psi\right]\!(\bar A) 
      = \bar A(\e)\, \Psi(\bar A).
  \label{def_h}
\end{equation}

The action of this operator on the spin network basis is not
immediate.  We obtain it by decomposing the state
$\hat{h}[\e]\Psi_{s}$ in the spin network basis, and this
has to be worked out explicitly using \eref{2.1}.  For
instance, if $\e$ and the graph $\gamma$ do not intersect,
then we have immediately that
$\hat{h}[\e]\Psi_{s}=\Psi_{(s\cup\e)}$, where $(s\cup\e)$ is
the spin network obtained by adding the (disconnected) edge
$\e$ to $\gamma$, and coloring it and its two open vertices
with the spin-1/2 representation.  As a second example,
assume that $\e$ sits entirely on the interior of one of the
edges of $s$, say with label $j$.  The decomposition of
$\hat{h}[\e]\Psi_{s}$ in spin network states can then be
obtained using the fact that the tensor product of two
$SU(2)$ representations is a sum of irreducible
representations.  Since $\bar A(\e)$ is in the 1/2
representation, we have in this case
\begin{equation}
\hat{h}[\e]\Psi_{s} = a_{+} \Psi_{s_{+}}+ a_{-} \Psi_{s_{-}}
	\label{def_h2}
\end{equation}
where $s_{\pm}$ are the spin networks obtained in the
following way.  Add two vertices to $s$, in correspondence
to the end points of $\e$.  Label the edge between these two
vertices by $j\pm 1/2$.  The representation $\rho$ on the
two new vertices is 1/2, and the intertwiner is the unique
one on the three representations $j, j\pm1, 1/2$.  The free
indices on the new vertices are the free indices of the
operator, and, finally, the coefficients $a_{\pm}$ are the
two Clebsch-Gordan coefficients of the expansion of
$j\otimes 1/2$. (See appendix C.)  More complicated cases
can be worked out in a similar way.

The rest of the elementary operators defined on $\Ho$
contain $\widetilde{E}^{a}_{i}$.  These can be defined using
the $T^n$ loop operators technology \cite{loops}, or using
left-invariant vector fields on the group
\cite{AshIsh,ALewKnots}.  Thiemann's HCO depends on the
volume, which is a composite operator based on such
elementary operators.

The volume operator has been constructed by Rovelli and
Smolin in \cite{RovelliSmolin95} using the $T^{n}$ loop
operators technology.  This construction is based on a
regularization and quantization of the classical volume.
(An important clarification role was played by Loll's work
\cite{Loll} using a lattice regularization
\cite{LollRETICOLO}.)  Later, Ashtekar and Lewandowski in
\cite{VolumeAL} used the left-invariant vector fields
technology to construct a ``mathematically natural''
operator $V_{AL}$. They presented it as ``closely related''
\cite{VolumeAL} to the construction of Rovelli and Smolin
and to the classical volume.  Lewandowski has clarified the
relation between the two operators, showing that the
difference between the two can be interpreted as due to
different regularization choices -- not to the technique
employed (both operators can be constructed in each
approach) \cite{Lewandowski97}.  See Appendix A.  The two
operators have the same action on generic trivalent vertices
(tangents of adjacent edges non coplanar), which is the only
case we consider here.  In the following we need the
explicit action of the volume operator.  This has been
worked out by DePietri and Rovelli in \cite{DePiRo96}. See
Appendix A.  Thiemann has also derived various matrix
elements of the volume operator in \cite{ThimVolume} using a
different technique and confirming the results of
\cite{DePiRo96}.


\section{Graphical binor representation
\label{sec:LoopRepr}}

In \cite{DePietri97} it was shown that computations in the
connection representation of quantum gravity can be
performed using the graphical Penrose binor calculus
\cite{binor} (as originally suggested by Smolin).  The main
ingredients of this calculus are summarized in
\ref{ap:Binor}.  Furthermore, it was shown in
\cite{DePietri97} that this graphical binor calculus is
fully equivalent to the loop representation developed in
\cite{loops,RovelliSmolin95,Rovelli95a,DePiRo96}.  These
results were obtained in the gauge-invariant context, but
there is no restriction to gauge invariant functions in the
rules of Penrose binor calculus, therefore they extend
immediately to the case of gauge non-invariant cylindrical
functions.  Let us review these results.

  \begin{figure}[t]
  \begin{indented}\item[]
  {\mbox{\epsfig{file=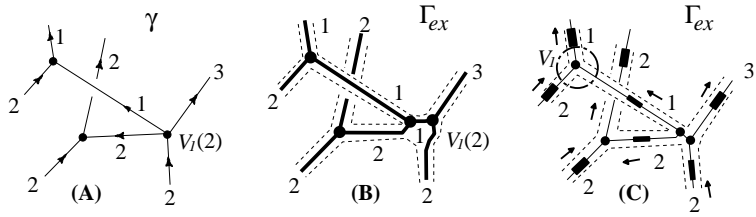,width=10cm}}}
  \end{indented}
  \caption{A spin network states and its graphical representations.
  \label{SpinNET}}
  \end{figure}

Following \cite{DePietri97}, we associate a graphical
representation {\it a' la\/} Penrose to each cylindrical
function $\Psi_{\gamma,\psi} $\footnote{More precisely we
define the graphical binor representation of polynomial
cylindrical functions, i.e, functions that are polynomial in
the $SU(2)$ matrices $g_\e = h_\e(\bar{A})$ associated to
each edge $\e$ of $\gamma$.}.  As a first step, we considers
a projection of the graph $\gamma$ on a plane and construct
its extended planar graph $\Gamma_{ex}$ (see
\fref{SpinNET}(B) and \cite{DePiRo96}).  The extended planar
graph is a two dimensional ``thickening out'' the graph: an
oriented two-dimensional surface with the topology of
$\gamma\times[0,1]$.  Next, we expand the cylindrical
function as a sum of products of irreducible representation
on the edges. For each term of the sum, we insert the
Penrose graphical representation of the matrices
$j_I(\bar{A}(\e_I))$ in the corresponding edges $\e_I$ of
the extended planar graph, using equation (\ref{eq:5}).
Next, we insert the graphical representation of the
contractors $c_\alpha$ in each vertex $v_\alpha$, using
equations\ (\ref{eq:6}) and (\ref{eq:7}).  Finally, lines
corresponding to summed dummy indices are connected, with an
additional factor $(-1)^p$ for each trace (closing a loop);
here $p$ is the color of the representation in which the
index lives (see \eref{eq:binorTRACE}).  It follows from
equation (\ref{eq:edgerevers}) that the planar graphical
representation is independent from the orientation of the
edges.  Finally, we notice that the Penrose convention of
indicating tensors by means of filled boxes is superfluous
in this context, because a colored line that runs along an
edge of the extended graphs always carries a parallel
propagator. Thus the information given by the boxes in
already contained in the fact that the line runs along an
edge, and we can drop the boxes.

In particular, we obtain in this manner a graphical
representation for each spin-network state.  Let us focus on
these.  First of all, in a spin network state we have a
single irreducible tensor on each edge.  Therefore each edge
carries a single colored line (recall: color is twice spin).
Second, at each vertex $v_\alpha$ we have an intertwining
tensor $c_\alpha$.  The graphical representation of each
invariant tensor can be given by means of a trivalent
tree-like graph (with open ends connected to the lines in
the adjacent edges).  This is a consequence of equation
\eref{eq:7}.  We denote the ``internal'' edges of this
trivalent decomposition as ``virtual edges''.  It is
important to keep in mind that virtual edges are drawn with
finite length on the extended planar graph, but they do not
represent finitely extended objects on the space manifold
$\Sigma$ -- they are simply a graphical representation of
the index pattern of combinations of the $SU(2)$ indices of
Clebsch-Gordan coefficients. They characterize an
intertwiner by giving its expansion in Clebsch-Gordan
coefficients.  Finally, at each vertex we have a
representation $\rho_\alpha$, and a vector $M_\alpha$ (or a
free index) sitting in this representation, contracted with
the invariant tensor.  Accordingly, in the graphical
representation of the spin network state we have one of the
virtual edges which is not attached to any external line.
This free end is missing if $\rho_\alpha$ is the trivial
representation, namely if the vertex is gauge invariant.

In \fref{SpinNET} we show, as an example, {\bf (A)}: a spin
network (in the connection representation) determined by an
oriented graph, and the labels of edges and vertices.  {\bf
(B)} A spin-network in the loop representation and its
extended planar graph (as in \cite{DePiRo96}).  ({\bf C})
The planar binor representation on the extended planar graph
of the cylindrical function given by the spin-network.

A simple advantage of this graphical representation is due
to the fact that planar graphs are easier to manipulate than
expressions with a large number of indices.  But the main
advantage is due to the peculiar nature of the tensors
involved, which are based on $SU(2)$ matrices.  These
tensors satisfy some basic relations that are translated
into simple graphical formulas.  The prototype of these is
equation (B.4), from which (as shown by Kauffman and Lins
\cite{Kauffman94}) all the other graphical relations can be
derived.  These graphical relations are the reason for the
great utility of graphical recoupling theory, well known and
much exploited in nuclear, atomic and molecular physics
(where $SO(3)$ plays a major role).  The important formulas
satisfied by the graphical representations of the quantum
states are summarized in Appendix C.  The calculations in
this paper rely heavily on these formulas.

As an example, consider the action of the holonomy operator
$\hat{h}[\e_I]$ on a spin network state $\Psi_{s}$, given in
(\ref{def_h}).  Consider the particular case in which $\e$
sits on the interior of an edge $\e'$ of $s$.  Following the
prescription given above, the graphical representation of
the r.h.s.~of (\ref{def_h}) is simply obtained by adding a
line colored 1, inside the (expanded) edge $e'$.  The
expansion of this state in spin network states is then
immediately given graphically by (\ref{eq:Recursion1}).

Notice that, following \cite{DePietri97}, we have fully
recovered in this way the ``pure'' loop representation of
quantum gravity, in which states and operators can be
defined without any reference to the Schr\"odinger-like
functions of connections. In the same fashion, the Dirac
energy-basis $|n\rangle$ representation of the harmonic
oscillator allows us to formulate and solve the theory
without reference to its Schr\"odinger representation.
However, here we have recovered the loop representation
starting from the gauge non-invariant connection
representation. Thus we have obtained the generalization of
the loop representation to gauge non-invariant states.  This
generalization is simply contained in the fact that the spin
networks can have (colored) open ends at the vertices.  Of
course, one might have avoided the detour through the
connection representation, and might have directly defined
the loop representation as a suitable representation of an
appropriate algebra of gauge non-invariant observables.
This can be done, but we have preferred here the long way,
in order to make the link with the connection representation
(using which Thiemann constructed the HCO) fully explicit.


\section{Thiemann's Hamiltonian constraint operator.
\label{sec:Thiemann}}

The Hamiltonian constraint of general relativity 
can be written as 
\begin{eqnarray}
  {\cal H} & = & {1 \over \sqrt{{\rm det}(q)}} 
{\rm Tr}((F_{ab} - 2[K_{a},K_{b}]) [E^{a},E^{b}]) 
 \nonumber \\
 & = & {\cal H}^{E} -{2 \over
 \sqrt{{\rm det}(q)}} \,   
{\rm Tr}[K_{a},K_{b}][E^{a},E^{b}].
\label{ham1} 
\end{eqnarray}   
$F_{ab}$ is the curvature of $A_{a}^{i}$, and $K_{a}$ is
related to the extrinsic curvature of the spatial manifold
$\Sigma$.  This is the form that the constraint of the
Lorentzian theory takes when written in terms of the real
connection.  Both $(A_{a}^{i}, \widetilde{E}^{a}_{i})$ and
$(\widetilde{E}^{a}_{i},K_{a}^{i})$ are canonical pairs of
variables. $K_{a}^{i}$ is related to $A_{a}^{i}$ by
$A_{a}^{i}=\Gamma_{a}^{i} -{\rm sgn}(e) K_{a}^{i}$, where
$\Gamma_{a}^{i}$ is the spin-connection annihilating the
triad and ${\rm sgn}(e)$ is the sign of the determinant of
undensitized triad $e_{a}^{i}$. There are two important
differences between this connection and the one introduced
by Ashtekar. First, all variables are real, while the
Ashtekar connection, $A_{a}^{i}=\Gamma_{a}^{i} + {\rm i}~
{\rm sgn}(e) K_{a}^{i}$, is complex.  Second, the constraint
(\ref{ham1}) contains the factor of one over the square root
of the metric. In Ashtekar's theory this factor is absorbed
into the lapse function.  The first term in the right hand
side of (\ref{ham1}) defines the hamiltonian constraint for
gravity with euclidean signature -- accordingly, ${\cal
H}^{E}$ is called the euclidean hamiltonian constraint.

In order to simplify the hamiltonian constraint, Thiemann
\cite{Thiemann} uses the following identities
\begin{equation}\label{id1}
  {[E^{a},E^{b}]^{i} \over \sqrt{{\rm det}q}} = \frac{2}{G}
  \epsilon^{abc}\{A_{c}^{i}, V\}
\end{equation}
and
\begin{equation}\label{id2}
K_{a}^{i} = \frac{1}{G} \{A_{c}^{i}, K\},
\end{equation}
where $V$ is the volume of $\Sigma$:
  \begin{figure}[t]
  \begin{indented}\item[]
  {~~~}{\mbox{\epsfig{file=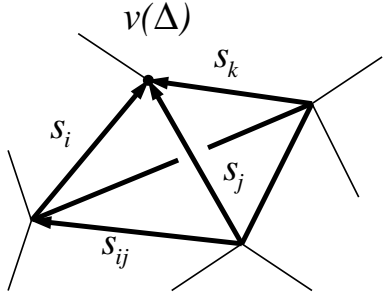,width=5cm}}}
  \end{indented}
  \caption{An elementary tetrahedron and the orientation
           choice of the edges.}
  \label{tetra}
  \end{figure}
\begin{equation}
V = \int_{\Sigma} d^{3}x \sqrt{{\rm det}q} = \int_{\Sigma}
d^{3}x \sqrt{\left|{1 \over 3!}\epsilon_{abc}\epsilon^{ijk}
\widetilde{E}_{i}^{a}\widetilde{E}_{j}^{b}\widetilde{E}_{k}^{c}\right|
},
\end{equation}   
and $K$ is the integrated trace of the densitized extrinsic
curvature of $\Sigma$:
\begin{equation}\label{excurv}
  K = \int_{\Sigma} d^{3}x\ K_{a}^{i}\ \widetilde{E}_{i}^{a}.
\end{equation}
Using the identities (\ref{id1}) and (\ref{id2}) we can write
the euclidean hamiltonian constraint as:
\begin{equation}\label{eucl}
  {\cal H}^{E}[N] = \frac{2}{G} \int_{\Sigma}d^{3}x\ N(x)\
  \epsilon^{abc}\ {\rm Tr} (F_{ab} \{A_{c},V\}),
\end{equation}
and the total hamiltonian constraint (\ref{ham1}) then takes the
form:
\begin{equation}\label{totham}
  {\cal H}[N] = {\cal H}^{E}[N] - \frac{8}{G^3}
  \int_{\Sigma}d^{3}x N(x) \epsilon^{abc}{\rm Tr}\left(
  \{A_{a},K\}\{A_{b},K\}\{A_{c},V\} \right).
\end{equation}
Another useful identity expresses the extrinsic curvature
(\ref{excurv}) in the form:
\begin{equation}\label{OPexcurv}
K = \frac{1}{G} \{V,{\cal H}^{E}[1]\},
\end{equation}
where ${\cal H}^{E}[1]$ is the euclidean hamiltonian
constraint from (\ref{eucl}) for lapse function set to one.

By plugging (\ref{eucl}) into (\ref{OPexcurv}) and then the
result into (\ref{totham}) we have an expression for the
lorentzian hamiltonian constraint defined entirely in terms
of the volume and the connection. It is then relatively
simple to turn this expression into an operator. In the
connection representation, the connection $A_{a}^{i}$ and
the curvature $F_{ab}^{k}$ act (essentially)
multiplicatively. The differential part of the HCO,
originating from the action of the triads, is captured by
the volume operator. But the volume operator has a well
understood action on the kinematical state space.

From now on, we concentrate on the euclidean term of the
constraint.  In the quantum theory we do not have an
operator correponding to the connection, but only an
operator corresponding to its parallel transport.
Accordingly, we regulate ${\cal H}^{E}$, expressing it in
terms of parallel transports.  We introduce a triangulation
$\cal T$ of $\Sigma$ formed by elementary tetrahedra
$\Delta$ with analytic edges. We denote by
$s_{i}^{v}(\Delta)$, or simply by $s_i, i=1,2,3$ the three
edges meeting at a vertex $v(\Delta)$ of a tetrahedron
$\Delta$. Also $\alpha_{ij}(\Delta) := s_{i}(\Delta)\circ
s_{ij}(\Delta)\circ s_{j}^{-1}(\Delta)$ is the loop, based
at $v(\Delta)$ and composed out of the edges $s_i$,
$s_{j}^{-1}$, and the corresponding edge $s_{ij}$ of the
tetrahedron (see \fref{tetra}).  To start with we write the
Hamiltonian constraint as a sum over all tetrahedra in the
triangulation
\begin{equation}\label{HamDel}
{\cal H}^{E}[N]= \sum_{\Delta \in {\cal T}} {\cal
H}^{E}_{\Delta}[N] = \sum_{\Delta \in {\cal T}}\int_{\Delta}
d^{3}x\ N(x)\, \epsilon^{abc} {\rm Tr} \left(F_{ab}\{A_{c},
V\}\right).
\end{equation}
In the limit in which a tetrahedron $\Delta$ shrinks to a
point $v(\Delta)$ the Hamiltonian constraint over $\Delta$
is
\begin{equation}\fl
   {\cal H}^{E}_{\Delta}[N] = 
   \left(-{2 \over 3 G}\right)N(v(\Delta))\epsilon^{ijk} {\rm Tr}
   \left(h_{\alpha_{ij}}h_{s_{k}}\{h^{-1}_{s_{k}},V\}\right) +
   {\cal O}(V(\Delta)^{4/3})
\label{hamTET}
\end{equation}
where $V(\Delta)$ is the volume of the tetrahedron $\Delta$.
In the above expression $h_{\alpha_{ij}}$ is the holonomy along
the closed loop $\alpha_{ij}$ and $h_{s_{k}}$ is the holonomy
along the open loop segment $s_{k}$. That the above limit gives
the correct expression (\ref{HamDel}) follows from the
expansions:
\begin{equation}
   \lim_{\Delta \rightarrow v(\Delta)} h_{\alpha_{ij}} = 1 + {1
   \over 2}F_{ab}s^{a}_{i}s^{b}_{j}
\end{equation}
and
\begin{equation}
   \lim_{\Delta \rightarrow v(\Delta)} h_{s_{k}} = 1 +
   A_{c}s^{c}_{k}.
\end{equation}

  \begin{figure}[t]\label{openloop}
  \begin{indented}\item[]
  {\mbox{\epsfig{file=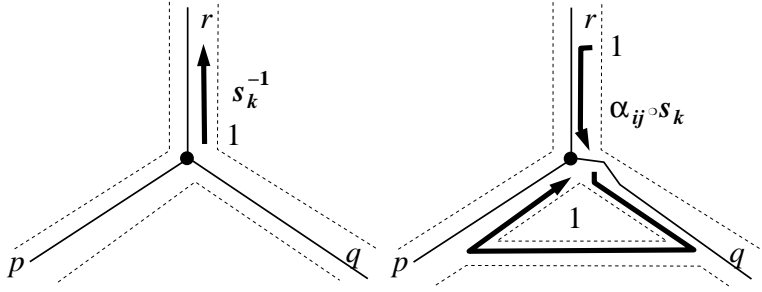,width=10cm}}}
  \end{indented}
  \caption{The ``small open loops'' of the Hamiltonian operator.
  Notice that in gauge 
  non-invariant states (states with open ends in the graphical
  representation) orientation matter. Two open
  ends can be joined only if their orientation is consistent.}
  \end{figure}

Thus, if we have a sequence $n\mapsto {\cal T}_{n}$ 
of finer and finer triangulations of $\Sigma$, we can write
\begin{equation}
   {\cal H}^{E}[N] = \lim_{n\to\infty} 
   \sum_{\Delta \in {\cal T}_{n}}{\cal H}^{E}_{\Delta}[N].
\end{equation}

Since the operators corresponding to the classical
quantities $h$ and $V$ are well understood, this equation
provides an expression for the constraint which is
appropriate for quantization. However, by simply turning
this expression into an operator, with an arbitrary chosen
sequence of triangulations, one obtains meaningless results.
Following Thiemann, we define the quantum Hamiltonian
operator by adapting the regularization to the state. For
every state $\Psi_{\gamma,\psi}$, we define
\begin{equation}\label{hamop}
   \hat{\cal H}^{E}[N]\ \Psi_{\gamma,\psi} = \lim_{n\to\infty} 
   \sum_{\Delta \in {\cal T}^{(\gamma)}_{n}}
   {\hat{\cal H}}^{E}_{\Delta}[N]\ \Psi_{\gamma,\psi}
\end{equation}
where
\begin{eqnarray} 
  \hat{\cal H}^{E}_{\Delta}[N]
    &=& N(v(\Delta)) {\hat{\cal H}}^{E}_{\Delta}
     =  N_{v}{\hat{\cal H}}^{E}_{\Delta}
\nonumber \\
\hat{\cal H}^{E}_{\Delta} 
 &=& 
  - {2 \over 3 {\rm i} l_0^{2}} 
  \epsilon^{ijk} {\rm Tr} \left(
  \frac{\hat{h}_{\alpha_{ij}} - \hat{h}_{\alpha_{ji}}}{2}
  \hat{h}_{s_{k}}[\hat{h}^{-1}_{s_{k}},V]\right), 
\label{hamfin}
\end{eqnarray}
and ${\cal T}^{(\gamma)}_n$ is a sequence of triangulations
satisfying the following condition: (for each $n$) there is
a vertex $v(\Delta)$ sitting on each of the vertices $v$ of
$\gamma$; and if the vertex $v$ is $k$-valent then the
corresponding vertex $v(\Delta)$ has precisely $k$ adjacent
edges, each contained in one of the $k$ edges of $\gamma$
adjacent to $v$.  The triangulation is arbitrary
elsewhere. Since the classical expression converges to the
correct hamiltonian constraint for any sequence of
triangulations, this choice does not spoil the formal
classical limit of the operator. On the other hand, this
choice yields a well defined, diffeomorphism-covariant and
non-trivial operator. This completes the definition of the
quantum operator.

Simple manipulations then show that the action of this
operator can be written as
\begin{equation}
   \hat{\cal H}_{\gamma}^{E}[N]\ \Psi_{\gamma,\psi}= \sum_{v \in
   {\cal V}(\gamma)} 8N_{v} \sum_{v(\Delta) = v}\ 
   \hat{H}_{\Delta}^{E}{{p}_{\Delta} \over
   {E}(v)} \ \Psi_{\gamma,\psi},
\end{equation}
where ${\cal V}(s)$ is the set of the vertices of $s$;
 $\hat{H}_{\Delta}^{E}$ is defined by (\ref{hamfin});
 ${p}_{\Delta}$ is one, whenever $\Delta$ is a tetrahedron
 with three edges coinciding with three edges of the spin
 network state joined at the vertex $v$, and zero
 otherwise. ${E}(v)$ is the number of non-coplanar triples
 of edges of $\gamma$ at the vertex $v$. In the case of an
 n-valent vertex having no triples of coplanar edges
 ${E}(v)=n(n-1)(n-2)/6$. Also $N_{v}$ is the value of the
 lapse function at the vertex (see \cite{Thiemann} for
 details).

Clearly ${\hat{\cal H}}^{E}_{\Delta}$ of \eref{hamfin} it is
not the only operator that has \eref{hamTET} for classical
limit.  Any other ordering of the operators
$\hat{h}_{\alpha_{ij}}$, $\hat{h}_{s_{k}}$ and
$[\hat{h}^{-1}_{s_{k}},V]$ gives \eref{hamTET}. We will
consider the action of Thiemann's Hamiltonian constraint
with the two orderings
\begin{eqnarray}
{\hat{\cal H}}^{E\uno}_{\Delta} 
 &=& 
  - {2 \over 3 {\rm i} l_0^{2}} 
  \epsilon^{ijk} {\rm Tr} \left(
  \frac{\hat{h}_{\alpha_{ij}} - \hat{h}_{\alpha_{ji}}}{2}
  \hat{h}_{s_{k}}[\hat{h}^{-1}_{s_{k}},V]\right) 
\label{hamfin1x} \\
{\hat{\cal H}}^{E\due}_{\Delta} 
 &=& 
  - {2 \over 3 {\rm i} l_0^{2}} 
  \epsilon^{ijk} {\rm Tr} \left(
  \hat{h}_{s_{k}}
  [\hat{h}^{-1}_{s_{k}},V]
  \frac{\hat{h}_{\alpha_{ij}} - \hat{h}_{\alpha_{ji}}}{2}
  \right) 
\label{hamfin2x}
\end{eqnarray}
which correspond to the cases analyzed by Thiemann
\cite{Thiemann}.  

To compute the action of \eref{hamfin1x} and
\eref{hamfin2x} on trivalent vertices, it is sufficient to
compute 
\begin{eqnarray}
  \hat{\cal H}^{E\uno}_{\Delta} ~\Psi_{\gamma,\psi}
  &=&   {2 {\rm i}  \over 3 l_0^{2}}
       \epsilon^{ijk} {\rm Tr}
       \bigg[
       \frac{\hat{h}_{\alpha_{ij}}-\hat{h}_{\alpha_{ji}}}{2}
       \hat{h}_{s_{k}}\hat{V}\hat{h}^{-1}_{s_{k}} \bigg]
       ~\Psi_{\gamma,\psi} ~,
\label{hamfin1} \\  
\hat{\cal H}^{E\due}_{\Delta} ~\Psi_{\gamma,\psi}
  &=&   {2 {\rm i}  \over 3 l_0^{2}}
       \epsilon^{ijk} {\rm Tr}
       \bigg[
       \hat{h}_{s_{k}} 
       \hat{V} \hat{h}^{-1}_{s_{k}} 
       \frac{\hat{h}_{\alpha_{ij}}-\hat{h}_{\alpha_{ji}}}{2}
       \bigg]
       ~\Psi_{\gamma,\psi}\ ;
\label{hamfin2}
\end{eqnarray}
because the volume annihilates the gauge-invariant trivalent
vertices and therefore the other term of the commutators
vanishes.  In section 5.3, we show that these two choices
exhaust all possible ``natural'' orderings.


\section{Calculation of the HCO matrix elements
\label{sec:Graph}}

The action of the HCO is local, namely it is a sum of
independent actions on each vertex. Therefore we can focus
on its action on a single vertex $v$.  Let us consider the
action of the operator $\hat{{\cal H}}_{\Delta}^{E}$ on a
single trivalent vertex with edges $\e_i, \e_j, \e_k$ with
colors $p, q, r$. We indicate the part of the spin network
state, containing only the vertex $v$ and its adjacent edges
as $\ket{v}$.

\subsection{First ordering choice
\label{sec:NonSym}}

We investigate the action of the Euclidean Hamiltonian
operator $\hat{{\cal H}}_{\Delta}^{E\uno}$ on $\ket{v}$
\begin{equation} 
  \hat{{\cal H}}_{\Delta}^{E\uno}\ket{v} = -{2 {\rm i} \over 3
 l_0^{2}}\epsilon^{ijk} \frac{\hat{h}[\alpha_{ij}]
 -\hat{h}[\alpha_{ji}]}{2}
 \hat{h}[s_k]\hat{V}\hat{h}[s^{-1}_k]\ket{v}
\end{equation}
   
As described in sections 2 and 3, $\hat{h}[s^{-1}_k]$ attaches an
open loop segment $s^{-1}_k$ to the edge $\e_k$ with color $r$.
It creates one additional vertex on $\e_{k}$, and alters
the color of the edge segment between the two vertices
of $s_k$.  From \eref{eq:Recursion1} we have 
\begin{equation}
\label{Tact} 
\fl \hat{h}[s_{k}^{-1}]
   \begin{array}{c}\mbox{\epsfig{file=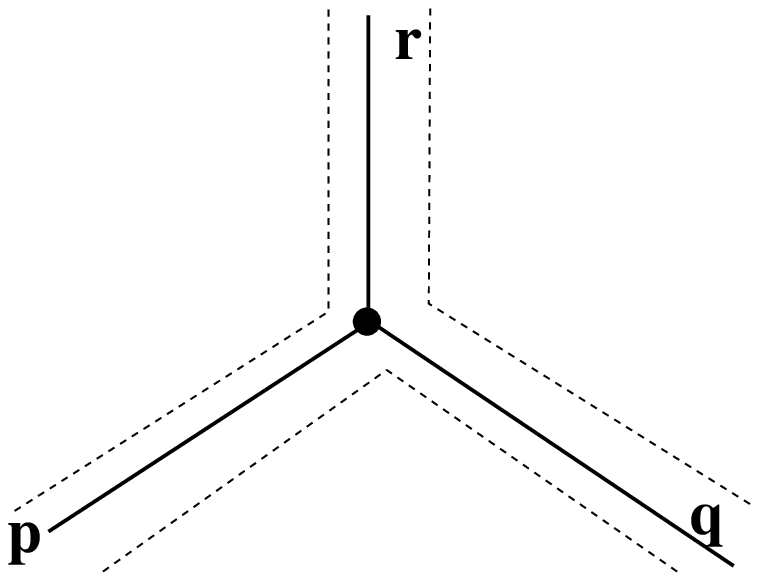,width=3.5cm}}
   \end{array}
\!\! =\sum_{\epsilon=\pm 1} a_\epsilon(r)
  \begin{array}{c}\mbox{\epsfig{file=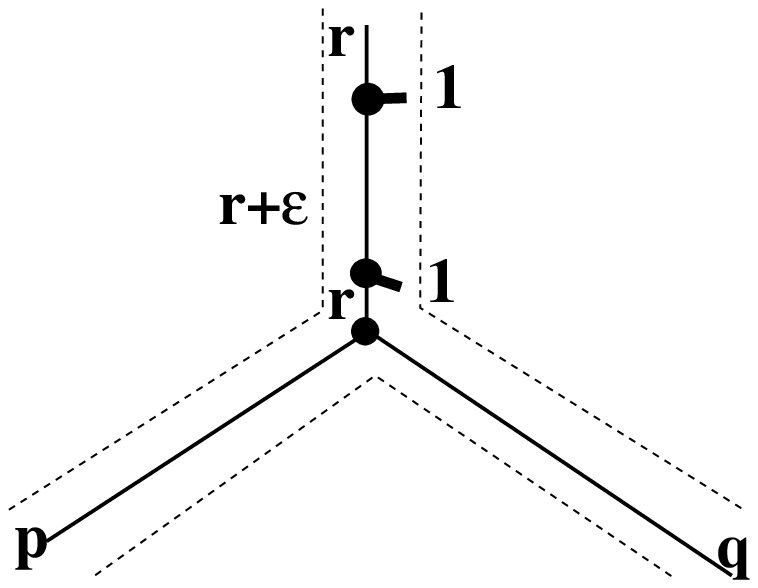,width=3.5cm}}
  \end{array}
  \label{openend} 
\end{equation} 
From Appendix C, the coefficients $a_{\pm}(r)$ are
\begin{equation}
  a_{+}(r)=1,\ \ \ a_{-}(r) = -
  \frac{r}{r+1} ~~.
\end{equation}
Notice that the (lower) open end in the right hand side of
(\ref{openend}) is located at the vertex.  Thus, the vertex
has now become gauge non-invariant: it carries the spin-1/2
representation, with a free index, and its interwiner is
determined by the the virtual edge of color $r$.

The volume operator acts on this vertex. The action of the
volume operator has been computed for gauge invariant states
only, but a moment of reflection shows that the results in
the gauge invariant case are relevant for the present case
as well. In fact (in both its definitions) the volume
operator acts on the (real!) edges adjacent to the vertex,
giving a sum of terms -- one term per triple of edges.  If
we act on a gauge-invariant 4-valent vertex, we obtain a sum
of four terms, one for each triple of edges. In the present
case, the volume acts on the three real edges, giving a
single term, but the action of this single term is precisely
equivalent to the action of one of the four terms of the
volume over a four-valent gauge invariant vertex (See
Appendix A).  The free index of the vertex behaves as the
fourth edge, but it is ``invisible'' for the operator.
Therefore the action of $\hat V$ can be immediately obtained
from the expression of $\hat V$ on 4-valent gauge-invariant
vertices, which was given in \cite{DePiRo96}, and is
recalled in Appendix A. In particular, we show in Appendix A
that for the combination of colors in (\ref{Tact}) this
action is diagonal.  Using \eref{VOL:3vertex}, we obtain
\begin{equation}\label{eq:STEP1} \fl
  \hat{V}\hat{h}[s_{k}^{-1}] 
  \!\!\!\!\!\!\!
  \begin{array}{c}\mbox{\epsfig{file=f2.eps,width=3.5cm}}
  \end{array} 
  \!\!\!\!\!\!\!
  = \frac{l_0^3}{4}
    \sum_{\epsilon=\pm1} a_\epsilon(r) 
    ~\sqrt{w(p,q,r+\epsilon,1)} \!\!\!\!
  \!\!\!\!\!\!\!
  \begin{array}{c}\mbox{\epsfig{file=f4.eps,width=3.5cm}}
  \end{array} 
  \!\!\!\!
\end{equation}
where $w(\ ,\ ,\ ,\ )$ is given in (A.10). 

The next step is to compute the action of
$\hat{h}[\alpha_{ij}]\hat{h}[s_{k}]$ on the right hand side
of (\ref{eq:STEP1}).  This operator attaches $\alpha_{ij}$
and $s_k$ to the open ends in (\ref{eq:STEP1}) according to
the prescription given by the orientation of the open
loops. Finally, the trace connects the free indices. This is
represented in the graphical computation by tying up the
open ends of the virtual edges (see
\fref{openloop}). Moreover, we have to add a minus sign due
to the trace prescription of the graphical representation
(see \sref{sec:LoopRepr}). We obtain
\begin{eqnarray} \label{eq:STEP2} \fl
\bigg[\frac{ \hat{h}[\alpha_{ij}] \hat{h}[s_k]
        -\hat{h}[\alpha_{ji}] \hat{h}[s_k] }{2} \bigg] 
  \!\! \!\!    
  \begin{array}{c}\mbox{\epsfig{file=f4.eps,width=3cm}}\end{array}
  =
\nonumber \\
\lo{=} (-1) \frac{1}{2}\left[ \!\! 
   \begin{array}{c}\mbox{\epsfig{file=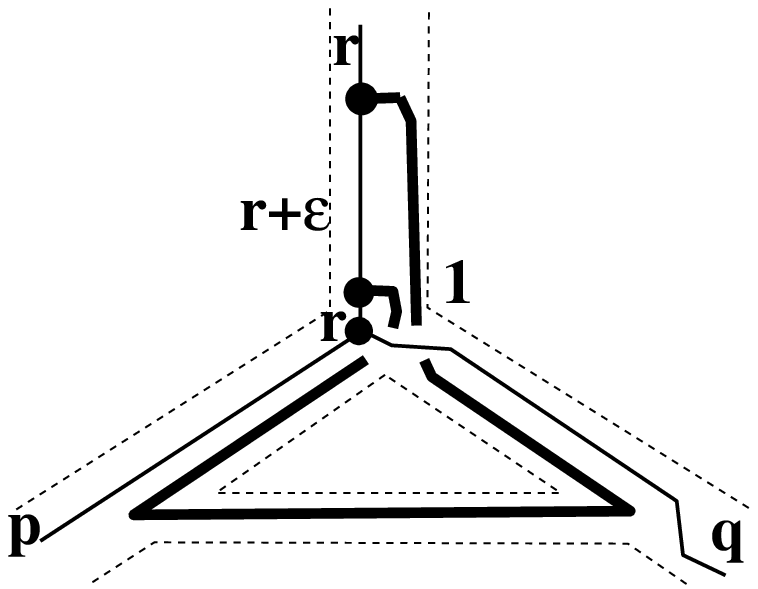,width=3cm}}\end{array}
 - \begin{array}{c}\mbox{\epsfig{file=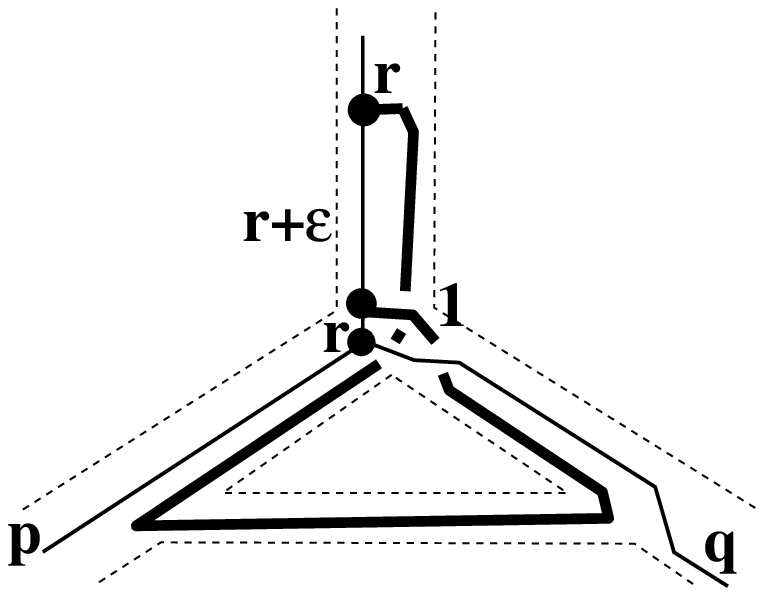,width=3cm}}\end{array}
  \!\!\!\! \right] =  
\nonumber \\ 
\lo{=} (-1) 
  \sum_{\bar{\epsilon}=\pm 1} \sum_{\tilde{\epsilon}=\pm 1}
   a_{\bar\epsilon}(p) a_{\tilde\epsilon}(q)
   \begin{array}{c}\mbox{\epsfig{file=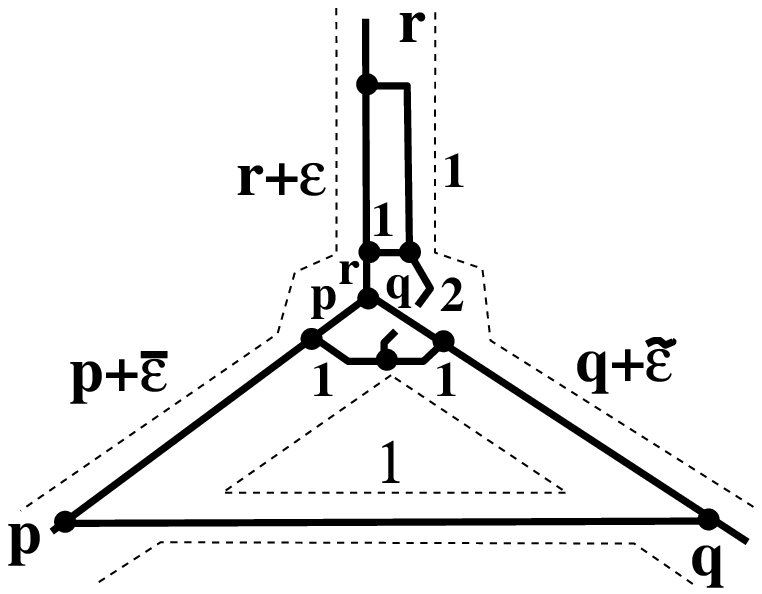,width=4cm}}\end{array}.
\end{eqnarray}
In the last equation we used the fact that the
anti-symmetrization of two loops is equivalent to two 3-vertices
of type (1,1,2) and we applied equation \eref{eq:Recursion1} to
the $p$ and $q$ edges.
Now, the tangle around the vertex in the right hand side of 
(\ref{eq:STEP2}) can be  
evaluated using recoupling theory.  We have (see Appendix C)
\begin{equation}\label{eval1}
 \begin{array}{c}\mbox{\epsfig{file=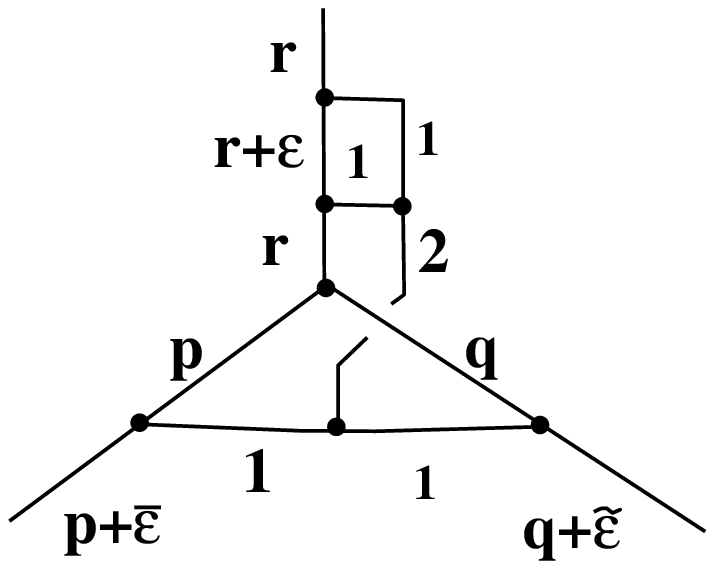,width=3cm}}\end{array}
 =  \frac{\begin{array}{c}\setlength{\unitlength}{.8 pt}
        \begin{picture}(55,40)
        \put( 0,15){\line(1,-1){15}} \put(0,22){${\scriptstyle r}$}
        \put( 0,15){\line(1, 1){15}} \put(0, 0){${\scriptstyle r}$}
        \put( 0,15){\circle*{3}}
        \put(30,15){\line(-1, 1){15}} \put(28,20){${\scriptstyle 1}$}
        \put(30,15){\line(-1,-1){15}} \put(28, 2){${\scriptstyle 1}$}
        \put(30,15){\circle*{3}}
        \put( 0,15){\line(1,0){30}} \put(12,16){${\scriptstyle 2}$}
        \put(15,30){\line(1,0){25}} \put(15,30){\circle*{3}}
        \put(15, 0){\line(1,0){25}} \put(15, 0){\circle*{3}}
        \put(40, 0){\line(0,1){30}} 
        \put(42,12){${\scriptstyle r+\epsilon}$}
        \end{picture}\end{array}
    }{ \begin{array}{c}\setlength{\unitlength}{.5 pt}
       \begin{picture}(40,40)
            \put(18,32){$\scriptstyle r$}
            \put(18,17){$\scriptstyle r$} 
            \put(18, 2){$\scriptstyle 2$} 
            \put(20,15){\oval(40,30)} \put( 0,15){\line(1,0){40}} 
            \put( 0,15){\circle*{3}}  \put(40,15){\circle*{3}}
       \end{picture}\end{array}
      } 
~\cdot~ 
\begin{array}{c}\setlength{\unitlength}{1 pt}
\begin{picture}(60,50)
       \put( 5, 5){\line( 1, 1){25}} \put( 0,0){$p+\bar{\epsilon}$}
       \put(55, 5){\line(-1, 1){25}} \put(40,0){$q+\tilde{\epsilon}$}
       \put(30,30){\line(0,1){25}}   \put(22,52){$r$}
       \put(30,30){\circle*{3}}      \put(22,32){$r$}
       \put(30,45){\circle*{3}} \put(30,45){\line(1,0){5}}
       \put(35,40){\oval(10,10)[tr]} \put(40,40){\line(0,-1){15}}
       \put(15,15){\circle*{3}}\put(45,15){\circle*{3}}
       \put(15,15){\line(1,0){30}} \put(30,15){\circle*{3}}
       \put(22,10){$\scriptstyle 1$}\put(37,10){$\scriptstyle 1$}
       \put(30,15){\line(1,1){10}} \put(23,17){$\scriptstyle 2$}
       \put(15,19){$\scriptstyle p$}
       \put(45,19){$\scriptstyle q$}
\end{picture}\end{array}
\end{equation}
In (\ref{eval1}) we have not shown the bottom link of
(\ref{eq:STEP2}). Using equations \eref{eq:MOVE2} end 
\eref{eq:TETred} we expand the right hand side of (\ref{eval1})
\begin{eqnarray}\label{eval2} \fl
\begin{array}{c}\setlength{\unitlength}{1 pt}
\begin{picture}(60,50)
       \put( 5, 5){\line( 1, 1){25}} \put( 0,0){$p+\bar{\epsilon}$}
       \put(55, 5){\line(-1, 1){25}} \put(40,0){$q+\tilde{\epsilon}$}
       \put(30,30){\line(0,1){25}}    \put(22,52){$r$}
       \put(30,30){\circle*{3}}       \put(22,32){$r$}
       \put(30,45){\circle*{3}} \put(30,45){\line(1,0){5}}
       \put(35,40){\oval(10,10)[tr]} \put(40,40){\line(0,-1){15}}
       \put(15,15){\circle*{3}}\put(45,15){\circle*{3}}
       \put(15,15){\line(1,0){30}} \put(30,15){\circle*{3}}
       \put(22,10){$\scriptstyle 1$}\put(37,10){$\scriptstyle 1$}
       \put(30,15){\line(1,1){10}} \put(23,17){$\scriptstyle 2$}
       \put(15,19){$\scriptstyle p$}
       \put(45,19){$\scriptstyle q$}
\end{picture}\end{array}
 =  \left[ 
 \frac{p}{r}
\begin{array}{c}\setlength{\unitlength}{1 pt}
\begin{picture}(60,50)
       \put( 5, 5){\line( 1, 2){20}} \put( 0,0){$p+\bar{\epsilon}$}
       \put(45, 5){\line(-1, 2){20}} \put(40,0){$q+\tilde{\epsilon}$}
       \put(25,45){\line(0,1){10}}    \put(17,52){$r$}
       \put(25,45){\circle*{3}}
       \put(10,15){\circle*{3}}\put(40,15){\circle*{3}}
       \put(10,15){\line(1,0){30}} \put(25,15){\circle*{3}}
       \put(17,10){$\scriptstyle 1$}\put(32,10){$\scriptstyle 1$}
       \put(25,15){\line(0,1){10}} \put(27,17){$\scriptstyle 2$}
       \put(15,25){\circle*{3}}\put(15,25){\line(1,0){10}}
       \put(15,19){$\scriptstyle p$}\put(20,30){$\scriptstyle p$}
       \put(35,30){$\scriptstyle q$}
\end{picture}\end{array}
-\frac{q}{r}
\begin{array}{c}\setlength{\unitlength}{1 pt}
\begin{picture}(60,50)
       \put( 5, 5){\line( 1, 2){20}} \put( 0,0){$p+\bar{\epsilon}$}
       \put(45, 5){\line(-1, 2){20}} \put(40,0){$q+\tilde{\epsilon}$}
       \put(25,45){\line(0,1){10}}    \put(17,52){$r$}
       \put(25,45){\circle*{3}}
       \put(10,15){\circle*{3}}\put(40,15){\circle*{3}}
       \put(10,15){\line(1,0){30}} \put(25,15){\circle*{3}}
       \put(17,10){$\scriptstyle 1$}\put(32,10){$\scriptstyle 1$}
       \put(25,15){\line(0,1){10}} \put(27,17){$\scriptstyle 2$}
       \put(35,25){\circle*{3}}\put(35,25){\line(-1,0){10}}
       \put(40,19){$\scriptstyle q$}\put(35,30){$\scriptstyle q$}
       \put(20,30){$\scriptstyle p$}
\end{picture}\end{array}
\right] =
\\ \fl ~~~ =
\left[ 
 \frac{p}{r}  \frac{\begin{array}{c}\setlength{\unitlength}{.8 pt}
        \begin{picture}(55,40)
        \put( 0,15){\line(1,-1){15}} \put(0,22){${\scriptstyle p}$}
        \put( 0,15){\line(1, 1){15}} \put(0, 0){${\scriptstyle p}$}
        \put( 0,15){\circle*{3}}
        \put(30,15){\line(-1, 1){15}} \put(28,20){${\scriptstyle 1}$}
        \put(30,15){\line(-1,-1){15}} \put(28, 2){${\scriptstyle 1}$}
        \put(30,15){\circle*{3}}
        \put( 0,15){\line(1,0){30}} \put(12,16){${\scriptstyle 2}$}
        \put(15,30){\line(1,0){25}} \put(15,30){\circle*{3}}
        \put(15, 0){\line(1,0){25}} \put(15, 0){\circle*{3}}
        \put(40, 0){\line(0,1){30}} 
        \put(42,12){${\scriptstyle p+\bar{\epsilon}}$}
        \end{picture}\end{array}
    }{ \begin{array}{c}\setlength{\unitlength}{.5 pt}
       \begin{picture}(40,40)
            \put(18,32){$\scriptstyle 1$}
            \put(18,17){$\scriptstyle p$} 
            \put(12, 2){$\scriptstyle p+\bar{\epsilon}$} 
            \put(20,15){\oval(40,30)} \put( 0,15){\line(1,0){40}} 
            \put( 0,15){\circle*{3}}  \put(40,15){\circle*{3}}
           \end{picture}\end{array}
      }
-\frac{q}{r}  \frac{\begin{array}{c}\setlength{\unitlength}{.8 pt}
        \begin{picture}(55,40)
        \put( 0,15){\line(1,-1){15}} \put(0,22){${\scriptstyle q}$}
        \put( 0,15){\line(1, 1){15}} \put(0, 0){${\scriptstyle q}$}
        \put( 0,15){\circle*{3}}
        \put(30,15){\line(-1, 1){15}} \put(28,20){${\scriptstyle 1}$}
        \put(30,15){\line(-1,-1){15}} \put(28, 2){${\scriptstyle 1}$}
        \put(30,15){\circle*{3}}
        \put( 0,15){\line(1,0){30}} \put(12,16){${\scriptstyle 2}$}
        \put(15,30){\line(1,0){25}} \put(15,30){\circle*{3}}
        \put(15, 0){\line(1,0){25}} \put(15, 0){\circle*{3}}
        \put(40, 0){\line(0,1){30}} 
        \put(42,12){${\scriptstyle q+\tilde{\epsilon}}$}
        \end{picture}\end{array}
    }{ \begin{array}{c}\setlength{\unitlength}{.5 pt}\begin{picture}(40,40)
            \put(18,32){$\scriptstyle 1$}
            \put(18,17){$\scriptstyle q$} 
            \put(12, 2){$\scriptstyle q+\tilde{\epsilon}$} 
            \put(20,15){\oval(40,30)} \put( 0,15){\line(1,0){40}} 
            \put( 0,15){\circle*{3}}  \put(40,15){\circle*{3}}
           \end{picture}\end{array}
      }
\right] ~\cdot~
\frac{\begin{array}{c}\setlength{\unitlength}{.8 pt}
        \begin{picture}(55,40)
        \put( 0,15){\line(1,-1){15}} \put(0,22){${\scriptstyle p}$}
        \put( 0,15){\line(1, 1){15}} 
        \put(-7, 0){${\scriptstyle p+\bar{\epsilon}}$}
        \put( 0,15){\circle*{3}}
        \put(30,15){\line(-1, 1){15}} \put(28,20){${\scriptstyle q}$}
        \put(30,15){\line(-1,-1){15}} 
        \put(22, 2){${\scriptstyle q+\tilde{\epsilon}}$}
        \put(30,15){\circle*{3}}
        \put( 0,15){\line(1,0){30}} \put(12,16){${\scriptstyle 1}$}
        \put(15,30){\line(1,0){25}} \put(15,30){\circle*{3}}
        \put(15, 0){\line(1,0){25}} \put(15, 0){\circle*{3}}
        \put(40, 0){\line(0,1){30}} \put(42,12){${\scriptstyle r}$}
        \end{picture}\end{array}
    }{ \begin{array}{c}\setlength{\unitlength}{.5 pt}
       \begin{picture}(40,40)
            \put(18,32){$\scriptstyle r$}
            \put(12,17){$\scriptstyle p+\bar{\epsilon}$} 
            \put(12, 2){$\scriptstyle q+\tilde{\epsilon}$} 
            \put(20,15){\oval(40,30)} \put( 0,15){\line(1,0){40}} 
            \put( 0,15){\circle*{3}}  \put(40,15){\circle*{3}}
           \end{picture}\end{array}
      }
~\cdot~ 
\begin{array}{c}\setlength{\unitlength}{1 pt}
\begin{picture}(50,50)
       \put(25,25){\line(-1, -1){20}} \put( 0,0){$p+\bar{\epsilon}$}
       \put(25,25){\line( 1, -1){20}} \put(40,0){$q+\tilde{\epsilon}$}
       \put(25,25){\line(0,1){20}}    \put(17,42){$r$}
       \put(25,25){\circle*{3}}
\end{picture}\end{array}
\nonumber
\end{eqnarray}
The chromatic evaluation of the closed nets in equations (\ref{eval1}) and
(\ref{eval2}) gives
\begin{eqnarray} \label{eq:defR1}
\fl R_1(r,\epsilon) = a_\epsilon(r) ~\cdot~
\frac{\begin{array}{c}\setlength{\unitlength}{.8 pt}
        \begin{picture}(55,40)
        \put( 0,15){\line(1,-1){15}} \put(0,22){${\scriptstyle r}$}
        \put( 0,15){\line(1, 1){15}} \put(0, 0){${\scriptstyle r}$}
        \put( 0,15){\circle*{3}}
        \put(30,15){\line(-1, 1){15}} \put(28,20){${\scriptstyle 1}$}
        \put(30,15){\line(-1,-1){15}} \put(28, 2){${\scriptstyle 1}$}
        \put(30,15){\circle*{3}}
        \put( 0,15){\line(1,0){30}} \put(12,16){${\scriptstyle 2}$}
        \put(15,30){\line(1,0){25}} \put(15,30){\circle*{3}}
        \put(15, 0){\line(1,0){25}} \put(15, 0){\circle*{3}}
        \put(40, 0){\line(0,1){30}} 
        \put(42,12){${\scriptstyle r+\epsilon}$}
        \end{picture}\end{array}
    }{ \begin{array}{c}\setlength{\unitlength}{.5 pt}
       \begin{picture}(40,40)
            \put(18,32){$\scriptstyle r$}
            \put(18,17){$\scriptstyle r$} 
            \put(18, 2){$\scriptstyle 2$} 
            \put(20,15){\oval(40,30)} \put( 0,15){\line(1,0){40}} 
            \put( 0,15){\circle*{3}}  \put(40,15){\circle*{3}}
        \end{picture}\end{array}
      } 
   = \epsilon {r \over r+1}
\\ \label{eq:defR2}
\fl R_2(p,\bar{\epsilon};q,\tilde{\epsilon};r) 
   =  a_{\bar{\epsilon}}(p) a_{\tilde{\epsilon}}(q)
\left[ 
 \frac{p}{r}  \frac{\begin{array}{c}\setlength{\unitlength}{.8 pt}
        \begin{picture}(55,40)
        \put( 0,15){\line(1,-1){15}} \put(0,22){${\scriptstyle p}$}
        \put( 0,15){\line(1, 1){15}} \put(0, 0){${\scriptstyle p}$}
        \put( 0,15){\circle*{3}}
        \put(30,15){\line(-1, 1){15}} \put(28,20){${\scriptstyle 1}$}
        \put(30,15){\line(-1,-1){15}} \put(28, 2){${\scriptstyle 1}$}
        \put(30,15){\circle*{3}}
        \put( 0,15){\line(1,0){30}} \put(12,16){${\scriptstyle 2}$}
        \put(15,30){\line(1,0){25}} \put(15,30){\circle*{3}}
        \put(15, 0){\line(1,0){25}} \put(15, 0){\circle*{3}}
        \put(40, 0){\line(0,1){30}} 
        \put(42,12){${\scriptstyle p+\bar{\epsilon}}$}
        \end{picture}\end{array}
    }{ \begin{array}{c}\setlength{\unitlength}{.5 pt}
       \begin{picture}(40,40)
            \put(18,32){$\scriptstyle 1$}
            \put(18,17){$\scriptstyle p$} 
            \put(12, 2){$\scriptstyle p+\bar{\epsilon}$} 
            \put(20,15){\oval(40,30)} \put( 0,15){\line(1,0){40}} 
            \put( 0,15){\circle*{3}}  \put(40,15){\circle*{3}}
        \end{picture}\end{array}
      }
-\frac{q}{r}  \frac{\begin{array}{c}\setlength{\unitlength}{.8 pt}
        \begin{picture}(55,40)
        \put( 0,15){\line(1,-1){15}} \put(0,22){${\scriptstyle q}$}
        \put( 0,15){\line(1, 1){15}} \put(0, 0){${\scriptstyle q}$}
        \put( 0,15){\circle*{3}}
        \put(30,15){\line(-1, 1){15}} \put(28,20){${\scriptstyle 1}$}
        \put(30,15){\line(-1,-1){15}} \put(28, 2){${\scriptstyle 1}$}
        \put(30,15){\circle*{3}}
        \put( 0,15){\line(1,0){30}} \put(12,16){${\scriptstyle 2}$}
        \put(15,30){\line(1,0){25}} \put(15,30){\circle*{3}}
        \put(15, 0){\line(1,0){25}} \put(15, 0){\circle*{3}}
        \put(40, 0){\line(0,1){30}} 
        \put(42,12){${\scriptstyle q+\tilde{\epsilon}}$}
        \end{picture}\end{array}
    }{ \begin{array}{c}\setlength{\unitlength}{.5 pt}
       \begin{picture}(40,40)
            \put(18,32){$\scriptstyle 1$}
            \put(18,17){$\scriptstyle q$} 
            \put(12, 2){$\scriptstyle q+\tilde{\epsilon}$} 
            \put(20,15){\oval(40,30)} \put( 0,15){\line(1,0){40}} 
            \put( 0,15){\circle*{3}}  \put(40,15){\circle*{3}}
           \end{picture}\end{array}
      }
\right] ~\cdot~
\frac{\begin{array}{c}\setlength{\unitlength}{.8 pt}
        \begin{picture}(45,40)
        \put( 0,15){\line(1,-1){15}} \put(0,22){${\scriptstyle p}$}
        \put( 0,15){\line(1, 1){15}} 
        \put(-7, 0){${\scriptstyle p+\bar{\epsilon}}$}
        \put( 0,15){\circle*{3}}
        \put(30,15){\line(-1, 1){15}} \put(28,20){${\scriptstyle q}$}
        \put(30,15){\line(-1,-1){15}} 
        \put(22, 2){${\scriptstyle q+\tilde{\epsilon}}$}
        \put(30,15){\circle*{3}}
        \put( 0,15){\line(1,0){30}} \put(12,16){${\scriptstyle 1}$}
        \put(15,30){\line(1,0){25}} \put(15,30){\circle*{3}}
        \put(15, 0){\line(1,0){25}} \put(15, 0){\circle*{3}}
        \put(40, 0){\line(0,1){30}} \put(42,12){${\scriptstyle r}$}
        \end{picture}\end{array}
    }{ \begin{array}{c}\setlength{\unitlength}{.5 pt}
       \begin{picture}(40,40)
            \put(18,32){$\scriptstyle r$}
            \put(12,17){$\scriptstyle p+\bar{\epsilon}$} 
            \put(12, 2){$\scriptstyle q+\tilde{\epsilon}$} 
            \put(20,15){\oval(40,30)} \put( 0,15){\line(1,0){40}} 
            \put( 0,15){\circle*{3}}  \put(40,15){\circle*{3}}
       \end{picture}\end{array}
      } 
\\ 
\lo{=} \left\{
\begin{array}{lcl}
{\displaystyle \frac{p-q}{2r} }
 &\mbox{~;~}& \bar{\epsilon}=+1,\tilde{\epsilon}=+1 \\[2mm]
{\displaystyle -\frac{(q-p+r)(2+p+q)}{4r(1+q)} }
 &\mbox{~;~}& \bar{\epsilon}=-1,\tilde{\epsilon}=+1 \\[2mm]
{\displaystyle \frac{(p-q+r)(2+p+q)}{4r(1+p)} }
 &\mbox{~;~}& \bar{\epsilon}=+1,\tilde{\epsilon}=-1 \\[2mm]
{\displaystyle \frac{(p-q)(p+q-r)(2+p+q+r)}{8r(1+p)(1+q)} }
 &\mbox{~;~}& \bar{\epsilon}=-1,\tilde{\epsilon}=-1 \\
\end{array}
\right.
\nonumber
\end{eqnarray}
We have obtained  
\[ \fl
  {\rm Tr}\left(
  \frac{\hat{h}_{\alpha_{ij}}-\hat{h}_{\alpha_{ji}}}{2}
  \hat{h}_{s_{k}}\hat{V}\hat{h}^{-1}_{s_{k}}
  \right)
  \!\!\!\! \!\!\!\! \!\!\!\!
  \begin{array}{c}\mbox{\epsfig{file=f2.eps,width=3cm}}\end{array} 
  \!\!\!\! \!\!\!\! \!\!\!\! \!\!
= - {l_0^3}
  \sum_{\bar{\epsilon},\tilde{\epsilon}=\pm 1} \!
  A^{\uno}(p,\bar{\epsilon};q,\tilde{\epsilon};r)
  \!\!\!\! \!\!
  \begin{array}{c}\mbox{\epsfig{file=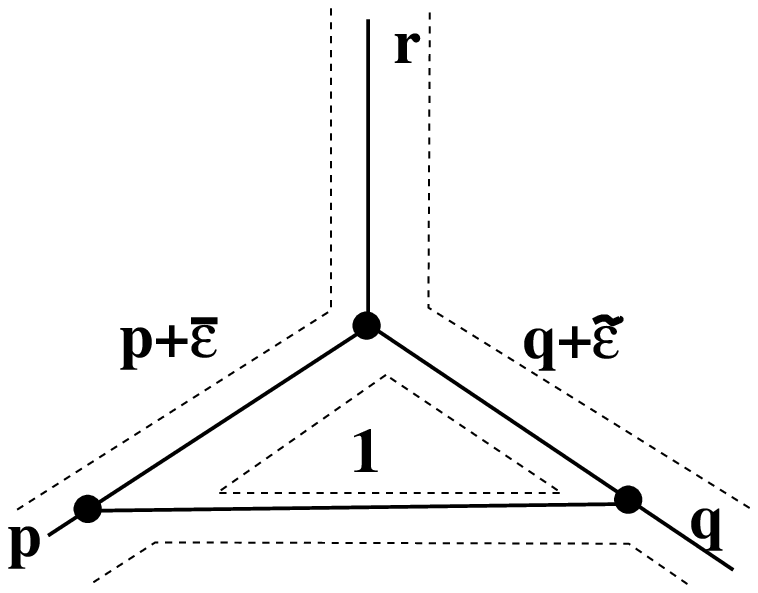,width=3cm}}\end{array}
  \!\!\!\!  .
\]
where 
\begin{equation}\fl
A^{\uno}(p,\bar{\epsilon};q,\tilde{\epsilon};r) =
\sum_{{\epsilon}=\pm 1} 
 \frac{1}{4}
 ~\sqrt{w(p,q,r+\epsilon,1)}
 ~R_1(r,\epsilon)  
 ~R_2(p,\bar{\epsilon};q,\tilde{\epsilon},r).
 \label{A}
\end{equation}

Collecting our results we obtain that the action of  
Thiemann's Euclidean HCO on a generic 3-vertex 
$| v(p,q,r) \rangle$ is given by:
\begin{eqnarray} \label{eq:ActionHe}  \fl
 \hat{{\cal H}}_{\Delta}^{E\uno}~ \big| v(p,q,r) \big\rangle 
 = {2 {\rm i} l_0 \over 3}
 \sum_{\bar{\epsilon}=\pm 1} \sum_{\tilde{\epsilon}=\pm 1} 
 \left[
 A^{\uno}(p,\bar{\epsilon};q,\tilde{\epsilon};r) 
    \bigg|\begin{array}{c} 
      \mbox{\epsfig{file=f9.eps,height=2.5cm,silent=}}
    \end{array}\!\!\bigg\rangle \right. +
\\ 
\fl ~~~~ \left.
+ A^{\uno}(q,\bar{\epsilon};r,\tilde{\epsilon};p)
    \bigg|\begin{array}{c} 
      \mbox{\epsfig{file=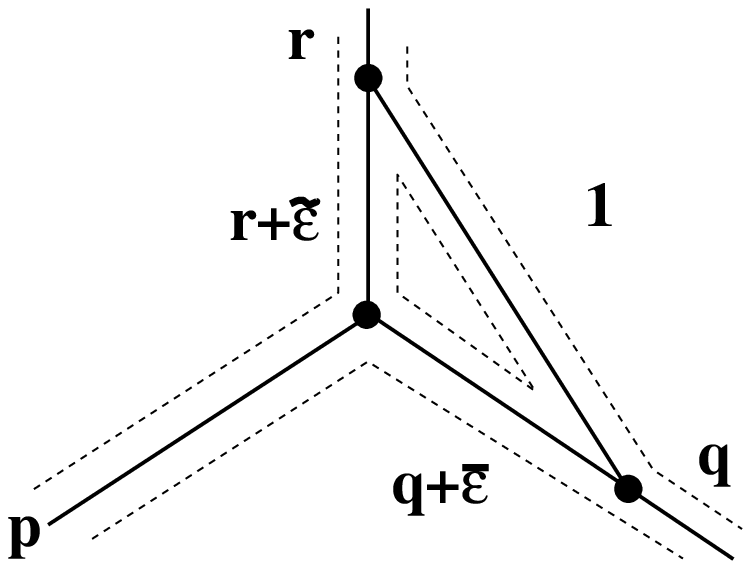,height=2.5cm,silent=}}
    \end{array}\!\!\bigg\rangle 
+ A^{\uno}(r,\bar{\epsilon};p,\tilde{\epsilon};q)
    \bigg|\begin{array}{c} 
      \mbox{\epsfig{file=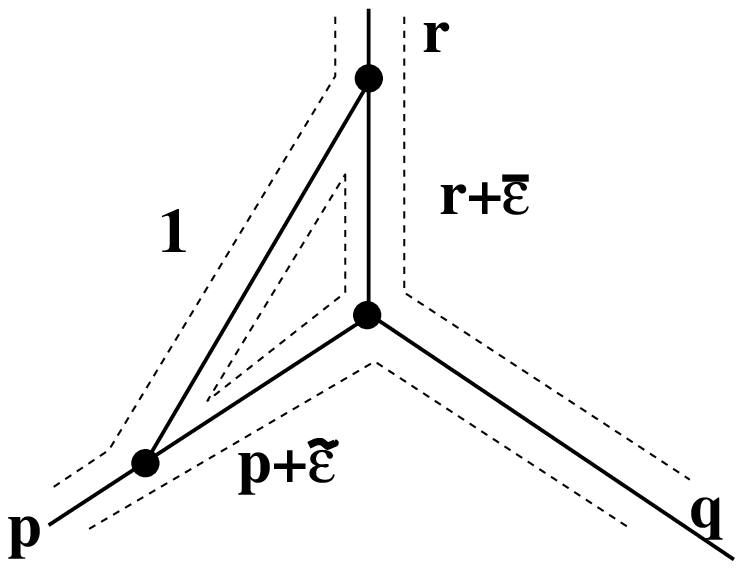,height=2.5cm,silent=}}
    \end{array}\!\!\bigg\rangle \right]
\nonumber 
\end{eqnarray}
where the coeficients $A^{1}$ are given by (5.10).


\subsection{Second ordering choice
\label{sec:Sym}}
 
In the case of the ordering choice given by \eref{hamfin2} we
have to perform the computations in the reverse order.  First we
compute the action of the elementary loop operator $
\hat{h}_{s_{k}}^{-1} \big[
\hat{h}_{\alpha_{ij}}-\hat{h}_{\alpha_{ji}} \big]$ on $\ket{v}$,
and then, the action of $\hat{h}_{s_{k}} \hat{V} $.  Using the
same technique as in the previous section we get:
\begin{eqnarray}
\fl
 \hat{h}_{s_{k}}^{-1} 
 \frac{ \hat{h}_{\alpha_{ij}}-\hat{h}_{\alpha_{ji}}}{2}
  \begin{array}{c}\mbox{\epsfig{file=f2.eps,width=3.5cm}}\end{array} 
= \begin{array}{c}\mbox{\epsfig{file=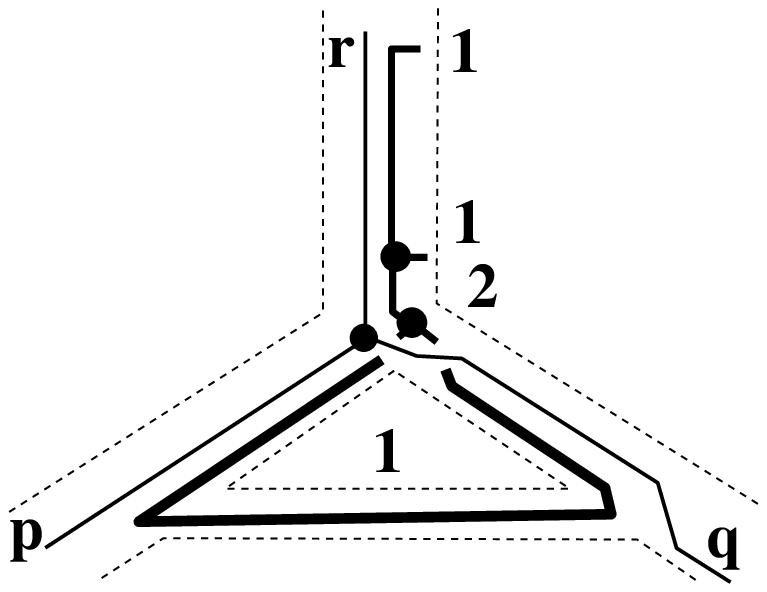,width=3.5cm}}\end{array} 
=
\\ \fl ~~~~~
= \sum_{\epsilon=\pm1} \sum_{\bar{\epsilon}=\pm 1} 
  \sum_{\tilde{\epsilon}=\pm 1} 
  a_\epsilon(r) a_{\bar\epsilon}(p) a_{\tilde\epsilon}(q)
  \begin{array}{c}\mbox{\epsfig{file=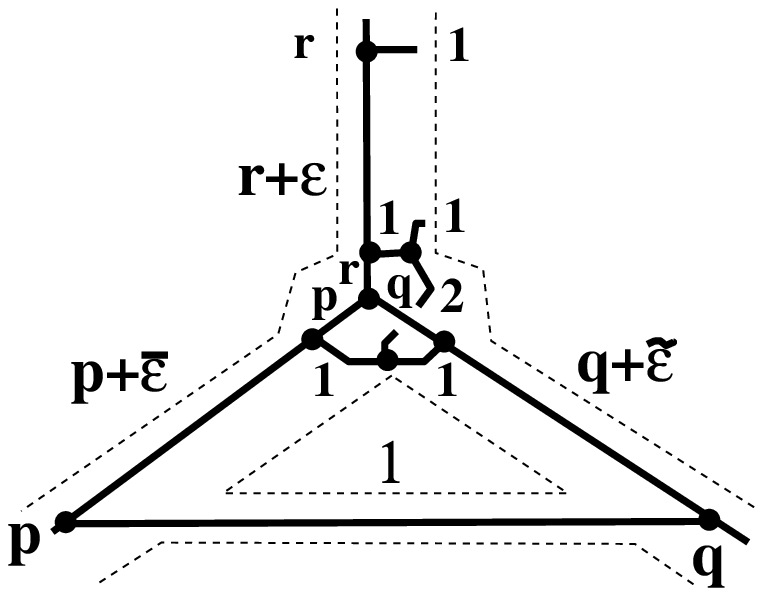,width=3.5cm}}\end{array} 
  = \nonumber \\ \fl ~~~~~
= \sum_{\epsilon,\bar{\epsilon},\tilde{\epsilon}=\pm1} 
  \sum_{J=r,r+2\epsilon} 
  a_\epsilon(r) a_{\bar\epsilon}(p) a_{\tilde\epsilon}(q)
  \left\{ \!\! \begin{array}{ccc}  
          r & \! r+\epsilon \!  & J \\ 1 & 2 & 1
  \end{array}\!\!\right\} \!\!\!\!
  \begin{array}{c}\mbox{\epsfig{file=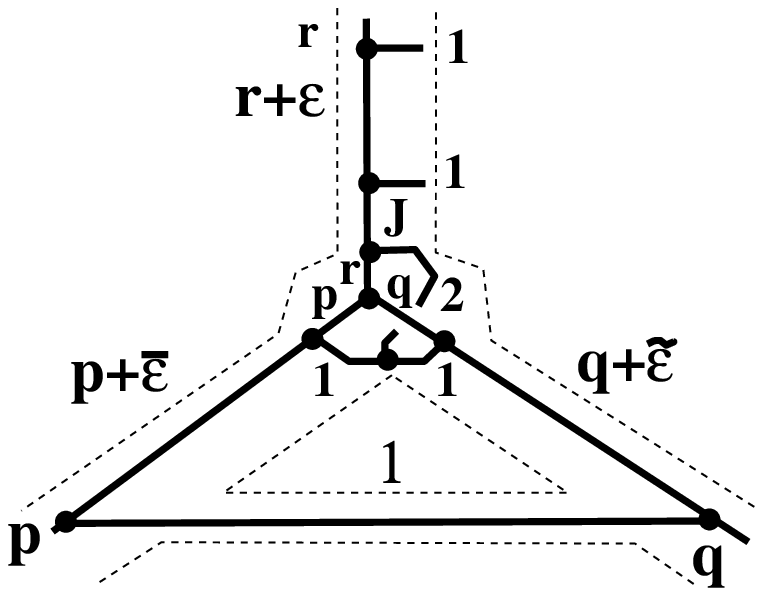,width=3.5cm}}\end{array} 
  =
\nonumber \\ \fl ~~~~~
= \sum_{\epsilon,\bar{\epsilon},\tilde{\epsilon}=\pm1} 
  \sum_{J=r,r+2\epsilon} 
  a_\epsilon(r) 
  \left\{ \!\! \begin{array}{ccc}  
          r & \! r+\epsilon \!  & J \\ 1 & 2 & 1
  \end{array}\!\!\!\!\right\} 
  \tilde{R}_2(p,\bar{\epsilon};q,\tilde{\epsilon};r,J) \!\!
  \begin{array}{c}\mbox{\epsfig{file=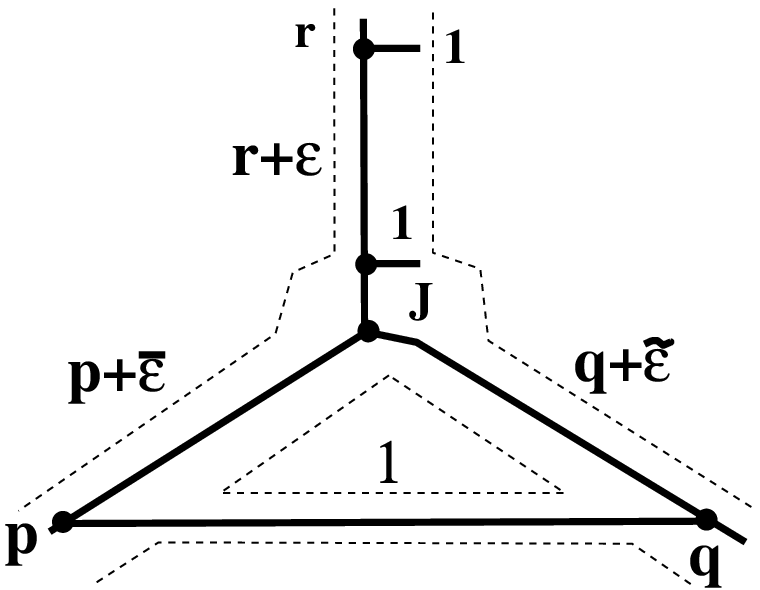,width=3.5cm}}\end{array} 
\!\!.
\nonumber
\end{eqnarray}
We first inserted the loop corresponding to the $\hat{h}[...]$
operator in the extended planar graph corresponding to the
3-vertex. Then we decomposed the result, using equation
\eref{eq:Recursion1}.  Then we have applied the recoupling
theorem \eref{eq:recTheorem} to the 4-vertex
$(r,r+\epsilon,1,2)$. Finally we have reduced the 3-vertex
$(p+\bar{\epsilon},q+\tilde{\epsilon},J)$ to its fundamental form
as in \eref{eq:STEP2}. In the 
last step we have defined:
\begin{equation}
\label{def:tildeR2} 
\begin{array}{c}\setlength{\unitlength}{1 pt}
\begin{picture}(60,65)
       \put( 5, 5){\line( 1, 1){25}} \put( 0,0){$p+\bar{\epsilon}$}
       \put(55, 5){\line(-1, 1){25}} \put(40,0){$q+\tilde{\epsilon}$}
       \put(30,30){\line(0,1){25}}    \put(22,52){$J$}
       \put(30,30){\circle*{3}}       \put(22,32){$r$}
       \put(30,45){\circle*{3}} \put(30,45){\line(1,0){5}}
       \put(35,40){\oval(10,10)[tr]} \put(40,40){\line(0,-1){15}}
       \put(15,15){\circle*{3}}\put(45,15){\circle*{3}}
       \put(15,15){\line(1,0){30}} \put(30,15){\circle*{3}}
       \put(22,10){$\scriptstyle 1$}\put(37,10){$\scriptstyle 1$}
       \put(30,15){\line(1,1){10}} \put(23,17){$\scriptstyle 2$}
       \put(15,19){$\scriptstyle p$}
       \put(45,19){$\scriptstyle q$}
\end{picture}\end{array}
 =  \frac{\tilde{R}_2(p,\bar{\epsilon};q,\tilde{\epsilon};r,J)}{ 
          a_{\bar{\epsilon}}(p) a_{\tilde{\epsilon}}(q)}
~\cdot~ 
\begin{array}{c}\setlength{\unitlength}{1 pt}
\begin{picture}(50,50)
       \put(25,25){\line(-1, -1){20}} \put( 0,0){$p+\bar{\epsilon}$}
       \put(25,25){\line( 1, -1){20}} \put(40,0){$q+\tilde{\epsilon}$}
       \put(25,25){\line(0,1){20}}    \put(17,42){$J$}
       \put(25,25){\circle*{3}}
\end{picture}\end{array}
\end{equation}
A direct comparison with equation \eref{eq:STEP2} shows that for $J=r$ 
\begin{eqnarray*} 
   \tilde{R}_2(p,\bar{\epsilon};q,\tilde{\epsilon};r,r)   
  = {R}_2(p,\bar{\epsilon};q,\tilde{\epsilon};r).
\end{eqnarray*} 
Computing the action of the volume and closing the trace with 
the $\hat{h}_{s_{k}}$ operator we obtain 
\begin{eqnarray*}
\fl  
  {\rm Tr}\left(
  \hat{h}_{s_{k}} \hat{V}
  \hat{h}_{s_{k}}^{-1}
  \frac{\hat{h}_{\alpha_{ij}}-\hat{h}_{\alpha_{ji}}}{2}
  \right)
 \begin{array}{c}\mbox{\epsfig{file=f2.eps,width=3cm}}\end{array} 
\\
\lo{=} - \frac{l_0^3}{4} 
     \sum_{\epsilon,\bar{\epsilon},\tilde{\epsilon}=\pm1} 
     \sum_{J=r,r+2\epsilon} 
  a_\epsilon(r) 
  \tilde{R}_2(p,\bar{\epsilon};q,\tilde{\epsilon};r,J) ~\times~
\\  ~\times~
\frac{\begin{array}{c}\setlength{\unitlength}{.5 pt}
     \begin{picture}(35,25)
        \put(15, 0){\line(0,1){10}}\put(20, 0){\line(0,1){10}}
        \put(15, 0){\line(1,0){ 5}}\put(15,10){\line(1,0){ 5}}
        \put(15,25){\line(1,0){5}}  \put(15,14){$\scriptstyle J$}
        \put(15,15){\oval(30,20)[l]}\put(20,15){\oval(30,20)[r]}
     \end{picture}\end{array}
     \begin{array}{c}\setlength{\unitlength}{.8 pt}
        \begin{picture}(55,40)
        \put( 0,15){\line(1,-1){15}} 
        \put(-3,29){${\scriptstyle r+\epsilon}$}
        \put( 0,15){\line(1, 1){15}} \put(0, 0){${\scriptstyle r}$}
        \put( 0,15){\circle*{3}}
        \put(30,15){\line(-1, 1){15}} \put(28,20){${\scriptstyle 1}$}
        \put(30,15){\line(-1,-1){15}} \put(28, 2){${\scriptstyle 2}$}
        \put(30,15){\circle*{3}}
        \put( 0,15){\line(1,0){30}} \put(12,16){${\scriptstyle 1}$}
        \put(15,30){\line(1,0){25}} \put(15,30){\circle*{3}}
        \put(15, 0){\line(1,0){25}} \put(15, 0){\circle*{3}}
        \put(40, 0){\line(0,1){30}} \put(42,12){${\scriptstyle J}$}
        \end{picture}\end{array}
    }{ \begin{array}{c}\setlength{\unitlength}{.5 pt}
       \begin{picture}(40,40)
            \put(18,32){$\scriptstyle r$}
            \put(18,17){$\scriptstyle J$} 
            \put(18, 2){$\scriptstyle 2$} 
            \put(20,15){\oval(40,30)} \put( 0,15){\line(1,0){40}} 
            \put( 0,15){\circle*{3}}  \put(40,15){\circle*{3}}
     \end{picture}\end{array}
     \begin{array}{c}\setlength{\unitlength}{.5 pt}
     \begin{picture}(40,40)
        \put(18,32){$\scriptstyle 1$}
        \put(18,17){$\scriptstyle J$} 
        \put(10, 2){$\scriptstyle r+\epsilon$} 
        \put(20,15){\oval(40,30)} \put( 0,15){\line(1,0){40}} 
        \put( 0,15){\circle*{3}}  \put(40,15){\circle*{3}}
     \end{picture}\end{array}} 
  ~~ \sqrt{w(p+\bar\epsilon,q+\tilde{\epsilon},r+\epsilon,1)} ~\times
\\  ~\times~
  \delta^r_J \frac{
     \begin{array}{c}\setlength{\unitlength}{.5 pt}
     \begin{picture}(40,40)
        \put(18,32){$\scriptstyle 1$}
        \put(18,17){$\scriptstyle r$} 
        \put(10, 2){$\scriptstyle r+\epsilon$} 
        \put(20,15){\oval(40,30)} \put( 0,15){\line(1,0){40}} 
        \put( 0,15){\circle*{3}}  \put(40,15){\circle*{3}}
     \end{picture}\end{array}}{     
     \begin{array}{c}\setlength{\unitlength}{.5 pt}
     \begin{picture}(35,25)
        \put(15, 0){\line(0,1){10}}\put(20, 0){\line(0,1){10}}
        \put(15, 0){\line(1,0){ 5}}\put(15,10){\line(1,0){ 5}}
        \put(15,25){\line(1,0){5}}  \put(15,15){$\scriptstyle r$}
        \put(15,15){\oval(30,20)[l]}\put(20,15){\oval(30,20)[r]}
     \end{picture}\end{array}
     }
  \begin{array}{c} 
      \mbox{\epsfig{file=f9.eps,height=3cm,silent=}}
  \end{array}
\end{eqnarray*} 
We define
\begin{eqnarray} \fl
A^{\due}(p,\bar{\epsilon};q,\tilde{\epsilon};r) =
\sum_{{\epsilon}=\pm 1} \frac{1}{4}
     ~\sqrt{w(p+\bar{\epsilon},q+\tilde{\epsilon},r+\epsilon,1)}
     ~\tilde{R}_2(p,\bar{\epsilon};q,\tilde{\epsilon};r,r) \times
\nonumber \\
   \times  ~a_\epsilon(r) 
   \frac{\begin{array}{c}\setlength{\unitlength}{.5 pt}
     \begin{picture}(35,25)
        \put(15, 0){\line(0,1){10}}\put(20, 0){\line(0,1){10}}
        \put(15, 0){\line(1,0){ 5}}\put(15,10){\line(1,0){ 5}}
        \put(15,25){\line(1,0){5}}  \put(15,14){$\scriptstyle r$}
        \put(15,15){\oval(30,20)[l]}\put(20,15){\oval(30,20)[r]}
     \end{picture}\end{array}
     \begin{array}{c}\setlength{\unitlength}{.8 pt}
        \begin{picture}(55,40)
        \put( 0,15){\line(1,-1){15}} 
        \put(-3,29){${\scriptstyle r+\epsilon}$}
        \put( 0,15){\line(1, 1){15}} \put(0, 0){${\scriptstyle r}$}
        \put( 0,15){\circle*{3}}
        \put(30,15){\line(-1, 1){15}} \put(28,20){${\scriptstyle 1}$}
        \put(30,15){\line(-1,-1){15}} \put(28, 2){${\scriptstyle 2}$}
        \put(30,15){\circle*{3}}
        \put( 0,15){\line(1,0){30}} \put(12,16){${\scriptstyle 1}$}
        \put(15,30){\line(1,0){25}} \put(15,30){\circle*{3}}
        \put(15, 0){\line(1,0){25}} \put(15, 0){\circle*{3}}
        \put(40, 0){\line(0,1){30}} \put(42,12){${\scriptstyle r}$}
        \end{picture}\end{array}
    }{ \begin{array}{c}\setlength{\unitlength}{.5 pt}
       \begin{picture}(40,40)
            \put(18,32){$\scriptstyle r$}
            \put(18,17){$\scriptstyle r$} 
            \put(18, 2){$\scriptstyle 2$} 
            \put(20,15){\oval(40,30)} \put( 0,15){\line(1,0){40}} 
            \put( 0,15){\circle*{3}}  \put(40,15){\circle*{3}}
     \end{picture}\end{array}
     \begin{array}{c}\setlength{\unitlength}{.5 pt}
     \begin{picture}(40,40)
        \put(18,32){$\scriptstyle 1$}
        \put(18,17){$\scriptstyle r$} 
        \put(10, 2){$\scriptstyle r+\epsilon$} 
        \put(20,15){\oval(40,30)} \put( 0,15){\line(1,0){40}} 
        \put( 0,15){\circle*{3}}  \put(40,15){\circle*{3}}
     \end{picture}\end{array}} 
   \frac{
     \begin{array}{c}\setlength{\unitlength}{.5 pt}
     \begin{picture}(40,40)
        \put(18,32){$\scriptstyle 1$}
        \put(18,17){$\scriptstyle r$} 
        \put(10, 2){$\scriptstyle r+\epsilon$} 
        \put(20,15){\oval(40,30)} \put( 0,15){\line(1,0){40}} 
        \put( 0,15){\circle*{3}}  \put(40,15){\circle*{3}}
     \end{picture}\end{array}}{     
     \begin{array}{c}\setlength{\unitlength}{.5 pt}
     \begin{picture}(35,25)
        \put(15, 0){\line(0,1){10}}\put(20, 0){\line(0,1){10}}
        \put(15, 0){\line(1,0){ 5}}\put(15,10){\line(1,0){ 5}}
        \put(15,25){\line(1,0){5}}  \put(15,15){$\scriptstyle r$}
        \put(15,15){\oval(30,20)[l]}\put(20,15){\oval(30,20)[r]}
     \end{picture}\end{array}
     } =
\label{def:As} \\
\lo{=}
\sum_{{\epsilon}=\pm 1} 
 \frac{1}{4}
 ~\sqrt{w(p+\bar{\epsilon},q+\tilde{\epsilon},r+\epsilon,1)}
 ~R_1(r,\epsilon)  
 ~R_2(p,\bar{\epsilon};q,\tilde{\epsilon},r).
\nonumber
\end{eqnarray}
We have finally obtained the action of  
${\hat{\cal H}}^{E\due}_{\Delta}{}$ on non planar 
3-vertices: 
\begin{eqnarray} \label{eq:ActionHeDAG}  \fl
 \hat{{\cal H}}_{\Delta}^{E\due}{}~ \big| v(p,q,r) \big\rangle 
 = {2 {\rm i} l_0 \over 3}
 \sum_{\bar{\epsilon}=\pm 1} \sum_{\tilde{\epsilon}=\pm 1} 
 \left[
 A^{\due}(p,\bar{\epsilon};q,\tilde{\epsilon};r) 
    \bigg|\begin{array}{c} 
      \mbox{\epsfig{file=f9.eps,height=2.5cm,silent=}}
    \end{array}\!\!\bigg\rangle \right. +
\\ 
\fl ~~ \left.
+ A^{\due}(q,\bar{\epsilon};r,\tilde{\epsilon};p)
    \bigg|\begin{array}{c} 
      \mbox{\epsfig{file=f10.eps,height=2.5cm,silent=}}
    \end{array}\!\!\bigg\rangle 
+ A^{\due}(r,\bar{\epsilon};p,\tilde{\epsilon};q)
    \bigg|\begin{array}{c} 
      \mbox{\epsfig{file=f11.eps,height=2.5cm,silent=}}
    \end{array}\!\!\bigg\rangle \right].
\nonumber 
\end{eqnarray}
where the coeficients $A^{2}$ are given by (5.14).
Notice that this is identical to the action of equation
\eref{eq:ActionHe} with
$A^{\uno}(p,\bar{\epsilon};q,\tilde{\epsilon};r)$ replaced by
$A^{\due}(p,\bar{\epsilon};q,\tilde{\epsilon};r)$.


\subsection{Other orderings
\label{sec:othSym}}

\begin{table}[t]
\caption{Matrix
   elements $A^{\uno}(p,\bar\epsilon;q,\tilde\epsilon;r)$ and
   $A^{\due}(p,\bar\epsilon;q,\tilde\epsilon;r)$ of the
   non-symmetric and symmetric euclidean Thiemann's hamiltonian
   constraints,  for the simplest 3-vertices. We have denoted by
   $(p^\prime=p\pm\bar\epsilon,q^\prime=q\pm\tilde\epsilon,r)$ the new
   colors of the 3-vertex.
   \label{table:1}}
\begin{indented}
\item[]\begin{tabular}{@{}lllrrccll}
\br
p & q & r & $\bar\epsilon$  & $\tilde\epsilon$ &
($p^\prime$,$q^\prime$,$r$) & 
          & $A^{\uno}(p,\bar\epsilon;q,\tilde\epsilon;r)$
          & $A^{\due}(p,\bar\epsilon;q,\tilde\epsilon;r)$ \\
\mr
1 & 0 & 1 & 1 & 1 & (2,1,1) && $0$ 
                            &  $1/16 \sqrt[4]{2}$ \\
  &   &   & 1 &-1 & ( NO )  && $0$ 
                            &  $0$  \\
  &   &   &-1 & 1 & (0,1,1)  && $0$  
                            &  $0$\\
  &   &   &-1 &-1 & ( NO ) &&  $0$ 
                            &  $0$ \\
\mr
1 & 1 & 0 & 1 & 1 & (2,2,0) && $0$ 
                            &  $0$ \\
  &   &   & 1 &-1 & ( NO )  && $0$ 
                            &  $0$ \\
  &   &   &-1 & 1 & ( NO )  && $0$  
                            &  $0$ \\
  &   &   &-1 &-1 & (0,0,0) && $0$ 
                            &  $0$ \\
\mr
2 & 0 & 2 & 1 & 1 & (3,1,2) && $0$ 
                            &  $1/24 \sqrt[4]{60}$ \\
  &   &   & 1 &-1 & ( NO )  && $0$ 
                            &  $0$  \\
  &   &   &-1 & 1 & (1,1,2) && $0$  
                            &  $-1/18 \sqrt[4]{12}$ \\
  &   &   &-1 &-1 & ( NO ) &&  $0$ 
                            &  $0$ \\
\mr
2 & 2 & 0 & 1 & 1 & (3,3,0) && $0$ 
                            &  $0$ \\
  &   &   & 1 &-1 & (3,1,0) && $0$ 
                            &  $0$ \\
  &   &   &-1 & 1 & (1,3,0) && $0$  
                            &  $0$ \\
  &   &   &-1 &-1 & (1,1,0) && $0$ 
                            &  $0$ \\
\mr
2 & 1 & 1 & 1 & 1 & (3,2,1) && $1/16 \sqrt[4]{2}$ 
                            &  $1/16 \sqrt[4]{5}$ \\
  &   &   & 1 &-1 & (3,0,1) && $0$ 
                            &  $0$  \\
  &   &   &-1 & 1 & (1,2,1) && $5/48 \sqrt[4]{2}$  
                            &  $5/48 \sqrt[4]{2}$\\
  &   &   &-1 &-1 & (1,0,1) && $1/32 \sqrt[4]{2}$ 
                            &  $0$ \\
\mr
1 & 2 & 1 & 1 & 1 & (2,3,1) && $-1/16 \sqrt[4]{2}$ 
                            &  $-1/16 \sqrt[4]{5}$ \\
  &   &   & 1 &-1 & (2,1,1) && $-5/48 \sqrt[4]{2}$  
                            &  $-5/48 \sqrt[4]{2}$\\
  &   &   &-1 & 1 & (0,3,1) && $0$ 
                            &  $0$  \\
  &   &   &-1 &-1 & (0,1,1) && $-1/32 \sqrt[4]{2}$ 
                            &  $0$ \\
\mr
1 & 1 & 2 & 1 & 1 & (2,2,2) && $0$  
                            &  $0$ \\
  &   &   & 1 &-1 & (2,0,2) && $1/24 \sqrt[4]{12}$  
                            &  $0$ \\
  &   &   &-1 & 1 & (0,2,2) && $-1/24 \sqrt[4]{12}$  
                            &  $0$ \\
  &   &   &-1 &-1 & (0,0,2) && $0$   
                            &  $0$ \\
\mr
2 & 2 & 2 & 1 & 1 & (3,3,2) && $0$   
                            &  $0$ \\
  &   &   & 1 &-1 & (3,1,2) && $-1/12 (\sqrt[4]{5} - \sqrt[4]{2})$ 
                            &  $-1/24 \sqrt[4]{30}$ \\
  &   &   &-1 & 1 & (1,3,2) && $1/12 (\sqrt[4]{5} - \sqrt[4]{2})$
                            &  $1/24 \sqrt[4]{30}$ \\
  &   &   &-1 &-1 & (1,1,2) && $0$  
                            &  $0$ \\
\br  
\end{tabular}
\end{indented}
\end{table}

Consider all the permutations of the four operator
$\hat{V}$, $\hat{h}_{s_{k}}$, $\hat{h}^{-1}_{s_{k}}$ and
$(\hat{h}_{\alpha_{ij}}-\hat{h}_{\alpha_{ji}})/2$. These are
the ``natural'' orderings of the hamiltonian operator.
Equations \eref{hamfin1} and \eref{hamfin2} exhaust the
orderings in which the volume (the only operator that
``grasps'' the cylindrical function) sits between
$\hat{h}_{s_{k}}$ and $\hat{h}^{-1}_{s_{k}}$. But these two
are the only non vanishing cases, for the following reason.
The $\hat{h}$ operators commute with each other and
therefore we have to consider only the permutations of the
three operator: $\hat{V}$, $\hat{O}_{(0)}=\hat{h}_{s_{k}}
\hat{h}^{-1}_{s_{k}}$ and
$\hat{O}_{(2)}=(\hat{h}_{\alpha_{ij}}-\hat{h}_{\alpha_{ji}})/2$.
By the retracting identity, the $\hat{O}_{(0)}$ operator is
a two-index operator in the spin $0$ representation
associated to the vertex $v$ while the $\hat{O}_{(2)}$
operator is a two-index operator in the spin $1$
representation associated to the same vertex $v$. But in the
hamiltonian constraint there is a trace, and thus in all the
orderings of the operators we have to take a trace in the
representation associated to the vertex. The volume operator
modifies only the recoupling in the vertex and not the
representation associated to it. Therefore we will always
have a trace between a spin $0$ and a spin $1$
representation, which gives zero. We can conclude that
\eref{hamfin1x} and \eref{hamfin2x} are the two only natural
orderings for the HCO and their action reduces to the
computation of equations \eref{hamfin1} and \eref{hamfin2}.

Following Thiemann \cite{Thiemann}, we consider the {\em
symmetric\/} euclidean HCO
\begin{equation}
\hat{\cal H}_{symm}^{E}[N]
  = \frac{1}{2} \left(
      \hat{\cal H}_{\gamma}^{E\uno}[N] 
    + \hat{\cal H}_{\gamma}^{E\uno}[N]^\dagger
   \right).
\end{equation}
Using the transformation properties of $\hat h$ under 
conjugation, we have 
\begin{equation}
\hat{\cal H}_{symm}^{E}[N]
  = \frac{1}{2} \left(
      \hat{\cal H}_{\gamma}^{E\uno}[N] 
    + \hat{\cal H}_{\gamma}^{E\due}[N]
   \right).
\end{equation}
The coefficients $A^{s}$ for the symmetric operators on
trivalent vertices are
\begin{equation}
  A^{s}(p,\bar{\epsilon};q,\tilde{\epsilon};r) = \frac{1}{3} 
    \bigg[A^{\uno}(p,\bar{\epsilon};q,\tilde{\epsilon};r)
        + A^{\due}(p,\bar{\epsilon};q,\tilde{\epsilon};r) \bigg]
\end{equation}
In \tref{table:1} we give the values of 
$A^{\uno}$ and $A^{\due}$ for the 
simplest 3-vertices.


\section{Conclusions}

We have computed matrix elements of Thiemann's euclidean
hamiltonian constraint, for its two natural orderings.  In
particular we have computed the action of the operator on
any trivalent vertex in closed algebraic form.

We collect here our results and express them in compact
form.  Following \cite{outline}, we define the operator
$\hat D_{vJK\bar\epsilon\tilde\epsilon}$ that acts on the
vertex $v$ by adding two new vertices on the edges $J$ and
$K$ adjacent to $v$, and altering the colors of the two
newly created edges by $\bar\epsilon,\tilde\epsilon=\pm 1$
\begin{equation}
\hat{D}_{vJK\bar{\epsilon}\tilde{\epsilon}}
  \left|
  \begin{array}{c}\setlength{\unitlength}{1 pt}
  \begin{picture}(65,50)
     \put( 0, 0){\line( 1,  1){20}} 
     \put( 0,20){\line( 1,  0){20}} 
     \put( 0,40){\line( 1, -1){20}}
     \put(20,20){\circle*{3}}
     \put(20,20){\line( 3,  2){30}} \put(52,42){$K$} 
     \put(20,20){\line( 3, -2){30}} \put(52, 2){$J$} 
     \put(45, 8){$p$}  \put(45,32){$q$} 
  \end{picture}\end{array}
  \right\rangle
= \left|
  \begin{array}{c}\setlength{\unitlength}{1 pt}
  \begin{picture}(65,50)
     \put( 0, 0){\line( 1,  1){20}} 
     \put( 0,20){\line( 1,  0){20}} 
     \put( 0,40){\line( 1, -1){20}}
     \put(20,20){\circle*{3}}
     \put(20,20){\line( 3,  2){30}} \put(52,42){$K$} 
     \put(20,20){\line( 3, -2){30}} \put(52, 2){$J$} 
     \put(45, 8){$p$}  \put(45,32){$q$} 
     \put(15, 2){$p\!+\!\bar{\epsilon}$}  
     \put(15,32){$q\!+\!\tilde{\epsilon}$}
     \put(35,10){\line(0,1){20}}
     \put(35,10){\circle*{3}} \put(35,30){\circle*{3}}
     \put(37,20){$1$}
  \end{picture}\end{array}
  \right\rangle.
\end{equation}
Using this operator, the action of Thiemann's
HCO over a generic spin-network 
state is  
\begin{equation} \fl
\hat{\cal H}^{E}[N] \Psi_s = {\rm i} l_0 \!
   \sum_{v\in{{\cal V}(s)}} N(x_v) \!\! \!\! \!\!
   \sum_{\begin{array}{c} 
         \scriptstyle {I,J,K\in {\cal E}(v)} \\[-1mm] 
         \scriptstyle {\bar{\epsilon},\tilde{\epsilon}=\pm 1}
         \end{array}} 
   \frac{1}{E(v)}\ 
   \hat{A}_{vIJK\bar{\epsilon}\tilde{\epsilon}}\ 
   \hat{D}_{vJK\bar{\epsilon}\tilde{\epsilon}}\ 
   \Psi_s
\end{equation}
where ${\cal V}(s)$ is the set of the vertices of $s$ and
${\cal E}(v)$ is the set of the edges adjacent to $v$. In
general, $\hat{A}_{vIJK\bar{\epsilon}\tilde{\epsilon}}$ are
matrices in the finite dimensional space of the contractors
at the vertex $v$, functions of the colors of the adjacent
edges.  On trivalent vertices, the space of these
contractors has dimension one, and therefore the matrices
$\hat{A}_{vIJK\bar{\epsilon}\tilde{\epsilon}}\ $ are
numbers.  Let $s$ have trivalent gauge-invariant vertices
only.  Denote as $(r,p,q)$ the colors associated to the
edges $I,J,K$ of a vertex $v$. Then
\begin{eqnarray} \fl
 \hat{{\cal H}}^{E\unodue}~ \Psi_s  
 =  {2 {\rm i} l_0 \over 3} \sum_{v\in{\cal V}(s)}
 \sum_{\bar{\epsilon}=\pm 1} \sum_{\tilde{\epsilon}=\pm 1} 
 \left[
 A^{\unodue}(p,\bar{\epsilon};q,\tilde{\epsilon};r) 
  \hat{D}_{vJK\bar{\epsilon}\tilde{\epsilon}}
  \right.
\\ 
\lo{} \left.
+ A^{\unodue}(q,\bar{\epsilon};r,\tilde{\epsilon};p)
  \hat{D}_{vKI\bar{\epsilon}\tilde{\epsilon}}
+ A^{\unodue}(r,\bar{\epsilon};p,\tilde{\epsilon};q)
  \hat{D}_{vIJ\bar{\epsilon}\tilde{\epsilon}}
 \right]\ \Psi_s,
\nonumber 
\end{eqnarray}
where $i=1,2$ for the two orderings we have studied. 
The explicit values of the matrix elements 
$A^{\uno}(p,\bar{\epsilon};q,\tilde{\epsilon};r)$ and
$A^{\due}(p,\bar{\epsilon};q,\tilde{\epsilon};r)$ are 
\begin{eqnarray}
\fl
A^{\uno}(p,\bar{\epsilon};q,\tilde{\epsilon};r) =
\sum_{{\epsilon}=\pm 1} 
 \frac{1}{4}
 ~\sqrt{w(p,q,r+\epsilon,1)}
 ~R_1(r,\epsilon)  
 ~R_2(p,\bar{\epsilon};q,\tilde{\epsilon},r).
\\ \fl
A^{\due}(p,\bar{\epsilon};q,\tilde{\epsilon};r) =
\sum_{{\epsilon}=\pm 1} 
 \frac{1}{4}
 ~\sqrt{w(p+\bar{\epsilon},q+\tilde{\epsilon},r+\epsilon,1)}
 ~R_1(r,\epsilon)  
 ~R_2(p,\bar{\epsilon};q,\tilde{\epsilon},r).
\end{eqnarray}
where $R_1(r,\epsilon)$, $R_2(p,\bar{\epsilon};q,\tilde{\epsilon},r)$
and $w(a,b,c,1)$ are explicitly given by  \eref{eq:defR1}
\eref{eq:defR2} and \eref{eq:vol3}.
We give some values of matrix elements in Table 1. 

The technique we have used can be applied in a
straightforward manner to vertices of any valence.  The only
limitation is given by the lack of an explicit algebraic
formula corresponding to equations \eref{VOL:3vertex} and
\eref{eq:vol3}, for vertices of valence greater than
four. The action of the volume operator is completly known;
however, explicit values of its matrix elements are so far
known throughout an algorithmic procedure \cite{DePiRo96}
only.  It is interesting to note that nothing in the action
of the operator seems to depend on the continuous moduli
studied in \cite{Grott}, which distinguish knots with high
valence intersections. This might indicate that these
parameters are superfluous and the theory lives actually in
a smaller space. In this regard, see \cite{Zapata}.
Finally, we expect that the technique used here can be
applied to the second term in (\ref{totham}) as well, and
therefore we expect that the total lorentzian hamiltonian
constraint can be analyzed along the lines of this work.


\ack 

We thank Thomas Thiemann for help in understanding his
construction, Jurek Lewandowski for valuable discussions,
criticisms and insights, Andreas Freund for pointing out a
mistake in an earlier version of this work and Luca Lusanna
for useful discussions on the quantization of constrained
dynamical system.  This work has been partially supported by
the NSF grants PHY-9515506, PHY-90-12099, PHY-9514240 and
PHY-5-3840400 and by the INFN grant ``Iniziativa specifica
FI-41''.


\appendix

\section{Explicit formulas for the volume operator.
\label{ap:VOLUME}}

A discussed in section 2.4, there exist two different versions of the
volume operator \cite{Lewandowski97}. They have the same action on 
generic trivalent vertices. 
The volume operator acts on a cylindrical function $\Psi_{\gamma,\psi}$
\cite{VolumeAL,Lewandowski97,ThimVolume} by:
\begin{eqnarray} \fl 
 \hat{V}\ \Psi_{\gamma,\psi}
  = l_0^{3}\sum_{v \in {\cal V}(\gamma)} \hat{V}_{v}\ \Psi_{\gamma,\psi} = 
      \left( l_0^{3} \sum_{v \in V(\gamma)}
        \sqrt{ \hat{V^2} } \right)
 \Psi_{\gamma,\psi} 
\label{Wdef}\\ 
\lo{=}   
  \left(l_0^{3}\sum_{v \in {\cal V}(\gamma)}\sqrt{\left|{i \over
16\cdot 3!}
  \sum_{\e_{I}\cup \e_{J} \cup \e_{K} = v}\epsilon(\e_{I},\e_{J},\e_{K})
  \hat{W}_{IJK} \right|}\right)
  \psi(g_{1},\ldots,g_{N}) ~~,
\nonumber 
\end{eqnarray}
where,  ${\cal V}(\gamma)$ is the set of all vertices of $\gamma$,
$g_{I}=j_I(\bar A(\e_{l}))$, and $X_I=X(g_{I})$ is the right invariant
vector field on $SU(2)$. All edges are chosen as outgoing.
Notice that the action of the operator is given as a sum over
terms in which three edges are ``grasped'' namely acted upon by
one of the left-invariant vector fields.

Here $\epsilon(\e_{I},\e_{J},\e_{K}) = {\rm sgn}({\rm
det}(\dot{e}_{I}(0), \dot{e}_{J}(0),\dot{e}_{K}(0)))$ and we have
defined
\begin{equation}
\hat{W}_{IJK}   = 2\ \epsilon_{ijk}\ X_{I}^{i}X_{J}^{j}X_{K}^{k}.   
\end{equation} 
($\hat{q}$ in \cite{VolumeAL,ThimVolume} is related to $\hat{W}$ in 
\cite{DePiRo96} by $\hat{W}_{[IJK]} = 2~ \hat{q}_{IJK}$.)
As noted in \cite{Lewandowski97}, the difference between 
the volume introduced by Rovelli and Smolin in
\cite{RovelliSmolin95} and the one later 
introduced by Ashtekar and Lewandowski 
\cite{VolumeAL} is given by 
\begin{eqnarray}
\hat{V}^2_{\rm RS} &=& \frac{1}{3!}
  \sum_{\e_{I}\cup \e_{J} \cup \e_{K} = v} 
  \left| \frac{\rm i}{16} 
  \hat{W}_{[IJK]} \right| ~~, \label{eq:volRS}\\
 \hat{V}^2_{\rm AL} &=& \left|
  \frac{\rm i}{16} \frac{1}{3!}
  \sum_{\e_{I}\cup \e_{J} \cup \e_{K} = v} 
  \epsilon(\e_{I},\e_{J},\e_{K})
  \hat{W}_{[IJK]} \right|\  . \label{eq:volAL}
\end{eqnarray}
This difference is due to a different regularization
procedure used to derive the operator from its classical
expression. From equations \eref{eq:volAL} and
\eref{eq:volRS} it follows that the action of the two
operators coincides when evaluated on generic (non coplanar)
trivalent vertices, since there is only one term in the sum.

Notice that the action of this operator is defined on gauge
non-invariant states as well, since it is defined only in
terms of left invariant vector field and therefore over the
whole $\bar{\cal A}$ space.  Furthermore, explicit
inspection of the definition of the operator shows that its
action on a gauge non-invariant n-valent vertex coincides
with one of the terms of its action on an a (n+1)-valent
vertex.

  \begin{figure}[t]
  \begin{indented}\item[]
  {\mbox{\epsfig{file=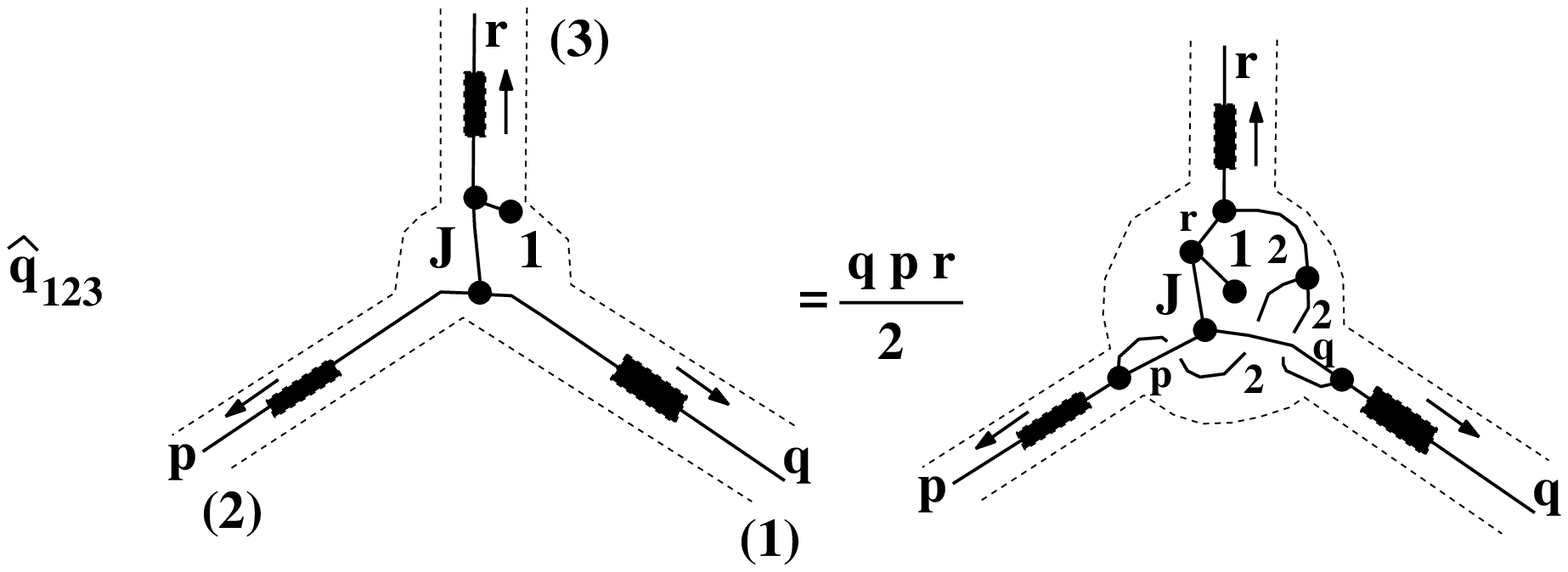,width=10cm}}}
  \end{indented}
  \caption{Graphical representation of the
  grasping of the volume operator in the case of a gauge 
  non-invariant 3-valent vertex. \label{graspings}}
  \end{figure}

To see this, consider the action of the volume on a 3-valent
gauge non-invariant vertex in which the three colors of the
adjacent edges are $p,q,r$, the color of the 
representation at the vertex is $t$, and the intertwiner is $c$. 
And consider
the action of the volume on a gauge-invariant 4-valent vertex
with edges colored $p,q,r,t$ and intertwiner $c$. The latter is
given by the sum of of four terms. In one of these four terms,
the three left invariant vector fields in (A.2) grasp the three
edges colored $p,q,r$.  It is obvious that the resulting algebra
is precisely the same as the one of the single term in the action
of the volume on the trivalent gauge non-invariant vertex.

Using this, we have, in graphical form, the key result on which
the computations of this paper are based::
\begin{equation}
  \hat{W}_{[012]} \begin{array}{c}\setlength{\unitlength}{1 pt}
   \begin{picture}(50,35)
          \put( 0,0){$q$}\put( 0,30){$p$}
          \put(45,30){$r$}
          \put( 5, 5){\line(1, 1){15}}
          \put( 5,35){\line(1,-1){15}}
          \put(30,20){\line(1, 1){15}}
          \put(30,20){\line(1,-1){10}}
          \put(39,11){\circle*{3}} \put(37,15){$t$}
          \put(20,20){\line(1,0){10}}\put(22,25){$J$}
          \put(20,20){\circle*{3}}\put(30,20){\circle*{3}}
          \bezier{20}( 5,20)( 5,35)(25,35)
          \bezier{20}(45,20)(45,35)(25,35)
          \bezier{20}( 5,20)( 5, 5)(25, 5)
          \bezier{20}(45,20)(45, 5)(25, 5)
   \end{picture}\end{array}  
  = \sum_I W_{[012]}^{(4)}{}_J{}^I(q,p,r,t)
\begin{array}{c}\setlength{\unitlength}{1 pt}
   \begin{picture}(50,35)
          \put( 0,0){$q$}\put( 0,30){$p$}
          \put(45,30){$r$}
          \put( 5, 5){\line(1, 1){15}}
          \put( 5,35){\line(1,-1){15}}
          \put(30,20){\line(1, 1){15}}
          \put(30,20){\line(1,-1){10}}
          \put(39,11){\circle*{3}} \put(37,15){$t$}
          \put(20,20){\line(1,0){10}}\put(22,25){$I$}
          \put(20,20){\circle*{3}}\put(30,20){\circle*{3}}
          \bezier{20}( 5,20)( 5,35)(25,35)
          \bezier{20}(45,20)(45,35)(25,35)
          \bezier{20}( 5,20)( 5, 5)(25, 5)
          \bezier{20}(45,20)(45, 5)(25, 5)
   \end{picture}\end{array},
\label{uffa}
\end{equation}
where we have indicated by $[012]$ the triple of edges
colored $q,p$ and $r$.  In the computations in the text, we
have the particular case in which $r$ is $r\pm1$ and
$t=1$. We write
\begin{equation} 
 \hat{W}_{[012]} \ket{v_J} 
   = \sum_I W_{[012]}^{(4)}{}_J{}^I(q,p,r\pm1,1) \ket{v_I}. 
\end{equation}
where $\ket{v_I}$ is a compact notation for the vertex
with internal edge colored $I$.
$W_{[012]}^{(4)}{}_J{}^I$ was explicitly computed in 
\cite{DePiRo96} in terms of 9-$j$ and 6-$j$ symbols: 
\begin{eqnarray}\fl
W^{(4)}_{[012]}{}_J{}^I(q,p,r\pm1,1)=
  \frac{q~p~(r\pm1)~  \Delta_I \ 
        \left\{\begin{array}{ccc}
           {q} & {p} & {I} \\
           {q} & {p} & {J} \\
           {2} & {2} & {2}
        \end{array}\right\}\ 
        {Tet\!\left[\begin{array}{ccc}
            J     & I     & 1 \\
            r\pm1 & r\pm1 & 2
        \end{array}\right]}
      }{\theta(q,p,I)\theta(r\pm1,1,I)\theta(2,J,I)} =
\nonumber \\[2 mm] 
\lo{=} q~p~(r\pm1) 
\frac{
     \begin{array}{c}\setlength{\unitlength}{.5 pt}
     \begin{picture}(35,25)
        \put(15, 0){\line(0,1){10}}\put(20, 0){\line(0,1){10}}
        \put(15, 0){\line(1,0){ 5}}\put(15,10){\line(1,0){ 5}}
        \put(15,25){\line(1,0){5}}  \put(15,14){$\scriptstyle I$}
        \put(15,15){\oval(30,20)[l]}\put(20,15){\oval(30,20)[r]}
     \end{picture}\end{array}
   }{\begin{array}{c}\setlength{\unitlength}{.5 pt}
     \begin{picture}(40,40)
        \put(18,34){$\scriptstyle q$}
        \put(18,19){$\scriptstyle p$} 
        \put(18, 2){$\scriptstyle I$} 
        \put(20,15){\oval(40,30)} \put( 0,15){\line(1,0){40}} 
        \put( 0,15){\circle*{3}}  \put(40,15){\circle*{3}}
     \end{picture}\end{array}
     \begin{array}{c}\setlength{\unitlength}{.5 pt}
     \begin{picture}(40,40)
        \put(18,34){$\scriptstyle r\pm1$}
        \put(18,19){$\scriptstyle 1$} 
        \put(18, 3){$\scriptstyle I$} 
        \put(20,15){\oval(40,30)} \put( 0,15){\line(1,0){40}} 
        \put( 0,15){\circle*{3}}  \put(40,15){\circle*{3}}
     \end{picture}\end{array}
   }
\begin{array}{c}
  \setlength{\unitlength}{1 pt}
  \begin{picture}(110,60)
  \put(15,40){\line(0,1){10}}  \put(17,38){$q$}
  \put(15,32){\oval(20,20)[tl]} 
  \put(15,42){\circle*{3}}    
  \put(40,30){\line(0,1){30}}  \put(42,38){$p$}
  \put(40,32){\oval(20,20)[tl]} 
  \put(40,42){\circle*{3}}     \put(40,30){\circle*{3}}
  \put(40,60){\circle*{3}}
  \put(65,30){\line(0,1){30}}  \put(67,38){$r\pm1$}
  \put(65,32){\oval(20,20)[tl]} 
  \put(65,42){\circle*{3}}     \put(65,30){\circle*{3}}
  \put(65,60){\circle*{3}}
  \put(90,40){\line(0,1){10}}  \put(92,38){$1$}
  \put( 5,32){\line(0,-1){12}} \put(15, 2){${\scriptstyle  2}$}
  \put(30,28){\line(0,-1){18}} \put(22,15){${\scriptstyle  2}$}
  \put(55,28){\line(0,-1){ 8}} \put(40, 2){${\scriptstyle  2}$}
  \put(30,20){\oval(50,20)[b]} \put(30,10){\circle*{3}}
  \put(40,40){\oval(50,20)[bl]}\put(65,40){\oval(50,20)[br]}
  \put(40,30){\line(1,0){20}}  \put(42,20){$J$} 
  \put(40,50){\oval(50,20)[tl]}\put(65,50){\oval(50,20)[tr]}
  \put(40,60){\line(1,0){20}}  \put(52,50){$I$}
\end{picture}\end{array}  
\end{eqnarray}
This is the action on non-normalized states. Inserting the 
vertex normalization (equation (7.28) of \cite{DePiRo96})
we get:
\begin{eqnarray*}
\fl \hat{W}_{[012]}
\sqrt{\frac{
     \begin{array}{c}\setlength{\unitlength}{.5 pt}
     \begin{picture}(40,40)
        \put(18,34){$\scriptstyle q$}
        \put(18,19){$\scriptstyle p$} 
        \put(18, 3){$\scriptstyle J$} 
        \put(20,15){\oval(40,30)} \put( 0,15){\line(1,0){40}} 
        \put( 0,15){\circle*{3}}  \put(40,15){\circle*{3}}
     \end{picture}\end{array}
     \begin{array}{c}\setlength{\unitlength}{.5 pt}
     \begin{picture}(40,40)
        \put(18,34){$\scriptstyle r\pm1$}
        \put(18,19){$\scriptstyle 1$} 
        \put(18, 3){$\scriptstyle J$} 
        \put(20,15){\oval(40,30)} \put( 0,15){\line(1,0){40}} 
        \put( 0,15){\circle*{3}}  \put(40,15){\circle*{3}}
     \end{picture}\end{array}
   }{\begin{array}{c}\setlength{\unitlength}{.5 pt}
     \begin{picture}(35,25)
        \put(15, 0){\line(0,1){10}}\put(20, 0){\line(0,1){10}}
        \put(15, 0){\line(1,0){ 5}}\put(15,10){\line(1,0){ 5}}
        \put(15,25){\line(1,0){5}}  \put(15,15){$\scriptstyle J$}
        \put(15,15){\oval(30,20)[l]}\put(20,15){\oval(30,20)[r]}
     \end{picture}\end{array}
   }}
   \ket{v_J}_{N} = 
\sum_I 
    {\rm i}~ W_{[012]}^{(4)}{}_J{}^I(p,q,r\pm1,1)
\sqrt{\frac{
     \begin{array}{c}\setlength{\unitlength}{.5 pt}
     \begin{picture}(40,40)
        \put(18,32){$\scriptstyle q$}
        \put(18,17){$\scriptstyle p$} 
        \put(18, 2){$\scriptstyle I$} 
        \put(20,15){\oval(40,30)} \put( 0,15){\line(1,0){40}} 
        \put( 0,15){\circle*{3}}  \put(40,15){\circle*{3}}
     \end{picture}\end{array}
     \begin{array}{c}\setlength{\unitlength}{.5 pt}
     \begin{picture}(40,40)
        \put(18,32){$\scriptstyle r\pm1$}
        \put(18,17){$\scriptstyle 1$} 
        \put(18, 2){$\scriptstyle I$} 
        \put(20,15){\oval(40,30)} \put( 0,15){\line(1,0){40}} 
        \put( 0,15){\circle*{3}}  \put(40,15){\circle*{3}}
     \end{picture}\end{array}
   }{\begin{array}{c}\setlength{\unitlength}{.5 pt}
     \begin{picture}(35,25)
        \put(15, 0){\line(0,1){10}}\put(20, 0){\line(0,1){10}}
        \put(15, 0){\line(1,0){ 5}}\put(15,10){\line(1,0){ 5}}
        \put(15,25){\line(1,0){5}}  \put(15,15){$\scriptstyle I$}
        \put(15,15){\oval(30,20)[l]}\put(20,15){\oval(30,20)[r]}
     \end{picture}\end{array}
   }}
   \ket{v_I}_{N}.
\end{eqnarray*}
Thus we can write the action of the (square of the) volume
operator on the normalized states as
\begin{equation}
\hat{W}_{[012]} \ket{v_J}_{N} =  
\sum_I {\rm i} \widetilde{W}_{[012]}^{(4)}{}_J{}^I(p,q,r\pm1,1)
 \ket{v_I}_{N},
\end{equation} 
where $\widetilde{W}_{[012]}^{(4)}{}_r{}^t(p,q,r\pm1,1)$ 
are the matrix elements of the $\hat{W}_{[012]}$ vertex 
operator in the normalized basis. Their values are explicitly given by:
\begin{eqnarray} \fl
\widetilde{W}_{[012]}^{(4)}{}_J{}^I(p,q,r\pm1,1) = 
\label{theWformula} \\
\lo{=} q~p~(r\pm1)
\sqrt{\frac{
     \begin{array}{c}\setlength{\unitlength}{.5 pt}
     \begin{picture}(35,25)
        \put(15, 0){\line(0,1){10}}\put(20, 0){\line(0,1){10}}
        \put(15, 0){\line(1,0){ 5}}\put(15,10){\line(1,0){ 5}}
        \put(15,25){\line(1,0){5}}  \put(15,15){$\scriptstyle J$}
        \put(15,15){\oval(30,20)[l]}\put(20,15){\oval(30,20)[r]}
     \end{picture}\end{array}
     \begin{array}{c}\setlength{\unitlength}{.5 pt}
     \begin{picture}(35,25)
        \put(15, 0){\line(0,1){10}}\put(20, 0){\line(0,1){10}}
        \put(15, 0){\line(1,0){ 5}}\put(15,10){\line(1,0){ 5}}
        \put(15,25){\line(1,0){5}}  \put(15,15){$\scriptstyle I$}
        \put(15,15){\oval(30,20)[l]}\put(20,15){\oval(30,20)[r]}
     \end{picture}\end{array}
   }{\begin{array}{c}\setlength{\unitlength}{.5 pt}
     \begin{picture}(40,40)
        \put(18,32){$\scriptstyle q$}
        \put(18,17){$\scriptstyle p$} 
        \put(18, 2){$\scriptstyle J$} 
        \put(20,15){\oval(40,30)} \put( 0,15){\line(1,0){40}} 
        \put( 0,15){\circle*{3}}  \put(40,15){\circle*{3}}
     \end{picture}\end{array}
     \begin{array}{c}\setlength{\unitlength}{.5 pt}
     \begin{picture}(40,40)
        \put(18,32){$\scriptstyle q$}
        \put(18,17){$\scriptstyle p$} 
        \put(18, 2){$\scriptstyle I$} 
        \put(20,15){\oval(40,30)} \put( 0,15){\line(1,0){40}} 
        \put( 0,15){\circle*{3}}  \put(40,15){\circle*{3}}
     \end{picture}\end{array}
     \begin{array}{c}\setlength{\unitlength}{.5 pt}
     \begin{picture}(40,40)
        \put(18,32){$\scriptstyle r\pm1$}
        \put(18,17){$\scriptstyle 1$} 
        \put(18, 2){$\scriptstyle J$} 
        \put(20,15){\oval(40,30)} \put( 0,15){\line(1,0){40}} 
        \put( 0,15){\circle*{3}}  \put(40,15){\circle*{3}}
     \end{picture}\end{array}
     \begin{array}{c}\setlength{\unitlength}{.5 pt}
     \begin{picture}(40,40)
        \put(18,32){$\scriptstyle r\pm1$}
        \put(18,17){$\scriptstyle 1$} 
        \put(18, 2){$\scriptstyle I$} 
        \put(20,15){\oval(40,30)} \put( 0,15){\line(1,0){40}} 
        \put( 0,15){\circle*{3}}  \put(40,15){\circle*{3}}
     \end{picture}\end{array}
   }}  \!
\begin{array}{c}
  \setlength{\unitlength}{1 pt}
  \begin{picture}(110,60)
  \put(15,40){\line(0,1){10}}  \put(17,38){$q$}
  \put(15,32){\oval(20,20)[tl]} 
  \put(15,42){\circle*{3}}   
  \put(40,30){\line(0,1){30}}  \put(42,38){$p$}
  \put(40,32){\oval(20,20)[tl]} 
  \put(40,42){\circle*{3}}     \put(40,30){\circle*{3}}
  \put(40,60){\circle*{3}}
  \put(65,30){\line(0,1){30}}  \put(67,38){$r\pm1$}
  \put(65,32){\oval(20,20)[tl]} 
  \put(65,42){\circle*{3}}     \put(65,30){\circle*{3}}
  \put(65,60){\circle*{3}}  
  \put(90,40){\line(0,1){10}}  \put(92,38){$1$}
  \put( 5,32){\line(0,-1){12}} \put(15, 2){${\scriptstyle  2}$}
  \put(30,28){\line(0,-1){18}} \put(22,15){${\scriptstyle  2}$}
  \put(55,28){\line(0,-1){ 8}} \put(40, 2){${\scriptstyle  2}$}
  \put(30,20){\oval(50,20)[b]} \put(30,10){\circle*{3}}
  \put(40,40){\oval(50,20)[bl]}\put(65,40){\oval(50,20)[br]}
  \put(40,30){\line(1,0){20}}  \put(42,20){$J$} 
  \put(40,50){\oval(50,20)[tl]}\put(65,50){\oval(50,20)[tr]}
  \put(40,60){\line(1,0){20}}  \put(52,50){$I$}
\end{picture}\end{array}  
\!\!\!\!\!\!\!\!
. \nonumber 
\end{eqnarray}
In this form it is obvious that
$\widetilde{W}_{[012]}^{(4)}{}_J{}^I(p,q,r\pm1,1)$ is
symmetric in $I$ and $J$.

An important observation is that $I$ and $J$ can take only
two values in order to be compatible with $r\pm 1$ and 1 at
the triple intersection; these are $r$ and $r+2$ for the
$(+)$ case; and $r$ and $r-2$ in the $(-)$ case. But since
the diagonal entries of $\widetilde{W}_J{}^I$ are zero,
there is only one possibility, which is $I=r\pm 2$. Thus the
eigenvalues of the $2\times2$ matrix $\tilde{W}$ are $\pm
{\rm i} ~w(q,p,r\pm1,1)$ with
\begin{eqnarray}
 w(p,q,r\pm1,1) 
 =\sqrt{\left|\tilde{W}^{(4)}_{[012]}{}_r{}^{r\pm2}(p,q,r\pm1,1)
        \right|}
\label{v} 
\end{eqnarray}
From the eigenvalues of the $\tilde{W}$ matrices we
immediately obtain the eigenvalues of the volume operator.
In fact, since the $W$ matrix is a $2\times2$ antisymmetric
matrix, its absolute value is a multiple of the
identity. Thus the volume acts diagonally in the case in
which we are interested, much simplifying the calculations.
Taking the square root of the absolute value yields
\begin{equation}
 \hat{V}_v \left|
   \begin{array}{c}\setlength{\unitlength}{1 pt}
   \begin{picture}(60,35)
          \put( 0,0){$q$}\put( 0,30){$p$}
          \put(45,30){$r\scriptstyle\pm1$}
          \put( 5, 5){\line(1, 1){15}}
          \put( 5,35){\line(1,-1){15}}
          \put(30,20){\line(1, 1){15}}
          \put(30,20){\line(1,-1){10}}
          \put(39,11){\circle*{3}} \put(37,15){$\scriptstyle 1$}
          \put(20,20){\line(1,0){10}}\put(22,25){$J$}
          \put(20,20){\circle*{3}}\put(30,20){\circle*{3}}
          \bezier{20}( 5,20)( 5,35)(25,35)
          \bezier{20}(45,20)(45,35)(25,35)
          \bezier{20}( 5,20)( 5, 5)(25, 5)
          \bezier{20}(45,20)(45, 5)(25, 5)
   \end{picture}\end{array}  
 \right\rangle = \frac{l_0^3}{4} \sqrt{w(q,p,r\pm1,1)}\left|  
   \begin{array}{c}\setlength{\unitlength}{1 pt}
   \begin{picture}(60,35)
          \put( 0,0){$q$}\put( 0,30){$p$}
          \put(45,30){$r\scriptstyle\pm1$}
          \put( 5, 5){\line(1, 1){15}}
          \put( 5,35){\line(1,-1){15}}
          \put(30,20){\line(1, 1){15}}
          \put(30,20){\line(1,-1){10}}
          \put(39,11){\circle*{3}} \put(37,15){$\scriptstyle 1$}
          \put(20,20){\line(1,0){10}}\put(22,25){$J$}
          \put(20,20){\circle*{3}}\put(30,20){\circle*{3}}
          \bezier{20}( 5,20)( 5,35)(25,35)
          \bezier{20}(45,20)(45,35)(25,35)
          \bezier{20}( 5,20)( 5, 5)(25, 5)
          \bezier{20}(45,20)(45, 5)(25, 5)
   \end{picture}\end{array}  
 \right\rangle
\label{VOL:3vertex} 
\end{equation} 

Let us compute the values of $w$.  Starting from
equation \eref{theWformula} the matrix elements of
$\tilde{W}^{(4)}_{[012]}{}_J{}^I$ are given by the chromatic
evaluation:
\begin{equation} \fl
\widetilde{W}^{(4)}_{[012]}(a,b,c,d){}_J{}^I =
a~b~c
\sqrt{\frac{
     \begin{array}{c}\setlength{\unitlength}{.5 pt}
     \begin{picture}(35,25)
        \put(15, 0){\line(0,1){10}}\put(20, 0){\line(0,1){10}}
        \put(15, 0){\line(1,0){ 5}}\put(15,10){\line(1,0){ 5}}
        \put(15,25){\line(1,0){5}}  \put(15,15){$\scriptstyle J$}
        \put(15,15){\oval(30,20)[l]}\put(20,15){\oval(30,20)[r]}
     \end{picture}\end{array}
     \begin{array}{c}\setlength{\unitlength}{.5 pt}
     \begin{picture}(35,25)
        \put(15, 0){\line(0,1){10}}\put(20, 0){\line(0,1){10}}
        \put(15, 0){\line(1,0){ 5}}\put(15,10){\line(1,0){ 5}}
        \put(15,25){\line(1,0){5}}  \put(15,15){$\scriptstyle I$}
        \put(15,15){\oval(30,20)[l]}\put(20,15){\oval(30,20)[r]}
     \end{picture}\end{array}
   }{\begin{array}{c}\setlength{\unitlength}{.5 pt}
     \begin{picture}(40,40)
        \put(18,32){$\scriptstyle a$}
        \put(18,17){$\scriptstyle b$} 
        \put(18, 2){$\scriptstyle J$} 
        \put(20,15){\oval(40,30)} \put( 0,15){\line(1,0){40}} 
        \put( 0,15){\circle*{3}}  \put(40,15){\circle*{3}}
     \end{picture}\end{array}
     \begin{array}{c}\setlength{\unitlength}{.5 pt}
     \begin{picture}(40,40)
        \put(18,32){$\scriptstyle a$}
        \put(18,17){$\scriptstyle b$} 
        \put(18, 2){$\scriptstyle I$} 
        \put(20,15){\oval(40,30)} \put( 0,15){\line(1,0){40}} 
        \put( 0,15){\circle*{3}}  \put(40,15){\circle*{3}}
     \end{picture}\end{array}
     \begin{array}{c}\setlength{\unitlength}{.5 pt}
     \begin{picture}(40,40)
        \put(18,32){$\scriptstyle c$}
        \put(18,17){$\scriptstyle d$} 
        \put(18, 2){$\scriptstyle J$} 
        \put(20,15){\oval(40,30)} \put( 0,15){\line(1,0){40}} 
        \put( 0,15){\circle*{3}}  \put(40,15){\circle*{3}}
     \end{picture}\end{array}
     \begin{array}{c}\setlength{\unitlength}{.5 pt}
     \begin{picture}(40,40)
        \put(18,32){$\scriptstyle c$}
        \put(18,17){$\scriptstyle d$} 
        \put(18, 2){$\scriptstyle I$} 
        \put(20,15){\oval(40,30)} \put( 0,15){\line(1,0){40}} 
        \put( 0,15){\circle*{3}}  \put(40,15){\circle*{3}}
     \end{picture}\end{array}
   }} 
\! \!
\begin{array}{c}
  \setlength{\unitlength}{1 pt}
  \begin{picture}(110,60)
  \put(15,40){\line(0,1){10}}  \put(17,38){$a$}
  \put(15,32){\oval(20,20)[tl]} 
  \put(15,42){\circle*{3}}      
  \put(40,30){\line(0,1){30}}  \put(42,38){$b$}
  \put(40,32){\oval(20,20)[tl]} 
  \put(40,42){\circle*{3}}     \put(40,30){\circle*{3}}
  \put(40,60){\circle*{3}}
  \put(65,30){\line(0,1){30}}  \put(67,38){$c$}
  \put(65,32){\oval(20,20)[tl]} 
  \put(65,42){\circle*{3}}     \put(65,30){\circle*{3}}
  \put(65,60){\circle*{3}}
  \put(90,40){\line(0,1){10}}  \put(92,38){$d$}
  \put( 5,32){\line(0,-1){12}} \put(15, 2){${\scriptstyle  2}$}
  \put(30,28){\line(0,-1){18}} \put(22,15){${\scriptstyle  2}$}
  \put(55,28){\line(0,-1){ 8}} \put(40, 2){${\scriptstyle  2}$}
  \put(30,20){\oval(50,20)[b]} \put(30,10){\circle*{3}}
  \put(40,40){\oval(50,20)[bl]}\put(65,40){\oval(50,20)[br]}
  \put(40,30){\line(1,0){20}}  \put(42,20){$J$} 
  \put(40,50){\oval(50,20)[tl]}\put(65,50){\oval(50,20)[tr]}
  \put(40,60){\line(1,0){20}}  \put(52,50){$I$}
\end{picture}\end{array}  
\!\!\!\!\!\!\!\!\!
\end{equation}
where the matrices $\widetilde{W}^{(4)}_{[012]}(a,b,c,d){}_J{}^I$
are real antisymmetric matrices and have elements different 
from zero only if $|J-I|=2$. Using equation (\ref{eq:TETred})
and applying (\ref{eq:MOVE2}) to the two vertices $v(a,2,a)$,
$v(a,b,I)$ we obtain:
\begin{eqnarray} \label{eqWred}
\fl \widetilde{W}^{(4)}_{[012]}(a,b,c,d){}_J{}^I =
\sqrt{\frac{
     \begin{array}{c}\setlength{\unitlength}{.5 pt}
     \begin{picture}(35,25)
        \put(15, 0){\line(0,1){10}}\put(20, 0){\line(0,1){10}}
        \put(15, 0){\line(1,0){ 5}}\put(15,10){\line(1,0){ 5}}
        \put(15,25){\line(1,0){5}}  \put(15,15){$\scriptstyle J$}
        \put(15,15){\oval(30,20)[l]}\put(20,15){\oval(30,20)[r]}
     \end{picture}\end{array}
     \begin{array}{c}\setlength{\unitlength}{.5 pt}
     \begin{picture}(35,25)
        \put(15, 0){\line(0,1){10}}\put(20, 0){\line(0,1){10}}
        \put(15, 0){\line(1,0){ 5}}\put(15,10){\line(1,0){ 5}}
        \put(15,25){\line(1,0){5}}  \put(15,15){$\scriptstyle I$}
        \put(15,15){\oval(30,20)[l]}\put(20,15){\oval(30,20)[r]}
     \end{picture}\end{array}
   }{\begin{array}{c}\setlength{\unitlength}{.5 pt}
     \begin{picture}(40,40)
        \put(18,32){$\scriptstyle a$}
        \put(18,17){$\scriptstyle b$} 
        \put(18, 2){$\scriptstyle J$} 
        \put(20,15){\oval(40,30)} \put( 0,15){\line(1,0){40}} 
        \put( 0,15){\circle*{3}}  \put(40,15){\circle*{3}}
     \end{picture}\end{array}
     \begin{array}{c}\setlength{\unitlength}{.5 pt}
     \begin{picture}(40,40)
        \put(18,32){$\scriptstyle a$}
        \put(18,17){$\scriptstyle b$} 
        \put(18, 2){$\scriptstyle I$} 
        \put(20,15){\oval(40,30)} \put( 0,15){\line(1,0){40}} 
        \put( 0,15){\circle*{3}}  \put(40,15){\circle*{3}}
     \end{picture}\end{array}
     \begin{array}{c}\setlength{\unitlength}{.5 pt}
     \begin{picture}(40,40)
        \put(18,32){$\scriptstyle c$}
        \put(18,17){$\scriptstyle d$} 
        \put(18, 2){$\scriptstyle J$} 
        \put(20,15){\oval(40,30)} \put( 0,15){\line(1,0){40}} 
        \put( 0,15){\circle*{3}}  \put(40,15){\circle*{3}}
     \end{picture}\end{array}
     \begin{array}{c}\setlength{\unitlength}{.5 pt}
     \begin{picture}(40,40)
        \put(18,32){$\scriptstyle c$}
        \put(18,17){$\scriptstyle d$} 
        \put(18, 2){$\scriptstyle I$} 
        \put(20,15){\oval(40,30)} \put( 0,15){\line(1,0){40}} 
        \put( 0,15){\circle*{3}}  \put(40,15){\circle*{3}}
     \end{picture}\end{array}
   }}  ~\cdot~ 
   \begin{array}{c}\setlength{\unitlength}{.5 pt}
   \begin{picture}(40,40)
            \put(18,32){$\scriptstyle I$}
            \put(18,17){$\scriptstyle 2$} 
            \put(18, 2){$\scriptstyle J$} 
            \put(20,15){\oval(40,30)} \put( 0,15){\line(1,0){40}} 
            \put( 0,15){\circle*{3}}  \put(40,15){\circle*{3}}
   \end{picture}\end{array}
   ~\times~ 
\\ \fl ~~ ~\times~
\bigg[ b ~
\frac{\begin{array}{c}\setlength{\unitlength}{.8 pt}
        \begin{picture}(55,40)
        \put( 0,15){\line(1,-1){15}} \put(0,22){${\scriptstyle b}$}
        \put( 0,15){\line(1, 1){15}} \put(0, 0){${\scriptstyle b}$}
        \put( 0,15){\circle*{3}}
        \put(30,15){\line(-1, 1){15}} \put(28,20){${\scriptstyle 2}$}
        \put(30,15){\line(-1,-1){15}} \put(28, 2){${\scriptstyle 2}$}
        \put(30,15){\circle*{3}}
        \put( 0,15){\line(1,0){30}} \put(12,16){${\scriptstyle 2}$}
        \put(15,30){\line(1,0){25}} \put(15,30){\circle*{3}}
        \put(15, 0){\line(1,0){25}} \put(15, 0){\circle*{3}}
        \put(40, 0){\line(0,1){30}} \put(42,12){${\scriptstyle b}$}
        \end{picture}\end{array}
    }{ \begin{array}{c}\setlength{\unitlength}{.5 pt}
       \begin{picture}(40,40)
            \put(18,32){$\scriptstyle b$}
            \put(18,17){$\scriptstyle b$} 
            \put(18, 2){$\scriptstyle 2$} 
            \put(20,15){\oval(40,30)} \put( 0,15){\line(1,0){40}} 
            \put( 0,15){\circle*{3}}  \put(40,15){\circle*{3}}
           \end{picture}\end{array}
      } 
 - I ~\frac{\begin{array}{c}\setlength{\unitlength}{.8 pt}
        \begin{picture}(55,40)
        \put( 0,15){\line(1,-1){15}} \put(0,22){${\scriptstyle I}$}
        \put( 0,15){\line(1, 1){15}} \put(0, 0){${\scriptstyle J}$}
        \put( 0,15){\circle*{3}}
        \put(30,15){\line(-1, 1){15}} \put(28,20){${\scriptstyle 2}$}
        \put(30,15){\line(-1,-1){15}} \put(28, 2){${\scriptstyle 2}$}
        \put(30,15){\circle*{3}}
        \put( 0,15){\line(1,0){30}} \put(12,16){${\scriptstyle 2}$}
        \put(15,30){\line(1,0){25}} \put(15,30){\circle*{3}}
        \put(15, 0){\line(1,0){25}} \put(15, 0){\circle*{3}}
        \put(40, 0){\line(0,1){30}} \put(42,12){${\scriptstyle I}$}
        \end{picture}\end{array}
    }{ \begin{array}{c}\setlength{\unitlength}{.5 pt}
       \begin{picture}(40,40)
            \put(18,32){$\scriptstyle I$}
            \put(18,17){$\scriptstyle 2$} 
            \put(18, 2){$\scriptstyle J$} 
            \put(20,15){\oval(40,30)} \put( 0,15){\line(1,0){40}} 
            \put( 0,15){\circle*{3}}  \put(40,15){\circle*{3}}
           \end{picture}\end{array}
      } 
\bigg] ~\cdot~
\bigg[ b ~
\frac{\begin{array}{c}\setlength{\unitlength}{.8 pt}
        \begin{picture}(55,40)
        \put( 0,15){\line(1,-1){15}} \put(0,22){${\scriptstyle J}$}
        \put( 0,15){\line(1, 1){15}} \put(0, 0){${\scriptstyle I}$}
        \put( 0,15){\circle*{3}}
        \put(30,15){\line(-1, 1){15}} \put(28,20){${\scriptstyle b}$}
        \put(30,15){\line(-1,-1){15}} \put(28, 2){${\scriptstyle b}$}
        \put(30,15){\circle*{3}}
        \put( 0,15){\line(1,0){30}} \put(12,16){${\scriptstyle 2}$}
        \put(15,30){\line(1,0){25}} \put(15,30){\circle*{3}}
        \put(15, 0){\line(1,0){25}} \put(15, 0){\circle*{3}}
        \put(40, 0){\line(0,1){30}} \put(42,12){${\scriptstyle a}$}
        \end{picture}\end{array}
    }{ \begin{array}{c}\setlength{\unitlength}{.5 pt}
       \begin{picture}(40,40)
            \put(18,32){$\scriptstyle I$}
            \put(18,17){$\scriptstyle 2$} 
            \put(18, 2){$\scriptstyle J$} 
            \put(20,15){\oval(40,30)} \put( 0,15){\line(1,0){40}} 
            \put( 0,15){\circle*{3}}  \put(40,15){\circle*{3}}
           \end{picture}\end{array}
      } 
\bigg] ~\cdot~ \bigg[ c ~
\frac{\begin{array}{c}\setlength{\unitlength}{.8 pt}
        \begin{picture}(55,40)
        \put( 0,15){\line(1,-1){15}} \put(0,22){${\scriptstyle I}$}
        \put( 0,15){\line(1, 1){15}} \put(0, 0){${\scriptstyle J}$}
        \put( 0,15){\circle*{3}}
        \put(30,15){\line(-1, 1){15}} \put(28,20){${\scriptstyle c}$}
        \put(30,15){\line(-1,-1){15}} \put(28, 2){${\scriptstyle c}$}
        \put(30,15){\circle*{3}}
        \put( 0,15){\line(1,0){30}} \put(12,16){${\scriptstyle 2}$}
        \put(15,30){\line(1,0){25}} \put(15,30){\circle*{3}}
        \put(15, 0){\line(1,0){25}} \put(15, 0){\circle*{3}}
        \put(40, 0){\line(0,1){30}} \put(42,12){${\scriptstyle d}$}
        \end{picture}\end{array}
    }{ \begin{array}{c}\setlength{\unitlength}{.5 pt}
       \begin{picture}(40,40)
            \put(18,32){$\scriptstyle I$}
            \put(18,17){$\scriptstyle 2$} 
            \put(18, 2){$\scriptstyle J$} 
            \put(20,15){\oval(40,30)} \put( 0,15){\line(1,0){40}} 
            \put( 0,15){\circle*{3}}  \put(40,15){\circle*{3}}
           \end{picture}\end{array}
      } 
\bigg] 
\nonumber 
\end{eqnarray}
This expression is evaluated using
(\ref{eq:EV1}),(\ref{eq:EV2}) and (\ref{eq:EV3}).  Defining
$t=(I-J)/2$ and $\epsilon=(I-J)/2$ we note that
$\widetilde{W}^{(4)}_{[012]}(a,b,c,d){}_J{}^I$ is different
from zero only if $\epsilon=\pm 1$ and all the 3-vertices in
equation (\ref{eqWred}) are admissible.  In this way, we
obtain the following explicit formula for the matrix
elements ($\epsilon=\pm 1$):
\begin{eqnarray} \fl
\widetilde{W}^{(4)}_{[012]}(a,b,c,d){}_{t-\epsilon}^{t+\epsilon} 
 &=& - \epsilon (-1)^{\frac{a+b+c+d}{2}} \bigg[
     \frac{1}{4 t (t+2)} \frac{a+b+t+3}{2}\frac{c+d+t+3}{2}
\nonumber \\ &&
     \frac{1+a+b-t}{2}\frac{1+a+t-b}{2}\frac{1+b+t-a}{2}
\\ &&
     \frac{1+c+d-t}{2}\frac{1+c+t-d}{2}\frac{1+d+t-c}{2}
     \bigg]^{\frac{1}{2}}  
\nonumber 
\end{eqnarray}
This formula can be used to obtain explicit expressions 
for the eigenvalues of the 
$| {\rm i} \hat{\widetilde{W}}{}^{(4)}_{[012]}(a,b,c,d){} |$ 
operator when they are $2\times 2$ matrices, as in the
case of equation \eref{VOL:3vertex}, 
\begin{eqnarray}
\fl w(a,a,1,1)       = \sqrt{\frac{a(a+2)}{4}} \\
\fl w(a,b,c,a+b+c-2) = 2 \sqrt{\frac{a~b~c~(a+b+c)}{16}} \\
\fl w(a,b,c,1) = \frac{1}{2} \sqrt{\frac{
           (a+b+c+3)(1+a+b-c)(1+c-a+b)(1+c+a-b)}{16}} 
\label{eq:vol3}
\end{eqnarray} 


\section{Graphical calculus in the connection representation.
\label{ap:Binor}}

Let us recall the basic elements of Penrose binor calculus
 \cite{binor}, which is the basis for expressing the
 cylindrical functions of generalized connections in graphical
 form \cite{DePietri97}.  The key idea of Penrose is to rewrite
 any tensor expression in which there are sums of dummy indices
 in a graphical way \cite{binor}. Penrose represents the basic
 elements of spinor calculus (i.e., tensor expression with
 indices $A,B,\ldots=1,2$) as
\begin{equation}
\begin{array}{rclcrclcrcl}
\delta_C^{~A} &=& 
   \begin{array}{c}\setlength{\unitlength}{1 pt}
   \begin{picture}(10,20)
            \put(5,5){\line(0,1){10}}
            \put(5,5){\circle*{3}}\put(5,15){\circle*{3}}
            \put(6,0){${\scriptstyle C}$}\put(6,14){${\scriptstyle A}$}
   \end{picture}\end{array} 
&&{\rm i} \epsilon_{AC} &=& 
   \begin{array}{c}\setlength{\unitlength}{1 pt}
   \begin{picture}(20,15)
               \put(5,5){\circle*{3}}\put(15,5){\circle*{3}}
               \put(10,5){\oval(10,20)[t]}
               \put(3,0){${\scriptstyle A}$}\put(12,0){${\scriptstyle C}$}
   \end{picture}\end{array}
&&{\rm i} \epsilon^{AC} &=& 
   \begin{array}{c}\setlength{\unitlength}{1 pt}
   \begin{picture}(20,15)
         \put(10,10){\oval(10,20)[b]}
         \put(5,10){\circle*{3}}\put(15,10){\circle*{3}}
         \put(3,11){${\scriptstyle A}$}\put(12,11){${\scriptstyle C}$}
   \end{picture}\end{array} \\
\eta_{A} &=& 
   \begin{array}{c}\setlength{\unitlength}{1 pt}
   \begin{picture}(20,15)
               \put(0,5){\framebox(10,10){$\eta$}}
               \put(6,-2){${\scriptstyle A}$}
               \put(5,0){\line(0,1){5}}\put(5,0){\circle*{3}}
   \end{picture}\end{array}
&&\eta^{A} &=& 
   \begin{array}{c}\setlength{\unitlength}{1 pt}
   \begin{picture}(20,15)
              \put(0,0){\framebox(10,10){$\eta$}}
              \put(6,13){${\scriptstyle A}$}
              \put(5,10){\line(0,1){5}}\put(5,15){\circle*{3}}
   \end{picture}\end{array}
&& X_{AB}^{C} &=& 
   \begin{array}{c}\setlength{\unitlength}{1 pt}
   \begin{picture}(20,25)
      \put(0,7){\framebox(20,10){${\scriptstyle X}$}}
      \put(6,0){${\scriptstyle A}$}\put(5,2){\line(0,1){5}}
      \put(5,2){\circle*{3}}
      \put(16,0){${\scriptstyle B}$}\put(15,2){\line(0,1){5}}
      \put(15,2){\circle*{3}}
      \put(11,20){${\scriptstyle C}$}\put(10,17){\line(0,1){5}}
      \put(10,22){\circle*{3}}
   \end{picture}\end{array}
\end{array}
\label{eq:binCONV}
\end{equation}
and assigns a minus sign to each crossing\footnote{The binor
representation is a graphical representation in the plane.}, i.e:
\begin{equation}
X^{AB}_{CD} = \delta_D^{~A}\delta_C^{~B} = - 
   \begin{array}{c}\setlength{\unitlength}{1 pt}
   \begin{picture}(20,20)
        \put(5,5){\line(1,1){10}}
        \put(5,5){\circle*{3}}\put(5,15){\circle*{3}}
        \put(6,0){${\scriptstyle C}$}\put(6,14){${\scriptstyle A}$}
        \put(15,5){\line(-1,1){10}}
        \put(15,5){\circle*{3}}\put(15,15){\circle*{3}}
        \put(16,0){${\scriptstyle D}$}\put(16,14){${\scriptstyle B}$}
   \end{picture}\end{array}
\label{eq:minus}
\end{equation}
Using this rule it is possible to represent any $SU(2)$
tensor expression in a graphical way as follows:
({\bf 1}) Define the  ``up'' direction in the plane;
({\bf 2}) Draw boxes with the name of the $SU(2)$ tensor
   (except for the $\delta$ and the $\epsilon$ that are simply
   represented by lines) with as many slots going up as the
   number of contravariant indices and as many slots going down as
   the number of covariant indices; 
({\bf 3}) For each dummy index connect with a line the
          corresponding slots of the boxes;
({\bf 4}) Assign a ``${\rm i}$'' factor to each minimum or maximum
          of the lines;
({\bf 5}) Assign a minus sign to each crossings of lines.  
Conversely, any curve can expressed as a product of $\delta$'s and
$\epsilon$'s. 

An unexpected feature of this graphical
calculus is its way of expressing the operation of taking the
trace.  Since any contravariant index of a tensor is represented
more ``up'' than any covariant index of the same tensor, there is no
way to write directly the operation of taking the trace unless
one uses the antisymmetric 2-tensors $\epsilon_{AB}$ and
$\epsilon^{AB}$ as ${\rm Tr} X^A_B = X^A_B \delta^B_A =
\epsilon_{A'A}\epsilon^{B'B} X^A_B \delta^{A'}_{B'}$. This
becomes, graphically, 
\begin{equation}
 {\rm Tr} X^A_{~B} = \delta^B_A X^A_{~B} 
 = - \begin{array}{c}\setlength{\unitlength}{1 pt}
     \begin{picture}(20,20)
        \put(10,15){\oval(10,10)[t]}
        \put(10, 5){\oval(10,10)[b]}
        \put(10, 5){\framebox(10,10){$\scriptstyle X$}}
        \put( 5, 5){\line(0,1){10}}
     \end{picture}\end{array}
\label{eq:binorTRACE}
~~.
\end{equation}
Moreover, the basic binor identity is graphically given by
\begin{equation}
\fl \quad
   \begin{array}{c}\setlength{\unitlength}{1 pt}
   \begin{picture}(10,15)
            \put( 0,0){\line( 2,3){10}}
            \put(10,0){\line(-2,3){10}}
   \end{picture}\end{array} 
+  \begin{array}{c}\setlength{\unitlength}{1 pt}
   \begin{picture}(10,15)
            \put( 0,0){\line(0,1){15}}
            \put(10,0){\line(0,1){15}}  
   \end{picture}\end{array}
+  \begin{array}{c}\setlength{\unitlength}{1 pt}
   \begin{picture}(10,15)
            \put(5, 0){\oval(10,10)[t]}
            \put(5,15){\oval(10,10)[b]} 
   \end{picture}\end{array}
= (-1)   ~\delta^{\cdot C}_{B}\ \delta^{\cdot D}_{A}  
   +      ~\delta^{\cdot C}_{A}\ \delta^{\cdot D}_{B}
   + (-1) ~\epsilon_{AB}\ \epsilon^{CD} = 0  , 
\label{eq:binor}
\end{equation}
perhaps the most relevant equation of loop quantum
gravity.

In particular, the irreducible representation of color $n$ 
can be constructed as the symmetrization $\Pi^{(\e)}_n$ (in
Penrose's graphical representation an anti-symmetrization because
of equation (\ref{eq:minus})) of the tensor product of $n$
fundamental representations 
$\begin{array}{c}\setlength{\unitlength}{1 pt}
 \begin{picture}(6,10)
    \put(3,0){\line(0,1){2}}\put(3,8){\line(0,1){2}}
    \put(0,2){\framebox(6,6){$\scriptscriptstyle g$}}
 \end{picture}\end{array}
{}_A^B = U(g)_A^B$ ($g$ being the group element).
Denoting by $\Pi^{(e)}_n$ the normalized symmetrizer
(graphically an anti-symmetrizer -- see \cite{DePietri97}), 
the color $n$ irreducible representation is given by:
\begin{eqnarray}
&&\Pi^{(e)}_n P_n = \frac{1}{n!} \sum_p
(-1)^{|p|}\ P^{(p)}_{n}
=  \begin{array}{c}\setlength{\unitlength}{1 pt}
   \begin{picture}(20,25)
     \put(10, 0){\line(0,1){10}}
     \put(0,10){\framebox(20,5){}}
     \put(10,15){\line(0,1){10}}\put(12,17){n}
   \end{picture}\end{array}
\label{projector}
\\
&& \j_I(n_I)   
=  \begin{array}{c}\setlength{\unitlength}{1 pt}
   \begin{picture}(15,30)
    \put(7,22){$\scriptstyle n_I$}
    \put(10,12){$\scriptstyle {\e_I}$}
    \put(2,5){\rule{6pt}{15pt}}
    \put(5,0){\line(0,1){5}}
    \put(5,20){\line(0,1){5}}
   \end{picture}\end{array}
=  \begin{array}{c}\setlength{\unitlength}{1 pt}
   \begin{picture}(40,40)
    \put(22,2){$\scriptstyle n_I$}
    \put(0,10){\framebox(40,2){}} \put(20,0){\line(0,1){10}}
    \put(5,12){\line(0,1){3}}\put(5,25){\line(0,1){3}}
    \put(-1,15){\framebox(12,10){$\scriptstyle g_{e_I}$}}
    \put(12,18){$\cdots$}
    \put(35,12){\line(0,1){3}}\put(35,25){\line(0,1){3}}
    \put(29,15){\framebox(12,10){$\scriptstyle g_{e_I}$}}
    \put(0,28){\framebox(40,2){}} \put(20,30){\line(0,1){10}}
    \put(22,34){$\scriptstyle n_I$}
   \end{picture}\end{array}
~, 
\label{eq:5}
\end{eqnarray}
($P_n$ represents $n$ parallel lines, $P_n^{(p)}$ is the
graphical representation in terms of lines of the
permutation $p$, $|p|$ is the parity of the permutation and a
line labeled by a positive integer $n$ represents $n$
non-intersecting parallel line). 

Analogously, it is possible to construct an explicit graphical
representation of the Clebsch-Gordan intertwining
matrix i.e., of the matrix that represents the coupling of
$3$ (or more) irreducible representations of the $SU(2)$ group.
 The representation of the Clebsch-Gordan
intertwining matrix in the binor formalism is given by the
special sum of ``tangles'' denoted as the $3$-vertex. Each
line of the vertex is labeled by a positive integer $a$, $b$
or $c$ and is defined as:
\begin{equation}
   \begin{array}{c}\setlength{\unitlength}{1 pt}
   \begin{picture}(30,40)
       \put(15,15){\line(-1, 1){10}} \put( 4,27){$a$}
       \put(15,15){\line( 1, 1){10}} \put(22,27){$b$}
       \put(15, 5){\line(0,1){10}}   \put(17,1){$c$}
       \put(15,15){\circle*{3}}
   \end{picture}\end{array}
\stackrel{def}{=}
   \begin{array}{c}\setlength{\unitlength}{1 pt}
   \begin{picture}(70,40) 
       \put(20,30){\line( 1, 0){30}} \put(28,32){$m$}
       \put(35,15){\line(-1, 1){15}} \put(20,17){$p$}
       \put(35,15){\line( 1, 1){15}} \put(44,17){$n$}
       \put(35, 1){\line(0,1){10}}   \put(42,1){$c$}
       \put(27,11){\framebox(16,4){}}
       \put(16,22){\framebox(4,16){}} 
       \put(6 ,30){\line(1,0){10}}\put(0,30){a}
       \put(50,22){\framebox(4,16){}}
       \put(54,30){\line(1,0){10}}\put(65,30){b}
   \end{picture}\end{array}
\;\;
\left\{\begin{array}{rcl} 
m &=& (a+b-c)/2 \\ n&=&(b+c-a)/2 \\ p&=&(c+a-b)/2
\end{array}\right.
\label{eq:6}
\end{equation}
where $m,n,p$ are positive integers.  This
last condition is called the {\it admissibility condition}
for the $3$-vertex.  By the Wigner-Eckart
theorem, or more precisely, by the version of it due to Yutsin,
Levinson and Vanagas \cite{GraphMethods}, any invariant tensorial
intertwining matrix representing the coupling of $n$
representations of a compact group can be expressed as the
product of Clebsch-Gordan coefficients or, which is the same,
in terms of a trivalent decomposition. It is easy to see
that any contractor $c_\alpha$ of a given vertex $v_\alpha$ with $n$
incoming (outgoing) edges of color $P_0$,\ldots,$P_{n-1}$, 
can be written in terms of a linear combination of the  
trivalent contractors obtained ordering the edges and assigning 
$n$-$3$ compatible integers, i.e. by a trivalent decomposition
of the vertex. Graphically:
\begin{equation}
\fl
\begin{array}{c}\setlength{\unitlength}{1 pt}
\begin{picture}(100,55)
     \put( 0, 0){$P_0$}\put(15, 0){\line(2,1){20}}
     \put( 0,20){$P_1$}\put(15,20){\line(1,0){20}}
     \put(10,40){$P_2$}\put(25,40){\line(1,-1){10}}
     \put(35,0){\framebox(25,30){$c_\alpha$}}
     \put(40,40){$\cdots$}        \put(40,33){$\ldots$} 
     \put(75,40){$P_{n-3}$}\put(60,30){\line(1,1){10}}
     \put(75,20){$P_{n-2}$}\put(60,20){\line(1,0){10}}
     \put(75, 0){$P_{n-1}$}\put(60,10){\line(1,-1){10}} 
   \end{picture}\end{array}
=  \sum_{i_2,\ldots,i_{n-2}} c(i_2,\ldots,i_{n-2})
   \begin{array}{c}\setlength{\unitlength}{1 pt}
   \begin{picture}(110,55)
     \put( 0, 0){$P_0$}\put(15, 0){\line(1,1){10}}
     \put( 0,20){$P_1$}\put(15,20){\line(1,0){10}}
     \put(10,40){$P_2$}\put(25,40){\line(1,-1){10}}
     \put(25,10){\line(0,1){10}}
     \put(25,20){\line(1,1){10}}\put(32,18){$\scriptstyle i_2$}
     \put(35,30){\line(1,0){10}}\put(40,22){$\scriptstyle i_3$}
     \put(25,20){\circle*{3}}   \put(35,30){\circle*{3}}
     \put(45,40){$\cdots$}       \put(45,30){$\ldots$} 
     \put(60,30){\line(1, 0){10}}
     \put(70,30){\line(1,-1){10}}
     \put(60,20){$\scriptstyle i_{n-2}$}
     \put(80,20){\line(0,-1){10}} 
     \put(80,20){\circle*{3}}\put(70,30){\circle*{3}}
     \put(85,40){$P_{n-3}$}  \put(70,30){\line(1,1){10}}
     \put(95,20){$P_{n-2}$} \put(80,20){\line(1,0){10}}
     \put(95, 0){$P_{n-1}$} \put(80,10){\line(1,-1){10}} 
   \end{picture}\end{array}
~,
\label{eq:7}
\end{equation}
where the sum is extended to all the integers $i_2,\ldots,i_{n-2}$
that satisfy the {\it admissibility conditions} for the 3-vertices.  

An important characteristic of the binor representation, related
to the existence of the natural ``metric'' (an 
invariant way to relate contravariant and covariant indices)
$\epsilon_{AB}$ in the 2-spinor space, it is its independence from 
the orientation chosen for the edges. In fact, the $SU(2)$
relation $U(g^{-1})_{~A}^B = \epsilon_{AC} \epsilon^{BD}
U(g)_{~D}^{C}$ is graphically represented (in the binor
representation) by:
\begin{equation}
\begin{array}{c}\setlength{\unitlength}{1 pt}
\begin{picture}(15,30)
  \put(8,0){\line(0,1){7}}\put(8,23){\line(0,1){7}}
  \put(0,7){\framebox(16,16){$\scriptstyle X^{\scriptscriptstyle -1}$}}
  \put(10,0){$\scriptstyle B$}\put(10,25){$\scriptstyle A$}
\end{picture}\end{array}
= 
\begin{array}{c}\setlength{\unitlength}{1 pt}
\begin{picture}(40,30)
    \put(5,0){\line(0,1){22}}\put(7,0){$\scriptstyle g$}
    \put(12,22){\oval(14,16)[t]}
    \put(12,8){\framebox(14,14){$\scriptstyle g$}}
    \put(26,8){\oval(14,16)[b]}
    \put(33,8){\line(0,1){22}} \put(35,24){$\scriptstyle A$}
\end{picture}\end{array}
,\;\; {\rm while} \;\;
\begin{array}{c}\setlength{\unitlength}{1 pt}
\begin{picture}(25,30)
    \put(7,22){$\scriptstyle n_i$}
    \put(10,12){${\e_i}^{\scriptscriptstyle -1}$}
    \put(2,5){\rule{6pt}{15pt}}
    \put(5,0){\line(0,1){5}}
    \put(5,20){\line(0,1){5}}
\end{picture}\end{array} 
=  
\begin{array}{c}\setlength{\unitlength}{1 pt}
\begin{picture}(45,30)
    \put(37,2){$\scriptstyle n_i$}
    \put(5,5) {\line(0,1){20}}
    \put(20,12){${\e_i}$}
    \put(12,5){\rule{6pt}{15pt}}
    \put(10, 5){\oval(10,10)[b]}
    \put(25,20){\oval(20,10)[t]}
    \put(35,0){\line(0,1){20}}
\end{picture}\end{array}
~~. 
\label{eq:edgerevers}
\end{equation}
is the related identity in case of the $n$-representation.

In this graphical notation the left invariant vector fields 
associate to an edge are represented by:
\begin{equation}
  X_{\e_I}^i 
   \begin{array}{c}\setlength{\unitlength}{1 pt}
   \begin{picture}(15,30)
    \put(7,22){$\scriptstyle n_I$}
    \put(10,12){$\scriptstyle {\e_I}$}
    \put(2,5){\rule{6pt}{15pt}}
    \put(5,0){\line(0,1){5}}
    \put(5,20){\line(0,1){5}}
   \end{picture}\end{array}
=  n_I
   \begin{array}{c}\setlength{\unitlength}{1 pt}
   \begin{picture}(25,40)
    \put( 7,37){$\scriptstyle n_I$}
    \put( 7, 2){$\scriptstyle n_I$}
    \put(10,27){$\scriptstyle {\e_I}$}
    \put( 2,20){\rule{6pt}{15pt}}
    \put( 5, 0){\line(0,1){20}}
    \put( 5,35){\line(0,1){5}}
    \put( 7,12){$\scriptstyle 2$} 
    \put( 5,10){\line(1,0){10}}  
    \put( 5,10){\circle*{3}}     \put(15,10){\circle*{3}}
    \put(15, 5){\line(0,1){10}}
    \put(20, 5){\oval(10, 5)[b]} \put(20,15){\oval(10, 5)[t]}
    \put(20, 5){\framebox(9,9){$\tau_i$}}
   \end{picture}\end{array} 
\end{equation}
From the identity (that can be proven in a straightforward way): 
\begin{equation}
\sum_{i,j,k=1}^3 \;\epsilon_{ijk}\;
\begin{array}{c}\setlength{\unitlength}{1 pt}
\begin{picture}(40,60)
    \put( 2,12){$\scriptstyle 2$} 
    \put( 0,10){\line(1,0){10}} \put(10,10){\circle*{3}}
    \put(10, 5){\line(0,1){10}}
    \put(15, 5){\oval(10, 5)[b]} \put(15,15){\oval(10, 5)[t]}
    \put(15, 5){\framebox(9,9){$\tau_i$}}
    \put( 2,32){$\scriptstyle 2$} 
    \put( 0,30){\line(1,0){10}}   \put(10,30){\circle*{3}}
    \put(10,25){\line(0,1){10}}
    \put(15,25){\oval(10, 5)[b]}  \put(15,35){\oval(10, 5)[t]}
    \put(15,25){\framebox(9,9){$\tau_j$}}
    \put( 2,52){$\scriptstyle 2$} 
    \put( 0,50){\line(1,0){10}}   \put(10,50){\circle*{3}}
    \put(10,45){\line(0,1){10}}
    \put(15,45){\oval(10, 5)[b]}  \put(15,55){\oval(10, 5)[t]}
    \put(15,45){\framebox(9,9){$\tau_k$}}
\end{picture}\end{array}
= \frac{1}{2}\;\;
\begin{array}{c}\setlength{\unitlength}{1 pt}
\begin{picture}(40,60)
    \put( 2,12){$\scriptstyle 2$} 
    \put( 0,10){\line(1,0){10}} \put(10,10){\circle*{3}}
    \put( 2,32){$\scriptstyle 2$} 
    \put( 0,30){\line(1,0){10}}   \put(10,30){\circle*{3}}
    \put( 2,52){$\scriptstyle 2$} 
    \put( 0,50){\line(1,0){10}}   \put(10,50){\circle*{3}}
    \put(10,30){\oval(20,40)[r]}
    \put(10,30){\line(1,0){10}}
    \put(20,30){\circle*{3}}
\end{picture}\end{array}
~~~.
\label{eq:3tau}
\end{equation}
we have the important result 
\begin{equation}\label{act_vol}
  \epsilon_{ijk} X_{\e_I}^i X_{\e_J}^j X_{\e_K}^k 
   \begin{array}{c}\setlength{\unitlength}{1 pt}
   \begin{picture}(55,30)
    \put( 7,22){$\scriptstyle n_I$}
    \put(10,12){$\scriptstyle {\e_I}$}
    \put( 2, 5){\rule{6pt}{15pt}}
    \put( 5, 0){\line(0,1){5}}
    \put( 5,20){\line(0,1){5}}
    \put(27,22){$\scriptstyle n_I$}
    \put(30,12){$\scriptstyle {\e_I}$}
    \put(22, 5){\rule{6pt}{15pt}}
    \put(25, 0){\line(0,1){5}}
    \put(25,20){\line(0,1){5}}
    \put(47,22){$\scriptstyle n_K$}
    \put(50,12){$\scriptstyle {\e_K}$}
    \put(42, 5){\rule{6pt}{15pt}}
    \put(45, 0){\line(0,1){5}}
    \put(45,20){\line(0,1){5}}
   \end{picture}\end{array}
= \frac{n_I n_J n_K}{2}
   \begin{array}{c}\setlength{\unitlength}{1 pt}
   \begin{picture}(55,50)
    \put( 7,42){$\scriptstyle n_I$}
    \put(10,32){$\scriptstyle {\e_I}$}
    \put( 2,25){\rule{6pt}{15pt}}
    \put( 5,20){\line(0,1){5}}
    \put( 5,15){\line(0,1){5}} \put( 5,20){\circle*{3}}
    \put( 5,15){\oval(10,10)[tr]}
    \put( 5,40){\line(0,1){5}}
    \put(27,42){$\scriptstyle n_I$}
    \put(30,32){$\scriptstyle {\e_I}$}
    \put(22,25){\rule{6pt}{15pt}}
    \put(25,20){\line(0,1){5}}
    \put(25,15){\line(0,1){5}} \put(25,20){\circle*{3}}
    \put(25,15){\oval(10,10)[tr]}
    \put(25,40){\line(0,1){5}}
    \put(47,42){$\scriptstyle n_K$}
    \put(50,32){$\scriptstyle {\e_K}$}
    \put(42,25){\rule{6pt}{15pt}}
    \put(45,20){\line(0,1){5}}
    \put(45,15){\line(0,1){5}} \put(45,20){\circle*{3}}
    \put(45,15){\oval(10,10)[tr]}
    \put(45,40){\line(0,1){5}} 
    \put(30,15){\oval(40,20)[b]}
    \put(30, 5){\line(0,1){10}}   \put(30, 5){\circle*{3}}   
   \end{picture}\end{array}
\end{equation}
which is the key ``bridge'' showing that the volume operator that
was defined in the loop representation in \cite{RovelliSmolin95},
based on the r.h.s.~of \eref{act_vol}, is indeed the same
operation as the volume that was defined in the connection
representation in \cite{VolumeAL}, based on the
l.h.s.~of \eref{act_vol}.


\section{Tangle-theoretic recoupling theory.}

One of the main results of the recoupling theory of colored knots
and links with trivalent vertices \cite{Kauffman94} is the
computation of the Kauffman bracket for framed spin networks. The
framing refers to the fact that in the computations one keeps track of
over- and under-crossing. In our context this distinction has
no meaning. Therefore we use a simplified version of the recoupling
theory, namely we replace the deformation parameter $A$ by its
``classical'' value $-1$, relevant to our case. This case
corresponds to the standard recoupling theory of the $SU(2)$ group.  
In the formulae from \cite{Kauffman94}, we thus replace the deformed (or
quantum) integers by ordinary integers. We list here the
basic formulae which we use. For more details 
see \cite{Kauffman94}. This beauty of the theory is that
it is completely determined by
the following two tangle properties:
\begin{eqnarray} &&
 {\big\langle} \begin{array}{c}{\setlength{\unitlength}{1 pt}
\begin{picture}(25,15) 
        \put( 0, 0){\line( 5, 3){10}}
        \put(25,15){\line(-5,-3){10}}
        \put(25, 0){\line(-5, 3){25}}
\end{picture}}\end{array}
~{\big\rangle} = A
     {\big\langle} \begin{array}{c}{\setlength{\unitlength}{1 pt}
\begin{picture}(10,15) 
\put(5, 0){\oval(10,10)[t]}
\put(5,15){\oval(10,10)[b]} 
\end{picture}}\end{array}
 ~{\big\rangle} + A^{-1}
     {\big\langle}~\begin{array}{c}{\setlength{\unitlength}{1 pt}
\begin{picture}(10,15) 
\put( 0,0){\line(0,1){15}}
\put(10,0){\line(0,1){15}}
\end{picture}}\end{array}
 ~{\big\rangle},
\label{eq:KB1}
\label{SkeinId}
\\ && 
{\big\langle} \begin{array}{c}{\setlength{\unitlength}{1 pt}
\begin{picture}(10,10) \put(5,5){\oval(10,10)}
\end{picture}}\end{array}
  ~\cup~ {\bf K}~{\big\rangle} = 
   d~ {\big\langle} ~{\bf K}~ {\big\rangle},
\label{eq:KB2}
\end{eqnarray}
where $d = - A^2 - A^{-2}$ and ${\bf K}$ is any diagram that
does not intersect the added loop. The graphical
representation of the states that we have defined in section
3 does satisfy these two equations (with $A=-1$) and
therefore all recoupling theory equations can be applied.

There is a subtle point that needs to be clarified here.
When tangle theoretical recoupling theory is defined on a
pane, it corresponds, via Penrose binor calculus to the
algebra of $\epsilon_{AB}$ and $\delta_{A}^{B}$ tensors
only. The states in the connection representation are of
course not given just by these tensors, because they contain
the parallel propagators. In Penrose binor calculus, these
tensors are represented as boxes.  Thus, tangle theoretical
identities can be applied only in between these boxes -- for
instance, a closed line cannot be replaced by a factor $d$,
as in \eref{eq:KB2}, if it includes a box.  However, in the
graphical representation of section 3, we have coded the
information of the presence of the parallel propagators in
the fact that a line runs along an edge.  A short reflection
will convince the reader that if one applies recoupling
theory to the tangles defined on the extended graph (which
is not a plane, but rather an oriented 2-d surface with a
complicated topology) that is, one defines tangles as
equivalence classes under smooth deformations {\it on the
extended graph}, then this is equivalent to respecting the
presence of the boxes -- for instance, a generic line will
not be contractible to a circle, and therefore equation
\eref{eq:KB2} can be applied only within the vertices, and
not for a loop wrapped around the extended graph. This is
why recoupling theory can be rigorously applied to tangles
defined on the extended graphs.

One of the consequences of equation (\ref{eq:KB2}) is that
any closed loop that can be shrink to a point within the
extended graph can be replaced by its loop value, which is
equal to $-2$. Indeed, any closed tangle that shrinks to a
point reduces to a number and the operation of computing
this number is usually referred as the chromatic evaluation.
The most useful evaluations are the following:

\noindent {(1)} The symmetrizer
\begin{equation}
\Delta_n = \begin{array}{c}\setlength{\unitlength}{1 pt}
           \begin{picture}(35,25)
           \put(15, 0){\line(0,1){10}}\put(20, 0){\line(0,1){10}}
           \put(15, 0){\line(1,0){ 5}}\put(15,10){\line(1,0){ 5}}
           \put(15,25){\line(1,0){5}}  \put(15,15){$n$}
           \put(15,15){\oval(30,20)[l]}\put(20,15){\oval(30,20)[r]}
           \end{picture}\end{array} 
     =  (-1)^n (n+1). 
\label{eq:valSIM}
\end{equation}
\noindent {(2)} The $\theta$ net
\begin{eqnarray} \fl
\theta(a,b,c) = 
\begin{array}{c}\setlength{\unitlength}{1 pt}\begin{picture}(40,40)
        \put(18,32){$a$}
        \put(18,17){$b$} 
        \put(18, 2){$c$} 
        \put(20,15){\oval(40,30)} \put( 0,15){\line(1,0){40}} 
        \put( 0,15){\circle*{3}}  \put(40,15){\circle*{3}}
\end{picture}\end{array}
= \frac{ (-1)^{m+n+p} (m+n+p+1)! ~m!~n!~p!}{
                         a! ~b! ~c!}
\label{eq:valTheta}
\end{eqnarray}
where $m=(a+b-c)/2$, $n=(b+c-a)/2$, $p=(c+a-b)/2$. 

\noindent {(3)} The Tetrahedral net
\begin{equation}
\fl {{\rm Tet}\left[\begin{array}{ccc} 
            A & B & E \\
            C & D & F  
  \end{array}\right]}
= \begin{array}{c}\setlength{\unitlength}{1 pt}\begin{picture}(55,40)
        \put( 0,15){\line(1,-1){15}} \put(0,22){${\scriptstyle B}$}
        \put( 0,15){\line(1, 1){15}} \put(0, 0){${\scriptstyle A}$}
        \put( 0,15){\circle*{3}}
        \put(30,15){\line(-1, 1){15}} \put(28,20){${\scriptstyle C}$}
        \put(30,15){\line(-1,-1){15}} \put(28, 2){${\scriptstyle D}$}
        \put(30,15){\circle*{3}}
        \put( 0,15){\line(1,0){30}} \put(12,16){${\scriptstyle F}$}
        \put(15,30){\line(1,0){25}} \put(15,30){\circle*{3}}
        \put(15, 0){\line(1,0){25}} \put(15, 0){\circle*{3}}
        \put(40, 0){\line(0,1){30}} \put(42,12){${\scriptstyle E}$}
  \end{picture}\end{array} 
= \frac{{\cal I}}{{\cal E}} \sum_{m\leq S \leq M}
 \frac{ (-1)^{S} (S+1)!}{\prod_i  ~(S-a_i)!~\prod_j ~(b_j-S)! }
~~,
\end{equation}
where
$$
{\displaystyle\begin{array}{rclcrcl} 
   a_1&=&{\displaystyle\frac{A+D+E}{2}},
   &\qquad&b_1&=&{\displaystyle\frac{B+D+E+F}{2}}, 
\\[2 mm]
   a_2&=&{\displaystyle \frac{B+C+E}{2}},
   &\qquad&b_2&=&{\displaystyle \frac{A+C+E+F}{2}}, 
\\[2 mm]
   a_3&=&{\displaystyle \frac{A+B+F}{2}},
   &\qquad&b_3&=&{\displaystyle \frac{A+B+C+D}{2}}, 
\\[2 mm]
   a_4&=&{\displaystyle \frac{C+D+F}{2}},&\qquad&     & &      
\end{array}}
$$
$$
{\displaystyle \begin{array}{rclcrcl} 
   m &=&{\rm max}\{ a_i \},   &&      
   M &=&{\rm min}\{ b_j \}, \\[2 mm]
   {\cal E} &=& A! B! C! D! E! F!,
 &&{\cal I} &=& \prod_{ij} (b_j-a_i)! 
~. 
\end{array}}
$$

The most important result of the theory is the recoupling theorem
which states:
\begin{equation}
\begin{array}{c}\setlength{\unitlength}{1 pt}
\begin{picture}(50,40)
          \put( 0,0){$a$}\put( 0,30){$b$}
          \put(45,0){$d$}\put(45,30){$c$}
          \put(10,10){\line(1,1){10}}\put(10,30){\line(1,-1){10}}
          \put(30,20){\line(1,1){10}}\put(30,20){\line(1,-1){10}}
          \put(20,20){\line(1,0){10}}\put(22,25){$j$}
          \put(20,20){\circle*{3}}\put(30,20){\circle*{3}}
\end{picture}\end{array}
    = \sum_i  \left\{\begin{array}{ccc}
                      a  & b & i \\
                      c  & d & j  
                \end{array}\right\}
\begin{array}{c}\setlength{\unitlength}{1 pt}
\begin{picture}(40,40)
      \put( 0,0){$a$}\put( 0,40){$b$}
      \put(35,0){$d$}\put(35,40){$c$}
      \put(10,10){\line(1,1){10}}\put(10,40){\line(1,-1){10}}
      \put(20,30){\line(1,1){10}}\put(20,20){\line(1,-1){10}}
      \put(20,20){\line(0,1){10}}\put(22,22){$i$}
      \put(20,20){\circle*{3}}\put(20,30){\circle*{3}}
\end{picture}\end{array}
\label{eq:recTheorem}
\end{equation}
where the quantities $\{ {}^{abi}_{cdj} \}$ are the $su(2)$
$6-j$ symbols (normalized as in \cite{Kauffman94}) and their
explicit values, in terms of chromatic evaluations, are: 
\begin{equation}
 \left\{\begin{array}{ccc}
      a  & b & i \\
      c  & d & j  
 \end{array}\right\}
= \frac{ {\displaystyle \Delta_i ~ {\rm Tet}\left[\begin{array}{ccc}
                      a  & b & i \\
                      c  & d & j  
              \end{array}\right] }
}{\theta(a,d,i) \theta(b,c,i) }
~. 
\end{equation}

From the previous formulae and definitions other relations
can be easily derived. In particular, 
exchanging the ordering of the three line in a 
3-vertex yields a sign factor:
\begin{equation}
\begin{array}{c}
\setlength{\unitlength}{1 pt}
\begin{picture}(40,50)
       \put(30,25){\line(-3, 2){30}} \put( 0,35){$a$}
       \put( 0,25){\line( 3, 2){13}} 
       \put(30,45){\line(-3,-2){13}} \put(30,35){$b$}
       \put(15,25){\oval(30,20)[b]}  \put(15,15){\circle*{3}}
       \put(15, 5){\line(0,1){10}}   \put(17,1){$c$}
\end{picture}\end{array} 
      = \lambda^{ab}_c 
\begin{array}{c}\setlength{\unitlength}{1 pt}
\begin{picture}(40,40)
       \put(15,15){\line(-1, 1){10}} \put( 4,27){$a$}
       \put(15,15){\line( 1, 1){10}} \put(22,27){$b$}
       \put(15, 5){\line(0,1){10}}   \put(17,1){$c$}
       \put(15,15){\circle*{3}}
\end{picture}\end{array}
      \end{equation}
where $\lambda^{ab}_c = (-1)^{(a+b-c)/2} ~(-1)^{(a'+b'-c')/2}$,
and $x'=x(x+2)$;
\noindent The following reduction holds
\begin{eqnarray}
\label{eq:TETred}
\begin{array}{c}
  \setlength{\unitlength}{1 pt}
  \begin{picture}(40,40)  
  \put(10, 5){\line(1,0){10}}   \put(2, 5){$i$}
  \put(10,20){\line(1,0){10}}   \put(2,20){$a$}
  \put(10,35){\line(1,0){10}}   \put(2,35){$k$}
  \put(20, 5){\circle*{3}}      
  \put(20,20){\circle*{3}}      
  \put(20,35){\circle*{3}}      
  \put(20, 5){\line(0,1){15}}   \put(22,10){$\bar{c}$}
  \put(20,20){\line(0,1){15}}   \put(22,25){$c$}
  \put(20,20){\oval(20,30)[r]}  \put(32,25){$d$}
\end{picture}\end{array}  
&=& \frac{\begin{array}{c}\setlength{\unitlength}{.8 pt}
        \begin{picture}(55,40)
        \put( 0,15){\line(1,-1){15}} \put(0,22){${\scriptstyle k}$}
        \put( 0,15){\line(1, 1){15}} \put(0, 0){${\scriptstyle i}$}
        \put( 0,15){\circle*{3}}
        \put(30,15){\line(-1, 1){15}} \put(28,20){${\scriptstyle c}$}
        \put(30,15){\line(-1,-1){15}} \put(28, 2){${\scriptstyle \bar{c}}$}
        \put(30,15){\circle*{3}}
        \put( 0,15){\line(1,0){30}} \put(12,16){${\scriptstyle a}$}
        \put(15,30){\line(1,0){25}} \put(15,30){\circle*{3}}
        \put(15, 0){\line(1,0){25}} \put(15, 0){\circle*{3}}
        \put(40, 0){\line(0,1){30}} \put(42,12){${\scriptstyle d}$}
        \end{picture}\end{array}
    }{ \begin{array}{c}\setlength{\unitlength}{.5 pt}\begin{picture}(40,40)
            \put(18,32){$\scriptstyle k$}
            \put(18,17){$\scriptstyle a$} 
            \put(18, 2){$\scriptstyle i$} 
            \put(20,15){\oval(40,30)} \put( 0,15){\line(1,0){40}} 
            \put( 0,15){\circle*{3}}  \put(40,15){\circle*{3}}
           \end{picture}\end{array}
      } 
~\cdot~
\begin{array}{c}
  \setlength{\unitlength}{1 pt}
  \begin{picture}(20,40)  
  \put(2, 5){$i$}
  \put(10,20){\line(1,0){15}}   \put(2,20){$a$}
  \put(2,35){$k$}
  \put(25,20){\circle*{3}}      
  \put(10,20){\oval(30,30)[r]} 
\end{picture}\end{array}  
\end{eqnarray}

\noindent Another key formula that can be derived is the one
we use for evaluating the action of the parallel propagator
operator on spin network states: 
\begin{equation}\label{eq:Recursion1}
 \begin{array}{c}\setlength{\unitlength}{1 pt}
 \begin{picture}(20,25)
       \put( 5,0){\line(0,1){25}}\put( 7,17){$\scriptstyle n$}
       \put(15,0){\line(0,1){25}}\put(17,17){$\scriptstyle 1$}
 \end{picture}\end{array}
= \sum_{\epsilon=\pm1} a_\epsilon(n)
 \begin{array}{c}\setlength{\unitlength}{1 pt}
 \begin{picture}(20,30)
       \put( 5, 0){\line(0,1){10}}\put( 7, 2){$\scriptstyle n$}
       \put(15, 0){\line(0,1){10}}\put(17, 2){$\scriptstyle 1$}
       \put( 5,10){\line(0,1){10}}\put( 7,12){$\scriptstyle n+\epsilon$}
       \put( 5,10){\circle*{3}}\put( 5,10){\line(1,0){10}}
       \put( 5,20){\circle*{3}}\put( 5,20){\line(1,0){10}}
       \put( 5,20){\line(0,1){10}}\put( 7,22){$\scriptstyle n$}
       \put(15,20){\line(0,1){10}}\put(17,22){$\scriptstyle 1$}
 \end{picture}\end{array}
=\begin{array}{c}\setlength{\unitlength}{1 pt}
 \begin{picture}(20,30)
       \put( 5, 0){\line(0,1){10}}\put( 7, 2){$\scriptstyle n$}
       \put(15, 0){\line(0,1){10}}\put(17, 2){$\scriptstyle 1$}
       \put( 5,10){\line(0,1){10}}\put( 7,12){$\scriptstyle n+1$}
       \put( 5,10){\circle*{3}}\put( 5,10){\line(1,0){10}}
       \put( 5,20){\circle*{3}}\put( 5,20){\line(1,0){10}}
       \put( 5,20){\line(0,1){10}}\put( 7,22){$\scriptstyle n$}
       \put(15,20){\line(0,1){10}}\put(17,22){$\scriptstyle 1$}
 \end{picture}\end{array}
+ \frac{\Delta_{n-1}}{\Delta_n}
 \begin{array}{c}\setlength{\unitlength}{1 pt}
 \begin{picture}(20,25)
       \put( 5, 0){\line(0,1){10}}\put( 7, 2){$\scriptstyle n$}
       \put(15, 0){\line(0,1){10}}\put(17, 2){$\scriptstyle 1$}
       \put( 5,10){\line(0,1){10}}\put( 7,12){$\scriptstyle n-1$}
       \put( 5,10){\circle*{3}}\put( 5,10){\line(1,0){10}}
       \put( 5,20){\circle*{3}}\put( 5,20){\line(1,0){10}}
       \put( 5,20){\line(0,1){10}}\put( 7,22){$\scriptstyle n$}
       \put(15,20){\line(0,1){10}}\put(17,22){$\scriptstyle 1$}
 \end{picture}\end{array}.
\end{equation}
Also
\begin{equation}\label{eq:MOVE2}
r ~\cdot~
\begin{array}{c}
  \setlength{\unitlength}{1 pt}
  \begin{picture}(30,40)
  \put(15,20){\line(0,-1){10}}
  \put(15,20){\line( 1,1){10}}
  \put(15,20){\line(-1,1){10}}
  \put(15,20){\circle*{3}}
  \put( 3,32){${\scriptstyle p}$}
  \put(23,32){${\scriptstyle q}$}
  \put(13, 2){${\scriptstyle r}$}
  \put(15,15){\line(-1,0){7}}
  \put(15,15){\circle*{3}}
  \put( 6,16){${\scriptstyle 2}$}
\end{picture}\end{array}  
= p~\cdot~
\begin{array}{c}
  \setlength{\unitlength}{1 pt}
  \begin{picture}(30,40)
  \put(15,20){\line(0,-1){10}}
  \put(15,20){\line( 1,1){10}}
  \put(15,20){\line(-1,1){10}}
  \put(15,20){\circle*{3}}
  \put( 7,32){${\scriptstyle p}$}
  \put(23,32){${\scriptstyle q}$}
  \put(13, 2){${\scriptstyle r}$}
  \put(10,25){\line(-1,0){7}}
  \put(10,25){\circle*{3}}
  \put( 1,26){${\scriptstyle 2}$}
\end{picture}\end{array}  
+ q~\cdot~
\begin{array}{c}
  \setlength{\unitlength}{1 pt}
  \begin{picture}(30,40)
  \put(15,20){\line(0,-1){10}}
  \put(15,20){\line( 1,1){10}}
  \put(15,20){\line(-1,1){10}}
  \put(15,20){\circle*{3}}
  \put( 3,32){${\scriptstyle p}$}
  \put(23,32){${\scriptstyle q}$}
  \put(13, 2){${\scriptstyle r}$}
  \put(20,25){\line(-1,0){7}}
  \put(20,25){\circle*{3}}
  \put(11,26){${\scriptstyle 2}$}
\end{picture}\end{array}  
\end{equation}

\noindent Finally, using equations (C.6) and (C.7), we have 
the chromatic evaluations ($\epsilon=0,\pm 1$)
\begin{equation}\label{eq:EV1}
(t+\epsilon)~\cdot~ \frac{
  \begin{array}{c}\setlength{\unitlength}{1 pt}
  \begin{picture}(55,40)
  \put( 0,15){\line(1,-1){15}} \put(-8,22){${\scriptstyle \!t+\epsilon}$}
  \put( 0,15){\line(1, 1){15}} \put(-8, 0){${\scriptstyle \!t-\epsilon}$}
  \put( 0,15){\circle*{3}}
  \put(30,15){\line(-1, 1){15}} \put(28,20){${\scriptstyle 2}$}
  \put(30,15){\line(-1,-1){15}} \put(28, 2){${\scriptstyle 2}$}
  \put(30,15){\circle*{3}}
  \put( 0,15){\line(1,0){30}} \put(12,16){${\scriptstyle 2}$}
  \put(15,30){\line(1,0){25}} \put(15,30){\circle*{3}}
  \put(15, 0){\line(1,0){25}} \put(15, 0){\circle*{3}}
  \put(40, 0){\line(0,1){30}} \put(42,12){${\scriptstyle t+\epsilon}$}
  \end{picture}\end{array} }{
  \begin{array}{c}\setlength{\unitlength}{.8 pt}\begin{picture}(40,40)
      \put(18,32){$\scriptstyle 2$}
      \put(18,17){$\scriptstyle t+\epsilon$} 
      \put(18, 2){$\scriptstyle t-\epsilon$} 
      \put(20,15){\oval(40,30)} \put( 0,15){\line(1,0){40}} 
      \put( 0,15){\circle*{3}}  \put(40,15){\circle*{3}}
   \end{picture}\end{array}
   }
= -1 - \epsilon~ \bigg[\frac{t+1}{2}\bigg]
\end{equation}
\begin{equation} \label{eq:EV2}
b~\cdot~
  \begin{array}{c}\setlength{\unitlength}{1 pt}
  \begin{picture}(55,40)
  \put( 0,15){\line(1,-1){15}} \put(0,22){$\!\!\!{\scriptstyle t+1}$}
  \put( 0,15){\line(1, 1){15}} \put(0, 0){$\!\!\!{\scriptstyle t-1}$}
  \put( 0,15){\circle*{3}}
  \put(30,15){\line(-1, 1){15}} \put(28,20){${\scriptstyle b}$}
  \put(30,15){\line(-1,-1){15}} \put(28, 2){${\scriptstyle b}$}
  \put(30,15){\circle*{3}}
  \put( 0,15){\line(1,0){30}} \put(12,16){${\scriptstyle 2}$}
  \put(15,30){\line(1,0){25}} \put(15,30){\circle*{3}}
  \put(15, 0){\line(1,0){25}} \put(15, 0){\circle*{3}}
  \put(40, 0){\line(0,1){30}} \put(42,12){${\scriptstyle a}$}
  \end{picture}\end{array}
= \frac{1+a+b-t}{2} ~\cdot~
  \begin{array}{c}\setlength{\unitlength}{.8 pt}\begin{picture}(40,40)
      \put(18,32){$\scriptstyle a$}
      \put(18,17){$\scriptstyle b$} 
      \put(18, 2){$\! \scriptstyle t+1$} 
      \put(20,15){\oval(40,30)} \put( 0,15){\line(1,0){40}} 
      \put( 0,15){\circle*{3}}  \put(40,15){\circle*{3}}
  \end{picture}\end{array}
\end{equation}
\begin{equation}\label{eq:EV3}
\frac{
  \begin{array}{c}\setlength{\unitlength}{.8 pt}\begin{picture}(40,40)
      \put(18,32){$\scriptstyle a$}
      \put(18,17){$\scriptstyle b$} 
      \put(18, 2){$\! \scriptstyle t+1$} 
      \put(20,15){\oval(40,30)} \put( 0,15){\line(1,0){40}} 
      \put( 0,15){\circle*{3}}  \put(40,15){\circle*{3}}
  \end{picture}\end{array}
}{  \begin{array}{c}\setlength{\unitlength}{.8 pt}\begin{picture}(40,40)
      \put(18,32){$\scriptstyle a$}
      \put(18,17){$\scriptstyle b$} 
      \put(18, 2){$\! \scriptstyle t-1$} 
      \put(20,15){\oval(40,30)} \put( 0,15){\line(1,0){40}} 
      \put( 0,15){\circle*{3}}  \put(40,15){\circle*{3}}
  \end{picture}\end{array}
} = - \frac{(3+a+b+t)(1+a+t-b)(1+b+t-a)}{4 t (t+1)(1+a+b-t)}
\end{equation}
These formulas are sufficient for the computations in the paper.


\section*{References}\par

\end{document}